\documentclass[preprints,article,accept,pdftex,moreauthors]{mdpi}

\firstpage{1}
\pubvolume{1}
\issuenum{1}
\articlenumber{0}
\pubyear{2023}
\copyrightyear{2023}
\externaleditor{Academic Editor: Alexander Kramida}
\datereceived{6 February 2023}
\daterevised{14 March 2023} 
\dateaccepted{23 March 2023}
\datepublished{}
\hreflink{https://doi.org/} 

\usepackage{fancyvrb}
\newcommand{\feii}{Fe\,{\sc ii}}
\newcommand{\feiii}{Fe\,{\sc iii}}

\Title{Atomic Data Assessment with PyNeb: Radiative and Electron Impact Excitation Rates for [\feii] and [\feiii]}

\TitleCitation{Atomic Data Assessment with PyNeb: Radiative and Electron Impact Excitation Rates for [\feii] and [\feiii]}

\Author{Claudio Mendoza $^{1,}$*\orcidA{}, Jos\'e E. M\'endez-Delgado $^{2}$\orcidB{}, Manuel 
 Bautista $^{3}$\orcidC{}, Jorge Garc\'ia-Rojas $^{4,5}$\orcidD{} \linebreak and Christophe Morisset $^{6}$\orcidE{}}

\AuthorNames{Claudio Mendoza, Jos\'e E. M\'endez-Delgado, Manuel Bautista, Jorge Garc\'ia-Rojas and Christophe Morisset}

\AuthorCitation{Mendoza, C.; M\'endez-Delgado, J.~E.; Bautista, M.; Garc\'ia-Rojas, J.; Morisset, C.}

\address{%
$^{1}$ \quad Physics Center, Venezuelan Institute for Scientific Research (IVIC), Caracas 1020, Venezuela\\
$^{2}$ \quad Astronomisches Rechen-Institut, Zentrum f\"ur Astronomie der Universit\"at Heidelberg, D-69120 Heidelberg, Germany; jemd@uni-heidelberg.de\\
$^{3}$ \quad Department of Physics, Western Michigan University, Kalamazoo, MI 49008, USA; manuel.bautista@wmich.edu\\
$^{4}$ \quad Instituto de Astrof\'isica de Canarias (IAC), E-38205 La Laguna, Spain; astroyorch@gmail.com\\
$^{5}$ \quad Departamento de Astrof\'isica, Universidad de La Laguna, E-38206 La Laguna, Spain\\
$^{6}$ \quad Instituto de Astronom\'ia, Universidad Nacional Aut\'onoma de M\'exico, Ensenada 22860, Mexico; chris.morisset@gmail.com
}

\corres{Correspondence: claudiom07@gmail.com; Tel.: +1-269-370-8465}

\abstract{We use the \texttt{PyNeb} 1.1.16 Python package to evaluate the atomic datasets available for the spectral modeling of [\feii] and [\feiii], which list level energies, $A$-values, and effective collision strengths. Most datasets are reconstructed from the sources, and new ones are incorporated to be compared with observed and measured benchmarks. For [\feiii], we arrive at conclusive results that allow us to select the default datasets, while for [\feii], the conspicuous temperature dependency on the collisional data becomes a deterrent. This dependency is mainly due to the singularly low critical density of the $\mathrm{3d^7\ a\,^4F_{9/2}}$ metastable level that strongly depends on both the radiative and collisional data, although the level populating by fluorescence pumping from the stellar continuum cannot be ruled out. A new version of \texttt{PyNeb} (1.1.17) is released containing the evaluated datasets.}


\begin{document}

\section{Introduction}

\texttt{PyNeb}\endnote{\url{http://research.iac.es/proyecto/PyNeb/}}~\cite{lur12,lur15} is a widely used Python tool for the analysis of emission lines in nebular plasmas to determine the temperature, density, and~chemical abundances~\cite{pei17}. It essentially solves the equations of statistical equilibrium by addressing an extensive atomic database of radiative and collisional rates; therefore, {\tt PyNeb}'s diagnostic potential relies directly on the accuracy of this database. In~this respect, its data curation strategy is based on the selection of recommended default datasets and keeping access to a historical archive of radiative and collisional files rather than discarding them. This singular scheme allows modelers to estimate the uncertainties of the plasma diagnostics and chemical abundances due to the scatter of the atomic parameters. It also provides a versatile environment for atomic data assessment, which has been previously used productively to benchmark the accuracy of the radiative rates in N- and P-like ions and effective collision strengths in C-like ions~\cite{mor20} and to determine the atomic-data impact on density diagnostics~\cite{jua21}.

As a follow-up to~\cite{mor20}, we apply {\tt PyNeb}'s capabilities to assess the radiative and collisional rates associated with the transition arrays of [\feii] and [\feiii]. These ions give rise to forbidden lines in the optical and infrared spectra of H\,{\sc ii} regions, planetary nebulae, and~quasars that are used to determine plasma conditions and iron abundance, the~latter particularly in the context of depletion factors in the life cycle of dust grains~\cite{per99,del09, del16}. These iron transition arrays are also useful for studying the shocked plasma caused by outflows from newborn stars that form bright Herbig--Haro (HH) objects perturbing stellar accretion~\cite{mes09, men21a, men21b}. The~high-resolution spectra of such HH objects---e.g., HH\,202S~\cite{mes09} and HH\,204~\cite{men21b} in the Orion Nebula---taken with the Ultraviolet Visual Echelle Spectrograph (range 3100--10,400~\AA)~\cite{DOdorico:2000} present an observational opportunity to benchmark the \feii\ and \feiii\ atomic data. The~optical emission spectra of these two HH objects are dominated by photoionization from $\theta^1$ Ori C, the~main ionization source of the Orion Nebula, and~can therefore be treated as small-scale H\,{\sc ii} regions~\cite{men21a}. 

Due to the astrophysical importance of \feii, the~computation of radiative and collisional rates for the spectral modeling of this system has received considerable attention since the early semi-empirical computations of $A$-values ({radiative transition probabilities}) for the forbidden~\cite{gar62} and dipole-allowed~\cite{kur75} transitions and electron-impact collision strengths with the distorted wave method~\cite{nus80, nus81}. Our intention here is not to evaluate this vast body of data but to concentrate on complete and accurate datasets to model the spectra of [\feii] and [\feiii] observed in nebular~plasmas.

A landmark systematic effort to compute radiative and collisional data for the fine-structure spectral modeling of Fe ions was undertaken by the Iron Project\endnote{\url{http://cdsweb.u-strasbg.fr/topbase/testop/TheIP.html}} using the $R$-matrix method~\cite{hum93} and relativistic multi-configuration structure codes such as \textsc{superstructure}~\cite{eis74}, \textsc{civ3}~\cite{hib75}, and~\textsc{hfr} ~\cite{cow81}. For~\feii\ and  \feiii, $A$-values were computed for both allowed and forbidden lines~\cite{nah95, nah96, qui96b} and electron-impact effective collision strengths (ECS) for infrared, optical, and~ultraviolet transitions~\cite{zha95, zha96, bau96}. Although~these $R$-matrix calculations were originally performed in $LS$ coupling followed by algebraic transformations to intermediate coupling, more recent scattering calculations for these ionic species~\cite{bau10, bad14, bau15, tay18, smy19} have relied on $R$-matrix implementations based on the Breit--Pauli Hamiltonian (\textsc{bprm}~\cite{ber95}, \textsc{bsr}~\cite{zat06}) and Dirac--Fock Hamiltonian (\textsc{darc}~\cite{ait96}). In~some cases, e.g.,~the collisional excitation of levels within the Fe$^+$ ground configuration, low-temperature ECS have been computed with several $R$-matrix implementations---intermediate coupling frame transformation (\textsc{icft},~\cite{grif98}), \textsc{bprm}, and~\textsc{darc}---to obtain uncertainty estimates~\cite{wan19}. 

Extensive multi-configuration Breit--Pauli (\textsc{mcbp}) calculations of $gf$-values ({weighted oscillator strengths}) and $A$-values for allowed and forbidden transitions in  \feii\ and  \feiii\ 
\cite{deb08, deb09a, deb09b, deb10a, deb10b, deb11, deb14}  with the code \textsc{civ3} have shown that fine-tuning corrections to the Hamiltonian matrix by fitting to the experimental level energies lead to more accurate configuration expansions. However, significant $A$-value differences (greater than 20--30\%) are found for some transitions computed with comparable multi-configuration methods that also implement fine-tuning corrections (e.g.,~\cite{qui96a, qui96b, fiv16}), which are caused by strong configuration mixing, double-excitation representations, and~subtle core--valence correlation involving the closed 3p$^6$ and open 3d$^n$ subshells. 
 
Since the \feii\ radiative data display a large scatter, experimentalists and astronomical observers have been enticed to measure accurate radiative parameters such as lifetimes and branching fractions to determine $A$- and $gf$-values; however, most of them address allowed transitions involving odd-parity states. Lifetimes have been measured for 21 \feii\ levels with a time-resolved nonlinear laser-induced fluorescence technique to an accuracy of 2--3\%, and~by means of branching fractions, absolute $A$-values were reported to be within 6\% and 26\% for the strong and weak transitions, respectively~\cite{sch04}. Accurate $gf$-values for 142 \feii\ lines were determined from laboratory and solar spectroscopic multiplets and theoretical fine-structure ratios, assuring that oscillator-strength error no longer hinders stellar abundance studies~\cite{mel09}. Branching fractions have been measured for 121 ultraviolet \feii\ lines (2250--3280~\AA), and~through published lifetimes, $gf$-values have been derived and applied to iron abundance determinations in the Sun and the metal-poor star HD~84937~\cite{den19}. Good agreement is obtained with the Fe standard solar abundance. On~the other hand, the~only lifetime measurements of [\feii] even-parity metastable states to derive transition probabilities have been carried out by the Ferrum Project using a laser probing technique on a stored ion beam, the~branching fractions being derived from astrophysical spectra~\cite{ros01, har03, gur09}. Consistent differences between the observed intensity ratios and theoretical branching fractions have suggested that the [\feii] lines might be used to determine~reddening. 

In the present data assessment, we revise the \feii\ and \feiii\ atomic datasets---namely, energy levels, $A$-values, and~ECS---available in \texttt{PyNeb}~1.1.16. In most cases, we reconstruct the atomic models from the sources by mapping the rates to the NIST level structure. We renormalize the number of levels to a predefined value to enable one-to-one comparisons among the theoretical datasets and benchmarks with the HH\,202S and HH\,204 spectra and the aforementioned experimental data. We also extend the \texttt{PyNeb} database with additional new datasets, the~main objective being the default selection for a new version: \texttt{PyNeb}~1.1.17. 

We give a brief description of \texttt{PyNeb}'s file structure and methods in Section~\ref{pyneb}, and~carry out the data assessments and discussions of \feiii\ and \feii\ in Sections~\ref{feiii} and \ref{feii} followed by the conclusions in Section~\ref{conc}.
\section{PyNeb}\label{pyneb}

A brief description of the \texttt{PyNeb} package is given in~\cite{mor20}, in particular of its object-oriented architecture. 
The ion object is thereby represented by the \texttt{Atom} class, offering a set of methods through the \texttt{atomicData} manager. Regarding energy levels, $A$-values, and~ECS, {\tt PyNeb} has three file~types:
\begin{itemize}
    \item $xxx\_$\texttt{levels.dat}: lists NIST energy levels for ionic species $xxx$;
    \item $xxx\_$\texttt{atom}\_\textit{ref}.\texttt{dat}: lists $A$-values for transitions between energy levels of ion $xxx$ contained in the \textit{ref} dataset;
    \item $xxx\_$\texttt{coll}\_\textit{ref}.\texttt{dat}: lists ECS for transitions between energy levels of ion $xxx$ contained in the \textit{ref} dataset.
\end{itemize}

A printout of the NIST energy levels for a specific ion, \feiii\ say, may be obtained with the function \texttt{getLevelsNIST()}:
\begin{Verbatim}[frame=single]
# Call thePyNeb Python module
import pyneb as pn
# List NIST energy levels for Fe III
levels = pn.getLevelsNIST('Fe3')
print(levels)
\end{Verbatim}
and a list of all the \feiii\ \texttt{atom} and \texttt{coll} files is currently available with: 
\begin{Verbatim}[frame=single]
pn.atomicData.getAllAvailableFiles('Fe3')
\end{Verbatim}
\begin{Verbatim}
['* fe_iii_atom_Q96_J00.dat',
 '* fe_iii_coll_Z96.dat',
 'fe_iii_atom_BBQ10.dat',
 'fe_iii_atom_NP96.dat',
 'fe_iii_coll_BB14.dat',
 'fe_iii_coll_BBQ10.dat']
\end{Verbatim}
where the recommended default files are prefixed with an asterisk. Alternative files may be easily accessed by the user to replace the default files with the function
\begin{Verbatim}[frame=single]
pn.atomicData.setDataFile('fe_iii_atom_BBQ10.dat')
pn.atomicData.setDataFile('fe_iii_coll_BBQ10.dat')
\end{Verbatim}

We can now instantiate the \feiii\ ion with the new atomic datasets
\begin{Verbatim}[frame=single]
pn.atomicData.getDataFile('Fe3')
Fe3=pn.Atom('Fe',3,NLevels=34)
\end{Verbatim}
and apply a series of methods to determine some of its plasma properties:

\begin{Verbatim}[frame=single]
# Emissivity for transition 2-1 at T = 1.0e4 K and n_e = 1.0e4 cm-3
Fe3.getEmissivity(tem=1.0e4,den=1.0e4,lev_i=2,lev_j=1)
# Emissivity for transition with wavelength 4658.17 A at T = 1.0e4 K 
# and n_e = 1.0e4 cm-3
Fe3.getEmissivity(tem=1.0e4,den=1.0e4,wave=4658.17)
# Level critical densities at T = 1.0e4 K 
Fe3.getCritDensity(1.0e4)
# Level populations at T = 1.0e4 K and n_e = 1.0e4 cm-3
Fe3.getPopulations(tem=1.0e4, den=1.0e4)
\end{Verbatim}
\noindent
In the ion instantiating command above, the~option \texttt{NLevels} specifies the number of levels of the atomic model, which can be the total (default) or a reduced~set. 

A complete list of methods is included in the \texttt{PyNeb} manual. The~\texttt{Atom} class also addresses the ion atomic properties, e.g.,~the ECS at a selected temperature
\begin{Verbatim}[frame=single]
# Collision strength for transition 2-1 at T = 1.0e4 K 
Fe3.getOmega(1.0e4,2,1)
\end{Verbatim}
\noindent
This function is useful as it returns the interpolated value at the prescribed input temperature allowing comparisons between different ECS datasets with tabulations on diverse temperature ranges. The~original temperature and ECS arrays may be listed with the~commands

\begin{Verbatim}[frame=single]
# Temperature and collision strength original arrays 
Fe3.getTemArray()
Fe3.getOmegaArray()
\end{Verbatim}

\section{\feiii \label{feiii}}

\subsection{\feiii\ Atomic Datasets in~PyNeb \label{pyneb_data_fe3}}

\feiii\ atomic models in {\tt PyNeb}~1.1.16 comprise 34 fine-structure levels (energy $E<60,000$\,cm$^{-1}$) from the $\mathrm{3d^6}$ and $\mathrm{3d^54s}$ electron configurations, whose attributes were obtained from the NIST server (April 2017). Levels $\mathrm{3d^6}$ $\mathrm{b\,^1D_2}$ and $\mathrm{b\,^1S_0}$ at $E>77,000$\,cm$^{-1}$ are therefore not taken into account, and~from the $\mathrm{3d^54s}$ configuration, only the lowly lying $\mathrm{^7S_3}$ and $\mathrm{^5S_2}$ are included. Transitions among the levels of these two even-parity configurations are electric dipole (E1) forbidden and~thus occur via electric quadrupole (E2) and magnetic dipole (M1) operators. {Shortcomings in the spectral fits resulting from the number of levels in the atomic models, 34 levels in this case, will be examined by comparing with extended models of 144 levels (see Section~\ref{newdata_fe3}}).

Two fundamental interactions in the computation of accurate radiative ($A$-values) and collisional (ECS) rates for atomic ions are electron correlation (configuration interaction, CI) and relativistic coupling. We give below brief descriptions of the numerical methods and approximations that were used to compute the datasets available in {\tt PyNeb}~1.1.16.

\begin{description}[style=unboxed,leftmargin=0cm]

\item[atom\_NP96]---Contains $A$-values for E2 and M1 transitions computed in an extensive CI framework with the multi-configuration Breit--Pauli  ({\sc mcbp})  code {\sc superstructure}~\cite{nah96}.

\item[coll\_Z96]---ECS were calculated in an 83-term non-relativistic $R$-matrix calculation including the $\mathrm{3d^6}$, $\mathrm{3d^54s}$, and~$\mathrm{3d^54p}$ configurations. ECS for 219 fine-structure levels were then obtained through algebraic recoupling~\cite{zha96}.

\item[atom\_Q96\_J00]---The Pauli Hartree--Fock {\sc hfr} code was used to compute $A$-values for the radiative transitions within the $\mathrm{3d^6}$ configuration~\cite{qui96a}. Wave functions were generated with CI expansions adjusting the electrostatic and spin--orbit integrals to fit the spectroscopic level energies. $A$-values for the $\mathrm{3d^54s\ ^7S_3}$ and $\mathrm{^5S_2}$ levels were obtained independently with {\sc hfr} using empirically adjusted Slater parameters~\cite{joh00}.

\item[atom\_BBQ10, coll\_BBQ10]---A 36-configuration CI expansion including pseudo-orbitals ($\overline{4f}$, $\overline{5s}$, $\overline{5p}$, and~$\overline{5d}$) was used to compute $A$-values with the {\sc mcbp} {\sc autostructure} atomic structure code~\cite{bau10, bad11}. Term-energy corrections were introduced to fine-tune the wave functions. ECS were computed with the Dirac--Coulomb $R$-matrix package ({\sc darc}) based on a 3-configuration target constructed with the {\sc grasp92} multi-configuration Dirac--Hartree--Fock structure code~\cite{par96} under the extended average level~approximation.

\item[coll\_BB14]---ECS were computed with the \textsc{icft} $R$-matrix method using a 136-term (322 fine-structure levels) target representation~\cite{bad14}. This {\sc mcbp} 3-configuration target was generated with {\sc autostructure} using a Thomas--Fermi--Dirac--Amaldi model potential.
\end{description}

\subsection{\feiii\ Revised and New Atomic~Datasets \label{newdata_fe3}}

The present work intends to revise the datasets currently available in {\tt PyNeb}, incorporate new ones, and~implement procedures for data evaluation. For~such purposes, we have collected the data from the original sources, verified consistency, and~implemented new \texttt{atom} and \texttt{coll} files. To~begin with, we have introduced a new level file, {\tt fe\_iii\_levels.dat}, with~updated spectroscopic measurements (improved accuracy and new levels such as $\mathrm{3d^6\ b\,^1D_2}$) from the NIST database (July 2022)~\cite{kra21}. We list below the revised and new atomic datasets with brief descriptions; the suffix ``$\_n$'' is now appended to the dataset label, where $n$ is the number of levels in the atomic model. For~data comparisons, we have normalized all \feiii\ datasets  to 34 levels, but~to estimate model convergence, we have extended some models to 144~levels.

An important issue is how to deal with missing energy levels in both the theoretical and spectroscopic reports since the {\tt PyNeb} energy-level files are based on the NIST database, which is incomplete at the higher energies. In~spectrum modeling, levels with zero decay rates may lead to numerical problems; we have, therefore, limited the number of levels in the atomic models to those that have been both computed and measured. From~the user's point of view, modeling turnaround time is also an important issue; therefore, the~number of levels in the atomic models must take this point into~consideration.

\begin{description}[style=unboxed,leftmargin=0cm]
\item[atom\_Q96\_34]---The main differences with the atom\_Q96\_J00 dataset described in \linebreak Section~\ref{pyneb_data_fe3} are the $A$-values for transitions involving the $\mathrm{3d^54s\ ^7S_3}$ and $\mathrm{^5S_2}$ levels. The~radiative rates in atom\_Q96\_J00 for transitions decaying from the $\mathrm{^7S_3}$ upper level are set to $A=0.0$, while those from $\mathrm{^5S_2}$ were calculated with {\sc hfr} with empirically adjusted Slater parameters~\cite{joh00}. In~this dataset, radiative rates for transitions involving these levels have been updated with $A$-values from atom\_BB14\_34. This choice is to a certain extent arbitrary as inferred from the wide $A$-value scatter shown for the transitions $\mathrm{^5S_2}-\mathrm{^5D}_J$ in Table~\ref{3d54s}. We include in this comparison $A$-values computed with extensive CI using both {\sc hfr} and {\sc autostructure} (atom\_FBQh16 and atom\_FBQa16)~\cite{fiv16}; however, these datasets lack the completeness required to model collisionally excited nebular lines and, consequently, will not be further considered in the present data~assessment.
\end{description}

\begin{table}[H]
\caption{$A$-value comparison for the $\mathrm{3d^54s\ ^5S_2}-\mathrm{3d^6\ ^5D}_J$ transitions in [\feiii]. $A$-values in J00 were calculated with the semi-relativistic {\sc hfr} code using empirical Slater parameters~\cite{joh00} and included in the atom\_Q96\_J00 dataset (see Section~\ref{pyneb_data_fe3}). The~methods used to compute the radiative data in atom\_BBQ10\_34 and atom\_BB14\_34 are described below. $A$-values in atom\_FBQh16 and atom\_FBQa16 were, respectively, computed with {\sc hfr} and {\sc autostructure} with extensive CI~\cite{fiv16}. \label{3d54s}}
\newcolumntype{C}{>{\centering\arraybackslash}X}
\begin{tabularx}{\textwidth}{CCCCCCC}
\toprule
\boldmath{$J$} 
& \boldmath{$\lambda_\mathrm{air}$} \textbf{(\AA)} 
& \multicolumn{5}{c}{\boldmath{$A$}\textbf{-Value (s}\boldmath{$^{-1}$}\textbf{)}}\\
\midrule
&  & \textbf{J00} & \textbf{BBQ10\_34} & \textbf{BB14\_34} & \textbf{FBQh16} &\textbf{FBQa16} \\
\midrule
4 & 2438.28 & 22.2 & 25.7 & 38.4 & 32.0 & 31.6 \\
3 & 2464.48 & 16.0 & 18.7 & 28.0 & 23.3 & 23.1 \\
2 & 2483.01 & 11.0 & 12.8 & 19.1 & 15.9 & 15.8 \\
1 & 2495.01 & 6.4  & 7.43 & 11.1 & 9.29 & 9.22 \\
0 & 2500.93 & 2.0  & 2.44 & 3.66 &      &      \\
\bottomrule
\end{tabularx}
\end{table}

\begin{description}[style=unboxed,leftmargin=0cm]
\item[coll\_Z96\_34]---ECS listed in the supplementary data associated with the original publication~\cite{zha96} were downloaded directly from the CDS to build a new file (see Section~\ref{pyneb_data_fe3} for details of the calculation).

\item[coll\_Z96\_144]---The coll\_Z96\_34 dataset was extended to include 144 levels from the $\mathrm{3d^6}$, $\mathrm{3d^54s}$, and~$\mathrm{3d^54p}$ configurations with $E<121,500$\,cm$^{-1}$. For~higher energies, inconsistencies in the reported list of measured and computed levels begin to appear. Although~this model extension is not expected to contribute to the collisionally excited nebular lines, it may be relevant if the populations of the $\mathrm{3d^54p}$ levels generate E1~arrays.

\item[atom\_DH09\_34]---$A$-values for the E2 and M1 transitions between levels of the $\mathrm{3d^6}$ configuration have been computed with the {\sc mcbp}  code {\sc civ3}~\cite{hib75,hib91} in a CI scheme spanning single and double excitations~\cite{deb09a}. To~improve accuracy, the~diagonal elements of the Hamiltonian matrix were fine-tuned to fit the experimental energies. Similarly to atom\_Q96\_34, $A$-values for transitions involving the $\mathrm{3d^54s\ ^7S_3}$ and $\mathrm{^5S_2}$ levels are taken from atom\_BB14\_34.

\item[atom\_BBQ10\_34, coll\_BBQ10\_34]---$A$-values and ECS were downloaded from the {\sc xstar}  database~\cite{men21}, and the {\tt PyNeb} {\tt atom} and {\tt coll} files were~rebuilt.

\item[atom\_BB14\_34, coll\_BB14\_34]---Computed $A$-values and ECS were obtained from the \\ {\tt crlike\_nrb13\#fe2.dat} {\tt adf04} file downloaded from the {\sc open-adas}\endnote{\url{https://open.adas.ac.uk/}} database. Details of the ECS computation~\cite{bad14} are described in Section~\ref{pyneb_data_fe3}. No information is given in the {\tt adf04} file on the provenance of the radiative data; thus, we assume they are coproducts of the target~calculation.

\item[atom\_BB14\_144, coll\_BB14\_144]---The BB14\_34 model has been extended to include 144 levels from the $\mathrm{3d^6}$, $\mathrm{3d^54s}$, and~$\mathrm{3d^54p}$ configurations with $E<121,500$\,cm$^{-1}$.
\end{description}


\subsection{\feiii\ Observational~Benchmarks}
\subsubsection{$A$-Value Ratio~Benchmark}\label{aratio_fe3}

It is well known that the intensity ratio of two spectral lines arising from a common upper level can be expressed as
\begin{equation}\label{aratio}
\frac{I(j,i)}{I(j,i')} = \frac{A(j,i)}{A(j,i')}\times \frac{\lambda(j,i')}{\lambda(j,i)}
\end{equation}
enabling an observational benchmark of the theoretical $A$-value ratio  $A(j,i)/A(j,i')$. This measure provides indications of the accuracy of both the observed intensities and computed radiative rates, signposting unreliable lines to be excluded from the spectrum~fits.     

In Tables~\ref{fe3_HH202_lineratios} and \ref{fe3_HH204_lineratios}, we compare the observed line-intensity ratios from HH\,202S and HH\,204 ~\cite{mes09, men21b} with the $A$-value ratios from the BBQ10\_34, BB14\_34, DH09\_34, and~Q96\_34 \texttt{atom} files. The~relative uncertainties in the observed intensity ratios from HH\,202S are, in general, larger than those from HH\,204. Although~both spectra are of comparable quality, HH\,204 was observed under photometric conditions whereby the absolute calibration error is slightly smaller; moreover, the~HH\,204 integrated emission comprised twice the area of HH\,202S with longer exposure~times.

For both sources, the~observed intensity ratios for transitions arising from levels $\mathrm{3d^6\ a\,^3F}_J$ or lower and ending up at a level within the ground configuration $\mathrm{3d^6\ a\,^5D}$ with wavelengths $\lambda > 4600$~\AA\ are generally more accurate (better than 15\%). For~transitions arising from higher levels, uncertainties as large as 70\% have been listed. On~the other hand, most theoretical ratios agree to within 10\% except for significantly discrepant cases in atom\_BBQ10\_34: $I(\lambda 3286)/I(\lambda 8729)$; $I(\lambda 3366)/I(\lambda 9960)$; and $I(\lambda 3357)/I(\lambda 8838)$. As~a whole, the~theoretical ratios lie within the observed ratio error bars except for some questionable lines in HH\,202S ($\lambda 8729$, affected by telluric absorptions) and in HH\,204 ($\lambda\lambda 4080, 4097$ blended with O~\textsc{ii} lines and  $\lambda\lambda 7078, 7088, 9204$ being misidentifications).

\begin{table}[H]
\caption{Comparison of observed [\feiii] line intensity ratios in HH\,202S (Obs) for transitions sharing a common upper level with those estimated from the $A$-values in the BBQ10\_34, BB14\_34, DH09\_34, and~Q96\_34 \texttt{atom} files. The~error of the least significant figure of the observed intensity ratio is indicated in~brackets. \label{fe3_HH202_lineratios}}
	\begin{adjustwidth}{-\extralength}{0cm}
		\newcolumntype{C}{>{\centering\arraybackslash}X}
		\begin{tabularx}{\textwidth+\extralength}{CCCCCCCCCC}
		  \toprule
\multicolumn{3}{c}{\textbf{Line~1}} & \multicolumn{2}{c}{\textbf{Line~2}} & \textbf{Obs}& \textbf{BBQ10\_34}& \textbf{BB14\_34}&
          \textbf{DH09\_34}& \textbf{Q96\_34} \\
          \midrule
          \textbf{Upper}&\textbf{Lower}&\boldmath{$\lambda_\mathrm{air}$}\textbf{(\AA)}&\textbf{Lower}&\boldmath{$\lambda_\mathrm{air}$}\textbf{(\AA)}&\multicolumn{5}{c}{\textbf{Line Intensity Ratio}}\\
          \midrule
$\mathrm{3d^6\ ^3D_3}$   &$\mathrm{3d^6\ ^5D_4}$&3239.79&$\mathrm{3d^6\ ^5D_3}$   &3286.24&4(1)   &3.58&3.59&3.28&3.60\\
$\mathrm{3d^6\ ^3D_3}$   &$\mathrm{3d^6\ ^5D_3}$&3286.24&$\mathrm{3d^6\ a\,^3P_2}$&8728.84&2.0(5) &4.88&3.12&3.80&3.29\\
$\mathrm{3d^6\ ^3D_1}$   &$\mathrm{3d^6\ ^5D_1}$&3355.50&$\mathrm{3d^6\ ^5D_0}$   &3366.22&1.6(8) &1.13&1.12&1.16&1.14\\
$\mathrm{3d^6\ ^3D_1}$   &$\mathrm{3d^6\ ^5D_0}$&3366.22&$\mathrm{3d^6\ a\,^3P_1}$&9959.85&5(4)   &9.07&5.42&6.37&5.54\\
$\mathrm{3d^6\ ^3D_2}$   &$\mathrm{3d^6\ ^5D_2}$&3334.95&$\mathrm{3d^6\ ^5D_1}$   &3356.59&1.2(2) &1.14&1.15&1.20&1.18\\
$\mathrm{3d^6\ ^3D_2}$   &$\mathrm{3d^6\ ^5D_1}$&3356.59&$\mathrm{3d^6\ a\,^3P_2}$&8838.14&4.1(8) &6.01&4.28&4.39&4.19\\
$\mathrm{3d^54s\ ^7S_3}$ &$\mathrm{3d^6\ ^5D_4}$&3322.47&$\mathrm{3d^6\ ^5D_3}$   &3371.35&1.4(3) &1.38&1.37&&        \\
$\mathrm{3d^54s\ ^7S_3}$ &$\mathrm{3d^6\ ^5D_3}$&3371.35&$\mathrm{3d^6\ ^5D_2}$   &3406.11&2.2(5) &2.33&2.31&&        \\
$\mathrm{3d^6\ ^3G_3}$   &$\mathrm{3d^6\ ^5D_3}$&4046.49&$\mathrm{3d^6\ ^5D_2}$   &4096.68&2.3(8) &2.27&2.45&2.56&2.53\\
$\mathrm{3d^6\ ^3G_4}$   &$\mathrm{3d^6\ ^5D_4}$&4008.34&$\mathrm{3d^6\ ^5D_3}$   &4079.69&3.8(7) &3.15&3.63&3.62&3.92\\
$\mathrm{3d^6\ a\,^3F_2}$&$\mathrm{3d^6\ ^5D_3}$&4667.11&$\mathrm{3d^6\ ^5D_2}$   &4734.00&0.29(4)&0.28&0.28&0.29&0.28\\
$\mathrm{3d^6\ a\,^3F_2}$&$\mathrm{3d^6\ ^5D_2}$&4734.00&$\mathrm{3d^6\ ^5D_1}$   &4777.70&2.0(3) &2.06&2.07&2.09&2.08\\
$\mathrm{3d^6\ a\,^3F_3}$&$\mathrm{3d^6\ ^5D_4}$&4607.12&$\mathrm{3d^6\ ^5D_3}$   &4701.64&0.19(3)&0.18&0.18&0.19&0.17\\
$\mathrm{3d^6\ a\,^3F_3}$&$\mathrm{3d^6\ ^5D_3}$&4701.64&$\mathrm{3d^6\ ^5D_2}$   &4769.53&2.8(3) &2.89&2.90&2.93&2.94\\
$\mathrm{3d^6\ a\,^3F_4}$&$\mathrm{3d^6\ ^5D_4}$&4658.17&$\mathrm{3d^6\ ^5D_3}$   &4754.81&5.3(6) &5.28&5.40&5.32&5.49\\
$\mathrm{3d^6\ a\,^3P_1}$&$\mathrm{3d^6\ ^5D_2}$&5011.41&$\mathrm{3d^6\ ^5D_0}$   &5084.85&5.9(9) &5.72&6.12&5.97&5.95\\
$\mathrm{3d^6\ ^3H_4}$    &$\mathrm{3d^6\ ^5D_4}$&4881.07&$\mathrm{3d^6\ ^5D_3}$  &4987.29&5.4(7) &4.99&5.33&5.50&5.76\\
$\mathrm{3d^6\ a\,^3P_2}$&$\mathrm{3d^6\ ^5D_3}$&5270.57&$\mathrm{3d^6\ ^5D_1}$   &5412.06&11(2)  &10.2&11.4&11.4&11.0\\
        \bottomrule
		\end{tabularx}
	\end{adjustwidth}
\end{table}

\vspace{-6pt}

\begin{table}[H]
\caption{Comparison of observed [\feiii] line intensity ratios in HH\,204 (Obs) for transitions sharing a common upper level with those estimated from $A$-values in the BBQ10\_34, BB14\_34, DH09\_34, and~Q96\_34 \texttt{atom} files. The~error of the least significant figure of the observed intensity ratio is indicated in~brackets. \label{fe3_HH204_lineratios}}
	\begin{adjustwidth}{-\extralength}{0cm}
		\newcolumntype{C}{>{\centering\arraybackslash}X}
		\begin{tabularx}{\textwidth+\extralength}{CCCCCCCCCC}
		  \toprule
\multicolumn{3}{c}{\textbf{Line~1}} & \multicolumn{2}{c}{\textbf{Line~2}} & \textbf{Obs}& \textbf{BBQ10\_34}& \textbf{BB14\_34}&
          \textbf{DH09\_34}& \textbf{Q96\_34} \\
          \midrule
          \textbf{Upper}&\textbf{Lower}&\boldmath{$\lambda_\mathrm{air}$}\textbf{(\AA)}&\textbf{Lower}&\boldmath{$\lambda_\mathrm{air}$}\textbf{(\AA)}&\multicolumn{5}{c}{\textbf{Line Intensity Ratio}}\\
          \midrule
$\mathrm{3d^6\ ^3D_3}$   &$\mathrm{3d^6\ ^5D_4}$ &3239.79&$\mathrm{3d^6\ ^5D_3}$   &3286.24&3.6(9)&3.58&3.59&3.28&3.60\\
$\mathrm{3d^6\ ^3D_3}$   &$\mathrm{3d^6\ ^5D_3}$ &3286.24&$\mathrm{3d^6\ ^5D_2}$   &3319.27&1.0(4)&1.35&1.36&1.47&1.41\\
$\mathrm{3d^6\ ^3D_3}$   &$\mathrm{3d^6\ ^5D_2}$ &3319.27&$\mathrm{3d^6\ a\,^3P_2}$&8728.84&3.1(8)&3.62&2.30&2.59&2.35\\
$\mathrm{3d^6\ ^3D_1}$   &$\mathrm{3d^6\ ^5D_1}$ &3355.50&$\mathrm{3d^6\ ^5D_0}$   &3366.22&1.5(3)&1.13&1.12&1.16&1.14\\
$\mathrm{3d^6\ ^3D_2}$   &$\mathrm{3d^6\ ^5D_2}$ &3334.95&$\mathrm{3d^6\ ^5D_1}$   &3356.59&1.2(3)&1.14&1.15&1.20&1.18\\
$\mathrm{3d^6\ ^3D_2}$   &$\mathrm{3d^6\ ^5D_1}$ &3356.59&$\mathrm{3d^6\ a\,^3P_2}$&8838.14&5.3(7)&6.01&4.28&4.39&4.19\\
$\mathrm{3d^6\ ^1I_6}$   &$\mathrm{3d^6\ ^3H_6}$ &9701.87&$\mathrm{3d^6\ ^3H_5}$   &9942.38&1.4(2)&1.58&1.54&1.56&1.55\\
$\mathrm{3d^54s\ ^7S_3}$ &$\mathrm{3d^6\ ^5D_4}$ &3322.47&$\mathrm{3d^6\ ^5D_3}$   &3371.35&1.5(2)&1.38&1.37&     &     \\
$\mathrm{3d^54s\ ^7S_3}$ &$\mathrm{3d^6\ ^5D_3}$ &3371.35&$\mathrm{3d^6\ ^5D_2}$   &3406.11&1.7(2)&2.33&2.31&     &     \\
$\mathrm{3d^6\ ^3G_3}$   &$\mathrm{3d^6\ ^5D_3}$ &4046.49&$\mathrm{3d^6\ ^5D_2}$   &4096.68&3.4(9)&2.27&2.45&2.56&2.53\\
$\mathrm{3d^6\ ^3G_4}$   &$\mathrm{3d^6\ ^5D_4}$ &4008.34&$\mathrm{3d^6\ ^5D_3}$   &4079.69&4.4(4)&3.15&3.63&3.62&3.92\\
$\mathrm{3d^6\ a\,^3F_2}$&$\mathrm{3d^6\ ^5D_3}$ &4667.11&$\mathrm{3d^6\ ^5D_2}$   &4734.00&0.29(1)& 0.28&0.28&0.29&0.28\\
$\mathrm{3d^6\ a\,^3F_2}$&$\mathrm{3d^6\ ^5D_2}$ &4734.00&$\mathrm{3d^6\ ^5D_1}$   &4777.70&2.1(1)&2.06&2.07&2.09&2.08\\
$\mathrm{3d^6\ a\,^3F_3}$&$\mathrm{3d^6\ ^5D_4}$ &4607.12&$\mathrm{3d^6\ ^5D_3}$   &4701.64&0.18(1)&0.18&0.18&0.19&0.17\\
$\mathrm{3d^6\ a\,^3F_3}$&$\mathrm{3d^6\ ^5D_3}$ &4701.64&$\mathrm{3d^6\ ^5D_2}$   &4769.53&2.9(1)&2.89&2.90&2.93&2.94\\
$\mathrm{3d^6\ a\,^3F_4}$&$\mathrm{3d^6\ ^5D_4}$ &4658.17&$\mathrm{3d^6\ ^5D_3}$   &4754.81&5.3(2)&5.28&5.40&5.32&5.49\\
$\mathrm{3d^6\ a\,^3P_1}$&$\mathrm{3d^6\ ^5D_2}$ &5011.41&$\mathrm{3d^6\ ^5D_0}$   &5084.85&5.9(4)&5.72&6.12&5.97&5.95\\
$\mathrm{3d^6\ ^3H_4}$   &$\mathrm{3d^6\ ^5D_4}$ &4881.07&$\mathrm{3d^6\ ^5D_3}$   &4987.29&6.1(2)&4.99&5.33&5.50&5.76\\
$\mathrm{3d^6\ a\,^3P_2}$&$\mathrm{3d^6\ ^5D_3}$ &5270.57&$\mathrm{3d^6\ ^5D_1}$  &5412.06&10.8(5)&10.2&11.4&11.4&11.0\\
          \bottomrule
		\end{tabularx}
	\end{adjustwidth}
\end{table}

\clearpage

\subsubsection{[\feiii] Spectrum~Fits \label{specfit}}

Following~\cite{mor20}, we take advantage of {\tt PyNeb}'s methods for emission-line modeling to implement a further observational benchmark to assess the accuracy of the atomic datasets described in Sections~\ref{pyneb_data_fe3} and \ref{newdata_fe3}. For~this purpose, we again use the high-resolution spectra of the bright HH\,202S and HH\,204~sources.

The electron temperature and density of the HH object are obtained by a least-squares fit of all the reliable spectral lines as proposed by~\cite{sto13}. The~theoretical emissivity and observed intensity of each line are respectively normalized with respect to the emissivity and intensity sums of the transition array. The~fit measure is given in terms of
\begin{equation}
\chi^2 = \sum_{i=1}^N\,\frac{(I_i-\epsilon_i)^2}{N\sigma_{I_i}^2}\ ,
\end{equation}
where $N$ is the number of spectral lines, $I_i$ and $\epsilon_i$ are the normalized intensity and emissivity, respectively, and~$\sigma_{I_i}^2$ is the variance of the normalized intensity~\cite{soc12}. No explicit error is considered in the theoretical emissivities. The~least-squares minimization is carried out with the Python function {\tt scipy.optimize.minimize} finding no degeneracies in the $\chi^2$ convergence.
 
{\tt PyNeb} provides two methods to compute the emissivity of a spectral line (e.g., $\lambda 5270.57$ between levels $i=6$ and $j=2$ of \feiii) at $(T_e, n_e)$ by addressing, say, the~datasets atom\_Q96\_J00 and coll\_Z96. The~Python commands to run these two methods ({\tt e1} and {\tt e2}) are as follows:

\begin{Verbatim}[frame=single]
# Call the PyNeb Python module
import pyneb as pn
# Select the radiative and collisional datasets
pn.atomicData.setDataFile('fe_iii_atom_Q96_J00.dat')
pn.atomicData.setDataFile('fe_iii_coll_Z96.dat')
# Instantiate Fe III
Fe3=pn.Atom('Fe',3)
# Check the datafiles
print(Fe3)
# Determine and print the two emissivities
e1=Fe3.getEmissivity(tem=1.e4,den=1.0e4,wave=5270.57)
e2=Fe3.getEmissivity(tem=1.e4,den=1.0e4,lev_i=6,lev_j=2)
print(e1,e2)
\end{Verbatim}

In the present work, consistency between {\tt e1} and {\tt e2} for the whole spectrum was tested to avoid incorrect line identifications, but~we found that {\tt e2} avoided the problem of transitions with the same wavelength in atomic models with a large number of~levels.  

In the spectrum fits of HH\,202S and HH\,204, a~handful of lines are excluded due to the following problems:

\begin{itemize}
\item Misidentifications: $\lambda\lambda 7078, 7088$, 9204
\item Line blending or telluric contamination~\cite{mes09}: $\lambda\lambda  4080, 4097,4925$, 4931, 4987, 8729, 8838 
\item Small $A$-values ($\lesssim10^{-3}$~s$^{-1}$): $\lambda\lambda 4008$, 4047.
\end{itemize}

We first examine in Table~\ref{old_data} the spectral fits with the atomic datasets available in {\tt PyNeb}~1.1.16 (see Section~\ref{pyneb_data_fe3}). For~each \texttt{coll} dataset BBQ10, BB14, and~Z96, we run the fits in turn with three \texttt{atom} datasets: BBQ10, NP96, and~Q96\_J00. For~HH\,202S, we obtain an average temperature and density of $T_e=(7\pm 1)\times 10^3$\,K and $n_e=(6\pm 5)\times 10^4$\,cm$^{-3}$, which indicate large uncertainties due to the atomic data. In~particular, coll\_BB14 gives, on average, a questionable low temperature ($5.4\times 10^3$\,K) and high density ($1.3\times 10^5$\,cm$^{-3}$). The~quality of the fits is generally poor, the~best being atom\_BBQ10--coll\_Z96 with $\chi^2=1.74$. The~fits for HH\,204 are as unimpressive: $T_e=(8\pm 2)\times 10^3$\,K) and $n_e=(5\pm 4)\times 10^4$\,cm$^{-3}$, the~best ($\chi^2=16.9$) being with the file pair atom\_BBQ10--coll\_Z96.

\begin{table}[H]
\caption{Temperature and density fits of the [\feiii] observed spectra of HH\,202S and HH\,204 with the radiative ({\tt atom}) and collisional ({\tt coll}) datasets available in {\tt PyNeb}~1.1.16.\label{old_data}}
	\begin{adjustwidth}{-\extralength}{0cm}
		\newcolumntype{C}{>{\centering\arraybackslash}X}
		\begin{tabularx}{\textwidth+\extralength}{CCC>{\centering\arraybackslash}m{2cm}CC>{\centering\arraybackslash}m{2cm}C}
			\toprule
             \multicolumn{2}{c}{\textbf{Datasets}} & \multicolumn{3}{c}{\textbf{HH\,202S}} & \multicolumn{3}{c}{\textbf{HH\,204}}\\
            \midrule
			\textbf{Atom}&\textbf{Coll}& \boldmath{$T_e(10^3\,\textrm{K})$} & \boldmath{$n_e(10^4$} \textbf{cm}\boldmath{$^{-3})$} & \boldmath{$\chi^2$}& \boldmath{$T_e(10^3\,\textrm{K})$} & \boldmath{$n_e(10^4$} \textbf{cm}\boldmath{$^{-3}$})& \boldmath{$\chi^2$} \\
			\midrule
			BBQ10	 & BBQ10 & 7.99 & 1.92 & 12.0 & 9.77 & 1.53 & 66.9 \\
			NP96	 & BBQ10 & 8.37 & 3.79 & 13.9 & 10.5 & 2.24 & 57.2 \\
            Q96\_J00 & BBQ10 & 7.80 & 1.96 & 14.4 & 9.60 & 1.36 & 58.1 \\
			BBQ10	 & BB14	 & 5.26 & 13.1 & 14.9 & 5.95 & 11.3 & 88.7 \\
			NP96	 & BB14	 & 5.54 & 16.8 & 15.3 & 6.15 & 10.5 & 60.2 \\
            Q96\_J00 & BB14	 & 5.26 & 8.54 & 15.8 & 5.95 & 5.51 & 60.4 \\
			BBQ10	 & Z96	 & 7.34 & 2.52 & 1.74 & 8.53 & 1.70 & 16.9 \\
			NP96	 & Z96	 & 7.44 & 7.09 & 15.2 & 9.54 & 4.20 & 69.9 \\
            Q96\_J00 & Z96	 & 7.35 & 2.55 & 12.0 & 8.79 & 1.68 & 37.7 \\
			\bottomrule
		\end{tabularx}
	\end{adjustwidth}
\end{table}

In a similar fashion, we proceed to benchmark the revised and new datasets described in Section~\ref{newdata_fe3} by running a grid of fits with different {\tt atom--coll} file pairs: four radiative and three collisional datasets (see Table~\ref{new_data}). The~improvement is considerable; the~average temperature and density for HH\,202S is now $T_e=(7.9\pm 0.3)\times 10^3$\,K and $n_e=(2.1\pm 0.3)\times 10^4$\,cm$^{-3}$ and for HH\,204 $T_e=(8.9\pm 0.6)\times 10^3$\,K and $n_e=(1.5\pm 0.3)\times 10^4$\,cm$^{-3}$. Outstanding fits for HH\,202S are obtained with the atom\_DH09\_34--coll\_Z96\_34 and atom\_Q96\_34--coll\_Z96\_34 dataset pairs, and~the revised coll\_BB14\_34 ECS dataset now gives more reasonable temperatures ($\sim8\times 10^3$\,K) and densities ($\sim2\times 10^4$\,cm$^{-3}$). We find questionable ECS differences between coll\_BB14 and coll\_BB14\_34 for transitions involving levels $\mathrm{a\,^3P_J}$, $\mathrm{a\,^3F_J}$, and~$\mathrm{a\,^1G_4}$; furthermore, for~levels with $E>43,000$~cm$^{-1}$, the~ECS in coll\_BB14 are $\Upsilon(j,i)=0.0$. We also found ECS differences for transitions to level $\mathrm{3d^54s\ ^7S_3}$ in coll\_BBQ10 and coll\_BBQ10\_34 due to improvements performed by~\cite{bau10} after publication. On~the other hand, the~extended datasets atom\_BB14\_144, coll\_BB14\_144, and~coll\_Z96\_144, as~expected, make no difference in these plasma~conditions.

The temperature and density scatters are now within 6\% and 20\%, respectively, and~our average values compare satisfactorily with those adopted in the observational \linebreak papers~\cite{mes09,men21b}: $T_e=(8.760\pm 0.180)\times 10^3$\,K and $n_e=(1.354\pm 0.121)\times 10^4$\,cm$^{-3}$ for HH\,204 and $n_e=(1.743\pm 0.236)\times 10^4$\,cm$^{-3}$ for HH\,202S. No adopted temperature is quoted for the latter, but~our average value compares well with their $T_e(\mathrm{[S\,II]})= (8.25\pm 0.54)\times 10^3$\,K, $T_e(\mathrm{[Ar\,III]})= (8.26\pm 0.41)\times 10^3$\,K, and~$T_e(\mathrm{He\,I})= (7.95\pm 0.20)\times 10^3$\,K.


\begin{table}[H]
\caption{Temperature and density fits of the [\feiii] observed spectra in HH\,202S and HH\,204 with the improved or new radiative ({\tt atom}) and collisional ({\tt coll}) datasets.\label{new_data}}
	\begin{adjustwidth}{-\extralength}{0cm}
		\newcolumntype{C}{>{\centering\arraybackslash}X}
		\begin{tabularx}{\textwidth+\extralength}{CCC>{\centering\arraybackslash}m{2cm}CC>{\centering\arraybackslash}m{2cm}C}
			\toprule
             \multicolumn{2}{c}{\textbf{Datasets}} & \multicolumn{3}{c}{\textbf{HH\,202S}} & \multicolumn{3}{c}{\textbf{HH\,204}}\\
            \midrule
			\textbf{Atom}&\textbf{Coll}& \boldmath{$T_e(10^3\,\textrm{K})$} & \boldmath{$n_e(10^4$} \textbf{cm}\boldmath{$^{-3})$} & \boldmath{$\chi^2$}& \boldmath{$T_e(10^3\,\textrm{K})$} & \boldmath{$n_e(10^4$} \textbf{cm}\boldmath{$^{-3}$})& \boldmath{$\chi^2$} \\		
            \midrule
			BBQ10\_34 & BBQ10\_34 & 7.68 & 1.94 & 6.31 & 9.49 & 1.51 & 47.6 \\
			BB14\_34  & BBQ10\_34 & 8.24 & 1.80 & 3.84 & 9.91 & 1.44 & 28.1 \\
            DH09\_34  & BBQ10\_34 & 7.94 & 1.31 & 2.24 & 9.25 & 0.82& 21.8 \\
            Q96\_34   & BBQ10\_34 & 7.97 & 1.66 & 3.54 & 9.39 & 1.19 & 35.6 \\
			BBQ10\_34 & BB14\_34  & 7.83 & 2.17 & 5.57 & 9.01 & 2.16 & 50.0 \\
			BB14\_34  & BB14\_34  & 8.47 & 2.14 & 4.56 & 9.03 & 1.49 & 33.2 \\
            DH09\_34  & BB14\_34  & 8.18 & 1.78 & 3.23 & 9.00 & 1.16 & 9.95 \\
            Q96\_34   & BB14\_34  & 8.12 & 1.91 & 3.97 & 8.97 & 1.49 & 11.9 \\
			BBQ10\_34 & Z96\_34	  & 7.34 & 2.51 & 1.74 & 8.53 & 1.71 & 16.9 \\
			BB14\_34  & Z96\_34	  & 7.94 & 2.32 & 2.54 & 8.23 & 1.61 & 28.5 \\
            DH09\_34  & Z96\_34	  & 7.71 & 1.97 & 0.68 & 8.34 & 1.30 & 4.36 \\
            Q96\_34   & Z96\_34   & 7.68 & 2.24 & 1.00 & 8.11 & 1.73 & 4.62 \\
            BB14\_144 & BB14\_144 & 8.47 & 2.12 & 4.57 & 9.03 & 1.49 & 33.2 \\
            BB14\_144 & Z96\_144  & 7.94 & 2.33 & 2.54 & 8.24 & 1.61 & 28.5 \\
			\bottomrule
		\end{tabularx}
	\end{adjustwidth}
\end{table}

\subsubsection{Line-Ratio~Diagnostics}\label{diagnostics_fe3}

Line-ratio diagnostics are widely used in nebular astrophysics to determine the plasma thermodynamic properties (temperature, density, and~ionic abundances). They are based on the assumption that the observed line-intensity ratio is equivalent to the theoretical emissivity ratio
\begin{equation}
\frac{I(j,i)}{I(j,i')} = \frac{\epsilon(j,i)}{\epsilon(j,i')}\ .
\end{equation}

Using our benchmarked atomic datasets for [\feiii], we investigate the impact of the atomic data scatter on the line-ratio temperature and density~behaviors.

\begin{figure}[H]
\vspace{-6pt}
\includegraphics[width=13.86cm]{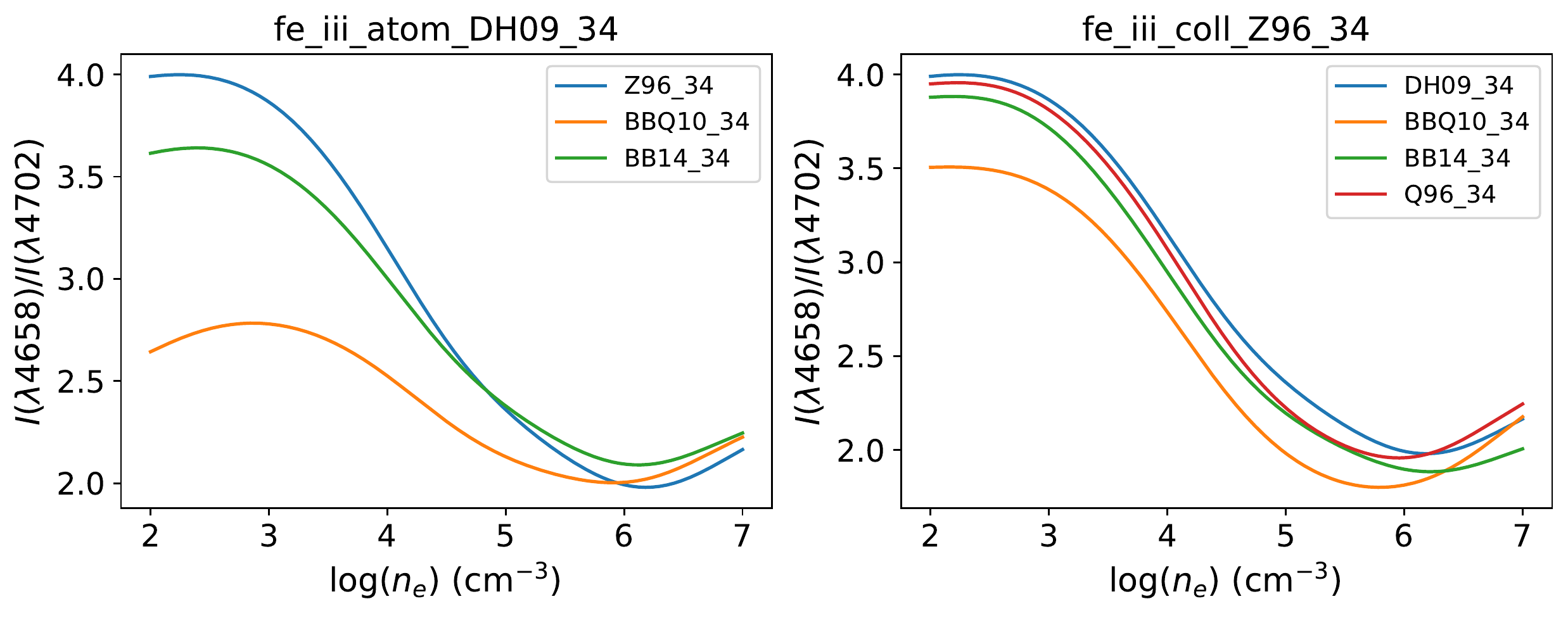}
\caption{Density behavior of the [\feiii] $I(\lambda4658)/I(\lambda4702)$ emissivity ratio at $T_e=7.9\times 10^3$\,K for different atomic datasets. \textbf{Left panel}: atom\_DH09\_34 dataset with three \texttt{coll} datasets ( Z96\_34, BBQ10\_34, and~BB14\_34). \textbf{Right panel}: coll\_Z96\_34 dataset with four \texttt{atom} datasets (DH09\_34, BBQ10\_34, BB14\_34, and~Q96\_34). \label{denratio}}
\end{figure}

A common [\feiii] emissivity ratio to determine the plasma density in the range $3\leq \log n_e\leq 6$ is $I(\lambda4658)/I(\lambda4702)$ (see, for~instance,~\cite{men21a,men21b}). In~Figure~\ref{denratio}, we show its dependency on the atomic data. In~the left panel, we select dataset atom\_DH09\_34 and plot $I(\lambda4658)/I(\lambda4702)$ as a function of density for three \texttt{coll} datasets: BBQ10\_34, BB14\_34, and~Z96\_34. Large discrepancies are found in the low-density regime ($\log n_e < 4$), particularly in coll\_BBQ10\_34 (40\% lower than coll\_Z96\_34 at $\log n_e < 3$). In~the right panel, we select dataset coll\_Z96\_34 and plot the emissivity ratio for four \texttt{atom} datasets: DH09\_34, BBQ10\_34, BB14\_34, and~Q96\_34. Differences are not as large, atom\_BBQ10\_34 being the more discordant ($\sim15\%$ lower at $\log n_e <3$).

\begin{figure}[H]
\includegraphics[width=6.8 cm]{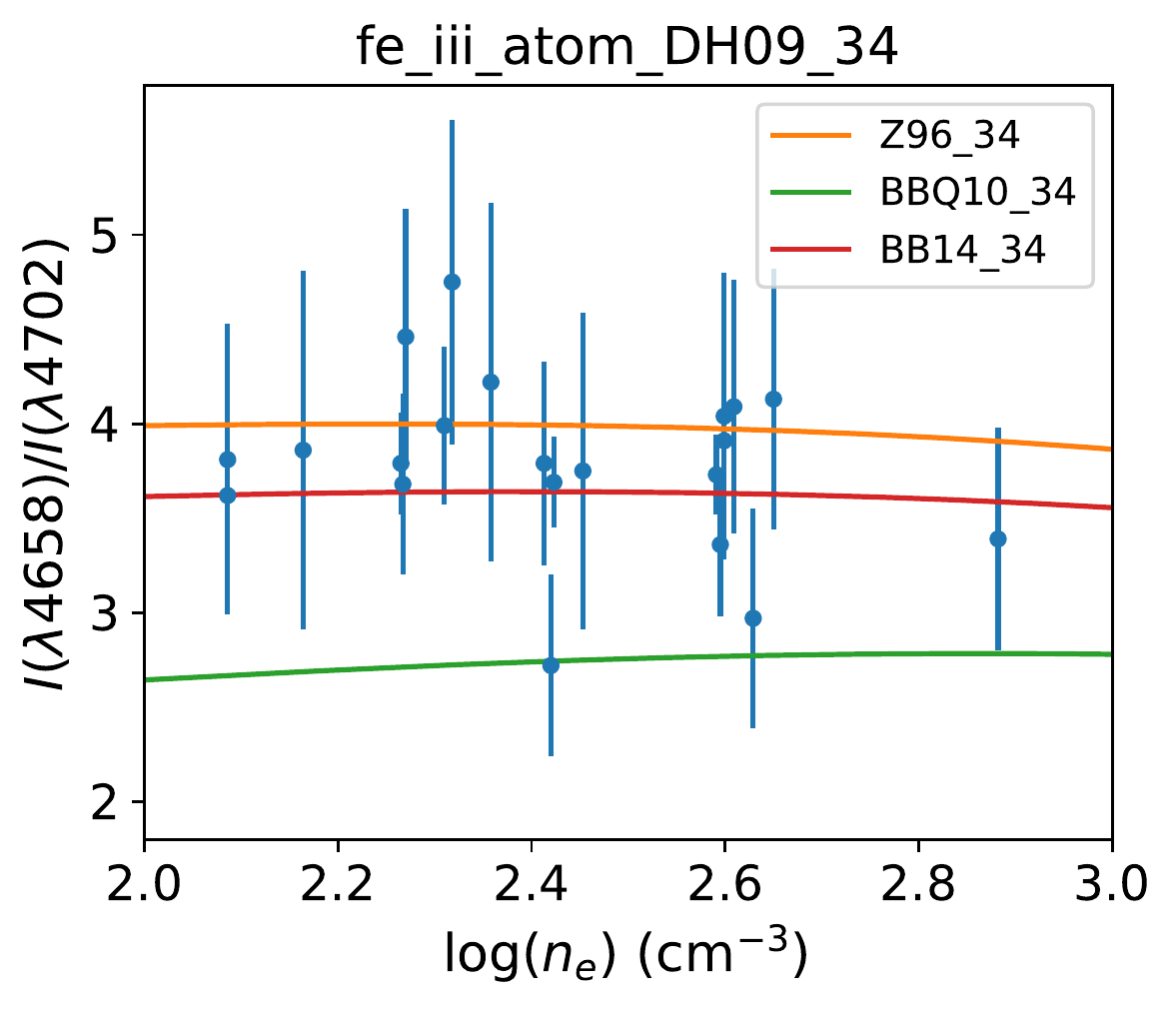}
\caption{Observed [\feiii] $I(\lambda4658)/I(\lambda4702)$ line-intensity ratio from several low-density sources as a function of the [S\,{\sc ii}] density. The~theoretical emissivity ratios determined by {\tt PyNeb} with dataset atom\_DH09\_34 and three \texttt{coll} datasets (BBQ10\_34, BB14\_34, and~Z96\_34) are also~depicted. \label{lowden}}
\end{figure}

For $\log n_e\leq 3$, the~emissivity ratio $I(\lambda4658)/I(\lambda4702)$ is practically density insensitive, being mostly dependent on the ECS of the involved atomic levels. To~discern a reliable magnitude in this regime in the light of the discrepancies due to the collisional datasets, we select from the DESIRED database (M\'endez-Delgado~et~al., unpublished) the following low-density sources: 30 Doradus~\cite{Peimbert:2003}; Sh\,2-311~\cite{GarciaRojas:2005}; M20~\cite{GarciaRojas:2006}; M17~\cite{GarciaRojas:2007}; HII-1, HII-2, UV-1~\cite{LopezSanchez:2007};  K932, NGC\,5461, NGC\,604, NGC\,2363, NGC\,4861, NGC\,1741-C, NGC\,5447, VS44, H1013~\cite{Esteban:2009}; NGC\,3125, POX\,4, TOL\,1924-416, TOL\,1457-262, NGC\,6822 (HV), NGC\,5408~\cite{Esteban:2014}; Sh\,2-100, Sh\,2-288~\cite{Esteban:2017}; NGC\,5471, NGC\,5455 ~\cite{Esteban:2020}, Sh\,2-152~\cite{Esteban:2018}; and N44C, N11B, N66A, NGC\,1714, IC\,2111, N81~\cite{DominguezGuzman:2022}. We then plot the observed line-intensity ratios in Figure~\ref{lowden} as a function of the [S\,{\sc ii}] density. In~spite of the scatter, there is a definite trend for $I(\lambda4658)/I(\lambda4702)> 3.5$, which would reinforce the validity of the coll\_BB14\_34 and coll\_Z96\_34~datasets.

\begin{figure}[H]
\includegraphics[width=13.86cm]{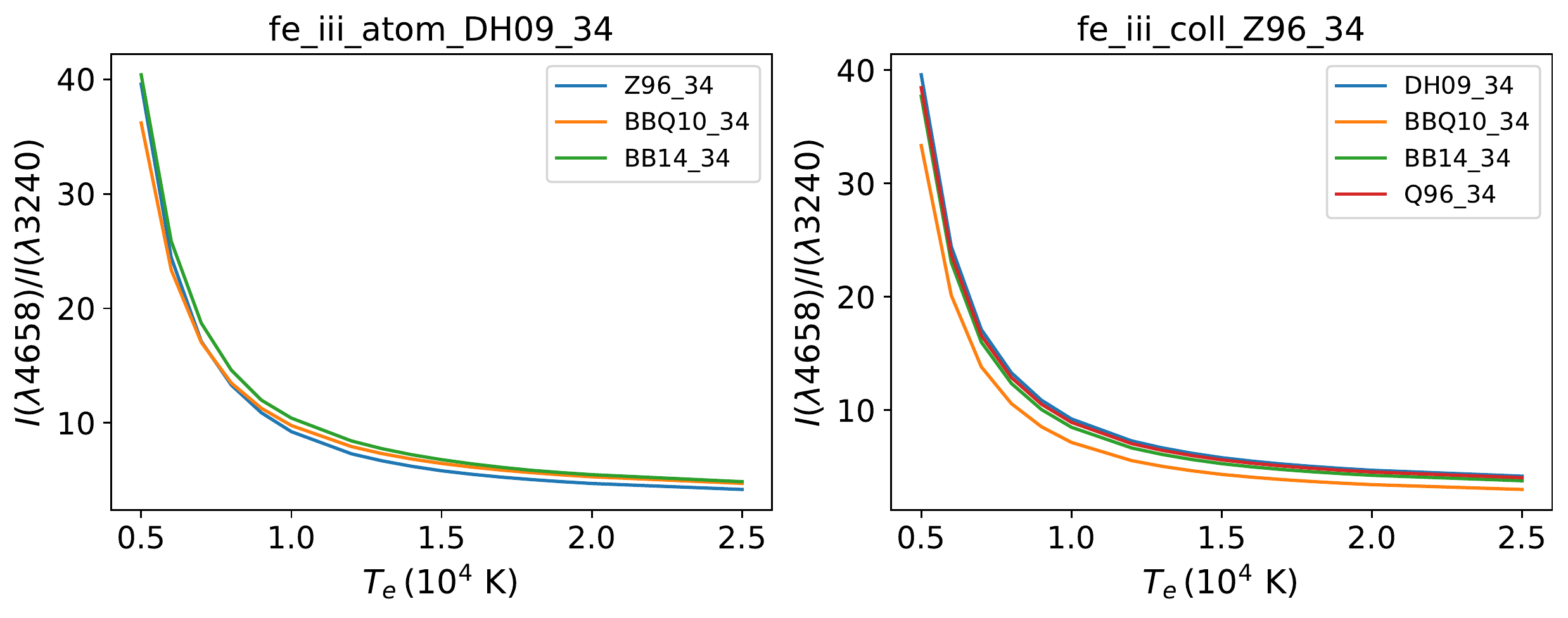}
\caption{Temperature behavior of the [\feiii] $I(\lambda4658)/I(\lambda3240)$ emissivity ratio at $n_e=2.1\times 10^4$\,cm$^{-3}$ for different atomic datasets. \textbf{Left panel}: dataset atom\_DH09\_34 with three \texttt{coll} datasets (Z96\_34, BBQ10\_34, and~BB14\_34). \textbf{Right panel}: dataset coll\_Z96\_34 with four \texttt{atom} datasets (DH09\_34, BBQ10\_34, BB14\_34, and~Q96\_34). \label{temratio}}
\end{figure}

Regarding a useful [\feiii] temperature diagnostic, we choose $I(\lambda4658)/I(\lambda3240)$ due to its small variations in density and atomic data. In~Figure~\ref{temratio}, we plot this emissivity ratio as a function of temperature for the same {\tt atom}--{\tt coll} grid used in the density diagnostic. The~curves show only a weak dependency on the choice of the collisional and radiative~datasets.

A relevant question is whether the density and temperature estimates obtained from the $I(\lambda4658)/I(\lambda4702)$ and $I(\lambda4658)/I(\lambda3240)$ line ratios, respectively, compare with those obtained from the spectral fits described in Section~\ref{specfit}. For~HH\,202S, as~an example, the~observed $I(\lambda4658)/I(\lambda4702)=2.81$ and $I(\lambda4658)/I(\lambda3240)=13.6$. In~the density diagnostic, we set the temperature at $T_e =7.9\times 10^3$\,K and obtain $n_e=2.34\times 10^4$\,cm$^{-3}$ for the pair atom\_DH09\_34--coll\_Z96\_34 and $n_e=1.86\times 10^4$\,cm$^{-3}$ for atom\_DH09\_34--coll\_BB14\_34, values that are fairly close to those obtained from the fits, respectively \linebreak $n_e=1.97\times 10^4$\,cm$^{-3}$ and $n_e=1.78\times 10^4$\,cm$^{-3}$. However, if~we designate the pair \linebreak  atom\_DH09\_34--coll\_BBQ10\_34, the~theoretical $I(\lambda4658)/I(\lambda4702)< 2.78$ impairs a density reading for HH\,202S. 
Furthermore, if~we set the density at $n_e=2.1\times 10^4$\,cm$^{-3}$, we obtain from the temperature diagnostic $7.89\times 10^3$\,K, $7.95\times 10^3$\,K, and~$8.33\times 10^3$\,K for the \texttt{atom}--\texttt{coll} pairs DH09\_34--Z96\_34, DH09\_34--BBQ10\_34, and~DH09\_34--BB14\_34, respectively, which compare well with $7.71\times 10^3$\,K, $7.94\times 10^3$\,K, and~$8.18\times 10^3$\,K from the spectral~fits.

\subsection{Radiative Lifetimes of $\mathrm{3d^6}$ and $\mathrm{3d^54s}$ Levels}\label{rlt_fe3}

As previously mentioned, the~[\feiii] nebular spectrum displays transitions between levels of the $\mathrm{3d^6}$ and $\mathrm{3d^54s}$ configurations. The~{\tt PyNeb} \feiii\ atomic models are based on the lower 34 levels with $E<60,000$~cm$^{-1}$; that is, the~cut is imposed just before the onset of the $\mathrm{3d^54s}$ level array. As~a result, in~the present lifetime analysis, we obviate the $\mathrm{3d^6\ b^1D_2}$ level at 77,044.67~cm$^{-1}$ and the unobserved $\mathrm{3d^6\ ^1S_0}$ estimated theoretically at $E > 99,000$~cm$^{-1}$. From~$\mathrm{3d^54s}$, only the lowly lying $\mathrm{^7S_3}$ and $\mathrm{^5S_2}$ levels are taken into~account.

\begin{figure}[H]
\includegraphics[width=6.8 cm]{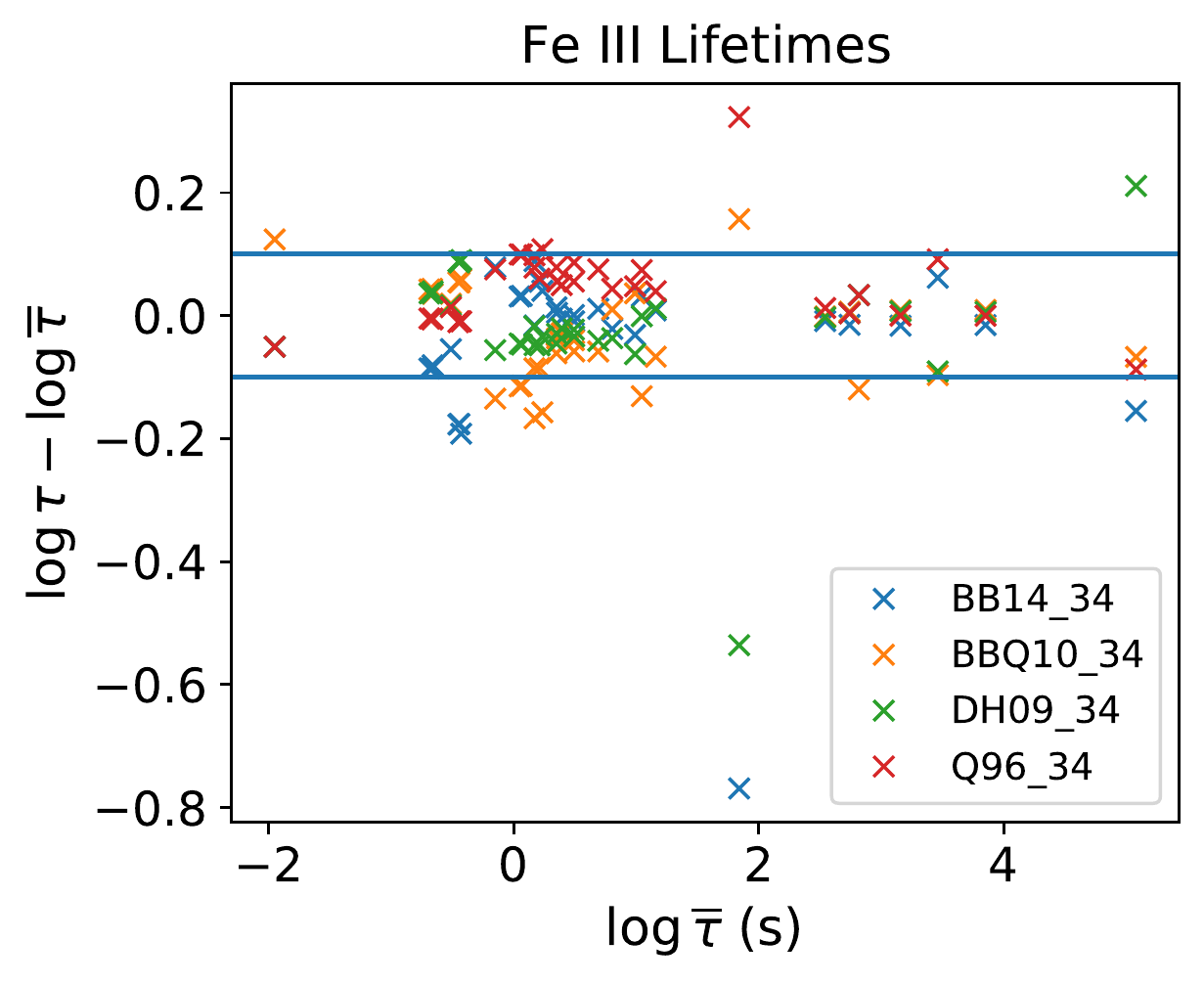}
\caption{Comparison of \feiii\ radiative lifetimes computed from the $A$-values in the \texttt{atom} datasets  BB14\_34, BBQ10\_34, DH09\_34, and~Q96\_34 \texttt{atom} files. For~each level, lifetime differences for these datasets with respect to the average value are plotted as a function of the average~value. \label{lt}}
\end{figure}

Radiative lifetimes derived from the $A$-values in the \texttt{atom} datasets BB14\_34, BBQ10\_34, DH09\_34, and~Q96\_34 are compared in Figure~\ref{lt}. For~each level, we determine the lifetime differences of these four datasets with respect to the average value. Such differences are, in general, within 0.1~dex except for five levels in atom\_BB14\_34, nine in atom\_BBQ10\_34, and~two in each atom\_DH09\_34 and atom\_Q96\_34. Notably large  discrepancies (as large as $\sim0.8$~dex) are found for the $\mathrm{3d^6\,^3H}_J$ levels, which would affect the physical conditions and Fe$^{2+}$ abundance derived from the $\lambda \lambda$4881, 4987, 4924, 5032, 4986 line intensities. Furthermore, the~$\mathrm{3d^6\ b\,^3P}_J$ levels in atom\_BB14\_34 and $\mathrm{3d^6\ a\,^3P_2}$ and $\mathrm{a\,^1G_4}$ in atom\_BBQ10\_34 have lifetimes almost 0.2~dex lower. Significant differences also stand out for the $\mathrm{3d^54s\,^7S_3}$ and $\mathrm{^5S_2}$ levels between these two datasets. A~reassuring outcome is the agreement between the atom\_DH09\_34 and atom\_Q96\_34 lifetimes: within 20\% except for the problematic $\mathrm{^3H}_J$ levels. Most of the conclusions from this comparison are in line with~\cite{bau10}.

\unskip

\subsection{Radiative Lifetimes of $\mathrm{3d^54p}$ Levels}\label{rlt_4p_fe3}

\begin{figure}[H]
\includegraphics[width=7.0 cm]{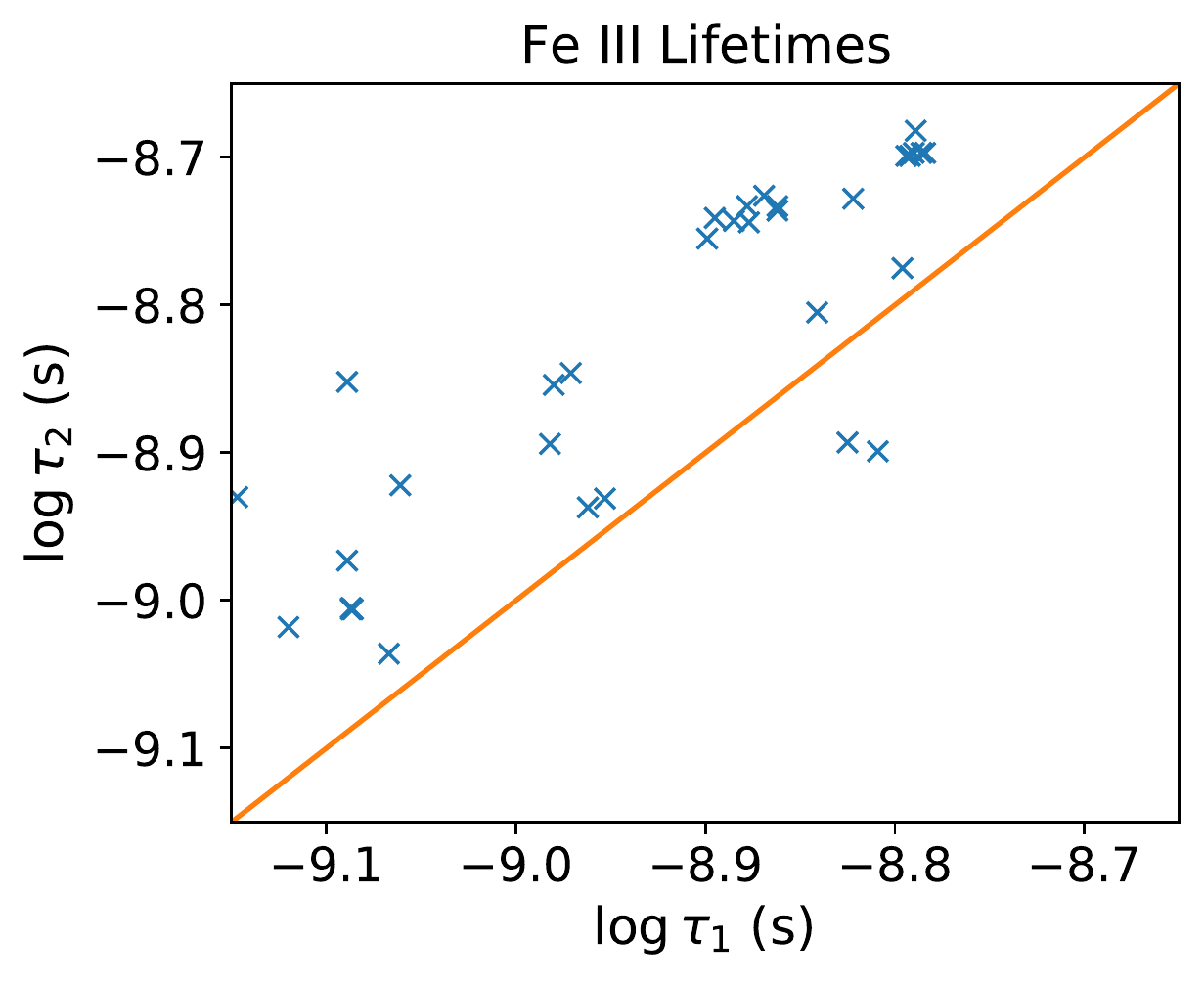}
\caption{Comparison of radiative lifetimes for 44 levels of the $\mathrm{3d^54p}$ configuration in \feiii\ derived from the $A$-values from atom\_BB14\_144 ($\tau_1$) and Ref.~\cite{deb09b} ($\tau_2$). \label{lt_4p}}
\end{figure}

Apart from even-parity levels, dataset atom\_BB14\_144 contains 44 odd-parity levels belonging to the $\mathrm{3d^54p}$ configuration. Their radiative lifetimes can be compared with those derived from the $A$-values computed with extensive CI (single and double excitations up to $n\leq 5$, $\ell\leq 3$, and~6p as well as selective promotions from the 3s and 3p subshells) with the \textsc{mcbp} code {\sc civ3} ~\cite{deb09b}. As~shown in Figure~\ref{lt_4p}, the~atom\_BB14\_144 lifetimes are on average 0.15~dex shorter. The~larger differences are found for levels belonging to the  $\mathrm{3d^5(^4P)4p\ ^3P^o}$, $\mathrm{3d^5(^4P)4p\ ^5P^o}$, and~$\mathrm{3d^5(^4D)4p\ ^5F^o}$ spectroscopic~terms.

\subsection{Effective Collision~Strengths}\label{ecs_fe3}

\begin{figure}[H]
\includegraphics[width=13.86cm]{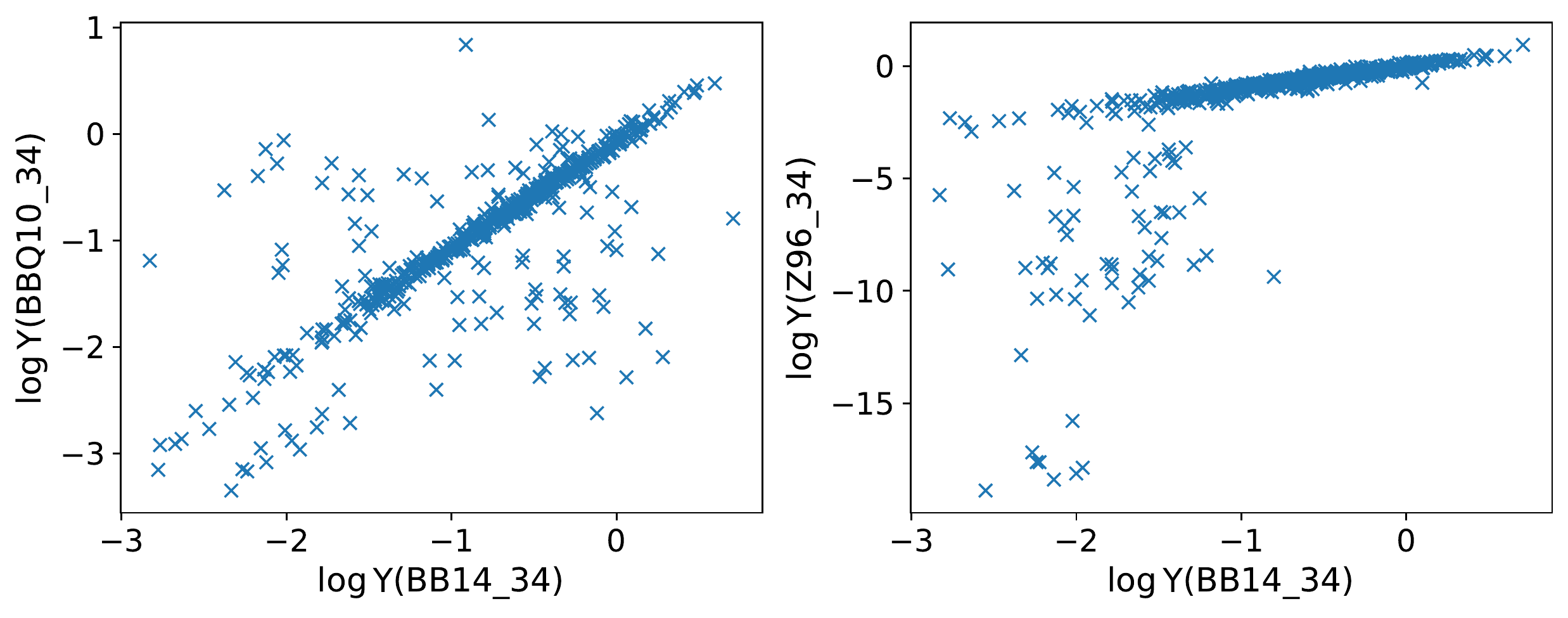}
\caption{Comparison of [\feiii] ECS at $T=10^4$\,K in the coll\_BBQ10\_34 and coll\_Z96\_34 datasets with respect to those in coll\_BB14\_34. \label{fe3_ecs}}
\end{figure}

As shown in Table~\ref{new_data}, we have three revised \texttt{coll} datasets containing ECS for the 34-level  atomic model of [\feiii]: Z96\_34, BBQ10\_34, and~BB14\_34. We compare these data at $T=10^4$\,K in Figure~\ref{fe3_ecs} using the latter as the reference. In~general, there is a reasonable agreement for most transitions, but~for several others, huge differences are encountered. Transitions involving the $\mathrm{3d^5(^6S)4s\ ^7S_3}$ level are problematic. For~instance, in~Figure~\ref{fe3_ecs} (right panel), we have excluded from coll\_Z96\_34 transitions of the type $\mathrm{3d^5(^6S)4s\ ^7S_3}- \mathrm{3d^6}\ ^1L_J$ as they appear with $\Upsilon=0.0$; furthermore, coll\_Z96\_34 transitions with $\Delta S=3/2$ have abnormally small ECS ($\log\Upsilon <-4$). These shortcomings may be due to the non-relativistic method used in~\cite{zha96}. On~the other hand, as~discussed in~\cite{bad14}, the~main differences between the ECS in coll\_BBQ10\_34 and coll\_BB14\_34 are due to a mixup by the former in the assignments of the $\mathrm{3d^5(^6S)4s\ ^7S_3}$ and $\mathrm{3d^6\ ^3G_3}$ adjacent levels. Large discrepancies are also found in transitions of the type $\mathrm{3d^6\ a\,^3P}_J- \mathrm{3d^6\ ^3H}_{J'}$ and $\mathrm{3d^6\ b\,^1G_4}- \mathrm{3d^6\ ^5D}_{J}$.

\unskip

\unskip

\subsection{\feiii\ Discussion}

Our initial step in the present atomic data assessment was the comparison of observed line-intensity ratios with the corresponding theoretical $A$-value ratios for transitions arising from a common upper level (see Section~\ref{aratio_fe3}). This procedure readily yields a first exposure of observational line misidentifications, blending, and~contamination (e.g., telluric or instrumental) and of inherent computational difficulties such as strong valence--core correlation, double-excitation couplings, and~strong CI mixing. Therefore, the~agreement between theoretical ratios, on~the one hand, and~between observed and theoretical ratios, on the other hand, are both~revealing. 

Although the spectra of HH\,202S and HH\,204 are of comparable quality, the~reported observational errors of the former are on average larger (36\%) than the latter (12\%). Comparisons among theoretical $A$-value ratios and with the observed counterparts lead to an overall accuracy level for the theoretical ratios of around 20--30\%, except~for odd theoretical outliers pinpointed in Section~\ref{aratio_fe3} believed to be caused by the aforementioned computational~difficulties.

The spectrum fits performed in Section~\ref{specfit} bring to the fore a frequent drawback in modeling codes: the error-prone mapping of radiative or collisional datasets onto an atomic model with different level numbering as illustrated with dataset coll\_BB14 that led to file deprecation. The~four revised \texttt{atom} datasets (BBQ10\_34, BB14\_34, DH09\_34, Q96\_34) and three \texttt{coll} datasets (BBQ10\_34, BB14\_34, Z96\_34) appear to give comparable temperatures and densities within error bands of 6\% and 20\%, respectively, which also match satisfactorily estimates obtained from [\feiii] line-ratio diagnostics such as $I(\lambda 4658)/I(\lambda 4702)$ and $I(\lambda 4658)/I(\lambda 3240)$ (see Section~\ref{diagnostics_fe3}) and from alternative ionic diagnostics reported in~\cite{mes09, men21b}. However, the~density-sensitive ratio $I(\lambda 4658)/I(\lambda 4702)$ computed with coll\_BBQ10\_34 does not match the ratios observed in low-density ($\log n_e\lesssim 3$)  nebular sources (see Figure~\ref{lowden}). 

As shown in Section~\ref{rlt_fe3}, radiative lifetimes for levels of the $\mathrm{3d^6}$ and $\mathrm{3d^54s}$ configurations computed with $A$-values from the BBQ10\_34, BB14\_34, DH09\_34, and~Q96\_34 \texttt{atom} datasets agree to within $\sim0.1$~dex except for $\mathrm{3d^6\ ^3H_J}$ and $\mathrm{3d^54s\ ^7S_J, ^5S_2}$. However, the~best agreement ($\sim20\%$) is found between $A$-values from structure calculations, namely atom\_DH09\_34 and atom\_Q96\_34, rather than those associated with scattering targets. This may be due to the convergence and accuracy of the CI expansions implemented in each calculation type. In~structure calculations of radiative lifetimes for transitions involving levels of the $\mathrm{3d^6}$ and $\mathrm{3d^54s}$ configurations in \feiii, for~instance, there is no need to include odd-parity configurations, while in a \feiii\ scattering target, a reduced set of configurations of both parity types must be considered. A~similar argument may be applied to the lifetimes of $\mathrm{3d^54p}$ levels (see Section~\ref{rlt_4p_fe3}), where the atom\_BB14\_144 lifetimes are on average shorter than those from a lengthy \textsc{mcbp} structure calculation~\cite{deb09b}.

The main finding of the ECS comparison in Section~\ref{ecs_fe3} is the large discrepancies displayed by coll\_Z96\_34 ECS with respect to coll\_BB14\_34 for some transitions with $\log\Upsilon < -1$ (see Figure~\ref{fe3_ecs}),  which may be due to the non-relativistic $R$-matrix method used by~\cite{zha96} in the computation of the former dataset. The~larger ECS differences between coll\_BBQ10\_34 and coll\_BB14\_34 are mainly due to level~misassignments.

Analyzing results from these data comparisons and spectral fits, we select as the [\feiii] default datasets in \texttt{PyNeb} 1.1.17 the atom\_BB14\_144 and coll\_BB14\_144 files, while files coll\_BBQ10 and coll\_BB14 are deprecated. Deprecated files are still available with the command \\ \verb|pn.atomicData.includeDeprecatedPath()|. Files atom\_BBQ10 and coll\_Z96 remain active, and files coll\_BBQ10\_34, atom\_BB14\_144, atom\_DH09\_34, and~atom\_Q96\_34 are~incorporated.

\section{\feii \label{feii}}

Similarly to the \feiii\ atomic data assessment, we implement observational benchmarks to evaluate the atomic datasets for \feii\  available in {\tt PyNeb} 1.1.16, transcribing them directly from the source and implementing new models. {NIST energy levels in this version of {\tt PyNeb} are the same as those listed in the database of July 2022}~\cite{kra21}. 

\subsection{\feii\ Atomic Datasets in~PyNeb \label{pyneb_fe2}}

The following datasets are available for \feii\ in {\tt PyNeb} 1.1.16:

\begin{description}[style=unboxed,leftmargin=0cm]
\item[atom\_VVKFHF99, coll\_VVKFHF99]---These datasets list radiative and collisional rates for transitions among 80 levels with energies $E< 44,760$\,cm$^{-1}$ compiled from different sources~\cite{ver99}. $A$-values for the dipole allowed and forbidden transitions are from~\cite{nah95} and \cite{qui96b}, respectively. ECS are mainly from~\cite{zha95}; however, the~source data are not available and $A$-values and ECS for transitions with spin change $\Delta S=2$ have been~obviated.

\item[atom\_B15, coll\_B15]---$A$-values and ECS for transitions among 52 levels belonging to the $\mathrm{3d^7}$, $\mathrm{3d^64s}$, and~$\mathrm{3d^54s^2}$ configurations ($E< 31,500$\,cm$^{-1}$)~\cite{bau15}. $A$-values were computed with the \textsc{mcbp} atomic structure code {\sc autostructure} using different expansions and orbital optimization strategies and a dipole correction to the Thomas–Fermi–Dirac–Amaldi potential. Fine-tuning was carried out by means of term-energy corrections adjusted by fitting to the experimental term energies. ECS were computed with the Dirac--Coulomb $R$-matrix package {\sc darc}~\cite{ait96} using a target generated with the fully relativistic atomic structure code {\sc grasp0}~\cite{par96} using a six-configuration expansion comprising 329~levels.

\item[atom\_SRKFB19, coll\_SRKFB19]---$A$-values and ECS for 250 levels from the $\mathrm{3d^7}$, $\mathrm{3d^64s}$, $\mathrm{3d^54s^2}$, $\mathrm{3d^64p}$, and~$\mathrm{3d^54s4p}$ configurations~\cite{smy19}. $A$-values were computed with {\sc grasp0} in a 20-configuration expansion and ECS with {\sc darc} using a 716-level target representation.
\end{description}

\subsection{\feii\ Revised and New Atomic~Datasets \label{newdata_fe2}}

For comparison purposes, models were normalized to 52 levels ($E < 31,500$\,cm$^{-1}$), i.e.,~the number of levels considered in atom\_B15 that includes the last even quartet term before the appearance of the $\mathrm{3d^64p}$ level array. Two datasets have been extended to comprise 173 levels ($E < 66,100$\,cm$^{-1}$),  a~cut before the onset of unmeasured levels in the NIST level list, and~225 levels ($E < 73,800$\,cm$^{-1}$) to discuss some assignment inconsistencies. Although~ECS from~\cite{zha95} are included in VVKFHF99, we have not considered their dataset individually as it excludes the doublet terms in their atomic~model. 

\begin{description}[style=unboxed,leftmargin=0cm]
\item[atom\_QDZa96, atom\_QDZh96]---E2 and M1 $A$-values calculated for transitions among the 63 levels of the configurations $\mathrm{3d^64s}$, $\mathrm{3d^7}$, and~$\mathrm{3d^54s^2}$~\cite{qui96b}. Extensive CI expansion was used with the Breit--Pauli {\sc superstructure}~\cite{eis74} (QDZa96) and Pauli {\sc hfr}~\cite{cow81} (QDZh96) atomic structure codes. These datasets are not complete enough to be used in \texttt{PyNeb} spectral modeling as they obviate transitions such as $\mathrm{a\,^4F-a\,^6D}$ and $\mathrm{a\,^2P-a\,^6D}$; therefore, they will only be used in radiative data~comparisons.

\item[atom\_VVKFHF99\_52, coll\_VVKFHF99\_52]---The VVKFHF99 datasets containing $A$-values and ECS described in Section~\ref{pyneb_fe2} are reduced to 52 levels from even-parity~configurations.

\item[atom\_DH11\_52]---E2 and M1 $A$-values for transitions between levels of the $\mathrm{3d^64s}$, $\mathrm{3d^7}$, and~$\mathrm{3d^54s^2}$ configurations were computed in a large-scale CI framework with the \textsc{mcbp} structure code {\sc civ3}~\cite{deb11}. Wave functions were fine-tuned to fit the experimental~energies.

\item[atom\_B15\_52, coll\_B15\_52]---Data ($A$-values and ECS) from the B15 calculation described in Section~\ref{pyneb_fe2} were downloaded directly from the CDS, and~the datasets were~reconstructed.

\item[atom\_TZ18\_52, coll\_TZ18\_52]---$A$-values and ECS were computed with the close-coupling $B$-spline Breit--Pauli $R$-matrix method for 340 levels of the  $\mathrm{3d^64s}$, $\mathrm{3d^7}$,  $\mathrm{3d^54s^2}$, $\mathrm{3d^64p}$, and~$\mathrm{3d^54s4p}$ configurations~\cite{tay18}.  The~CI target representation was constructed with a Hartree--Fock method. Semi-empirical fine-tuning in the structure and scattering calculations was introduced by fitting tothe experimental energies. The final Hamiltonian of the scattering calculation contains the spin--orbit~term.

\item[atom\_SRKFB19\_52, coll\_SRKFB19\_52]---Radiative and collisional data from the SRKFB19 calculation were described in Section~\ref{pyneb_fe2}. New files were constructed with data extracted from the {\tt adf04\_DARC716\_2.10.18} file (Cathy Ramsbottom, private communication) and reduced to 52~levels.

\item[atom\_SRKFB19\_173, coll\_SRKFB19\_173, atom\_TZ18\_173, coll\_TZ18\_173]---The SRKFB19 and TZ18 datasets have been extended to 173 levels ($E < 66,100$\,cm$^{-1}$) to test model~convergence.

\item[atom\_SRKFB19\_225, atom\_TZ18\_225]---The {theoretical} SRKFB19 and TZ18 datasets have been extended to 225 levels ($E < 73,800$\,cm$^{-1}$) to bring out term assignment inconsistencies involving levels with total orbital angular momentum quantum number $L=5$ (see Table~\ref{srkfb19}). With~respect to NIST and other theoretical datasets, the~term assignments of levels at $21,430.36$ and $21,581.62$~cm$^{-1}$ ($\mathrm{a\,^4H}$) and $26,170.18$ and $26,352.77$~cm$^{-1}$ ($\mathrm{b\,^2H}$) have been interchanged in atom\_SRKFB19\_225. Since the total angular momentum quantum number $J$ of each level coincides in the different datasets, the~incongruent term labeling would be inconsequential if the $A$-values for the transitions involving these levels were comparable. As~shown in Table~\ref{Htrans}, this is not the case, as discrepancies as large as an order of magnitude were encountered. A~similar mixup was found with the odd-parity $\mathrm{z\,^4H^o}$ and $\mathrm{y\,^2H^o}$ levels, which in this case are even energetically misplaced (see Table~\ref{srkfb19}). {Assignment discrepancies of this sort could be due to strong CI admixture in terms with high orbital angular momentum ($L>4$ say) in lowly ionized systems that the {\sc grasp0} multi-configuration Dirac--Hartree--Fock structure code has problems resolving}.
\end{description}

\begin{table}[H]
\caption{Comparison of NIST energies (cm$^{-1}$) for levels in \feii\ displaying a total orbital angular momentum quantum number $L=5$ with those listed in different \texttt{atom} datasets (see Section~\ref{newdata_fe2}). Questionable level assignments are found in atom\_SRKFB19\_225.\label{srkfb19}}
	\begin{adjustwidth}{-\extralength}{0cm}
		\newcolumntype{C}{>{\centering\arraybackslash}X}
		\begin{tabularx}{\textwidth+\extralength}{CCCCCCC}
			\toprule
		  \textbf{Level} & \textbf{NIST} & \textbf{SRKFB19\_225} & \textbf{TZ18\_225} & \textbf{DH11\_52} & \textbf{B15\_52} & \textbf{QDZh96} \\
		    \midrule
$\mathrm{3d^7\ a\,^2H_{11/2}}$ & 20,340.25 & 20,340.25 & 20,327.91 & 20,340.30 & 20,340.30 & 20,340 \\
$\mathrm{3d^7\ a\,^2H_{9/2}}$ & 20,805.76 & 20,805.76 & 20,824.99 & 20,805.77 & 20,805.77 & 20,806 \\
$\mathrm{3d^6(^3H)4s\ a\,^4H_{13/2}}$ & 21,251.58 & 21,251.58 & 21,283.52 & 21,251.61 & 21,251.61 & 21,252 \\
$\mathrm{3d^6(^3H)4s\ a\,^4H_{11/2}}$ & 21,430.36 & {26,170.17} & 21,433.70 & 21,430.36 & 21,430.36 & 21,430 \\
$\mathrm{3d^6(^3H)4s\ a\,^4H_{9/2}}$ & 21,581.62 & {26,352.77} & 21,560.17 & 21,581.64 & 21,581.64 & 21,582 \\
$\mathrm{3d^6(^3H)4s\ a\,^4H_{7/2}}$ & 21,711.90 & 21,711.90 & 21,671.63 & 21,711.92 & 21,711.92 & 21,712 \\
$\mathrm{3d^6(^3H)4s\ b\,^2H_{11/2}}$ & 26,170.18 & {21,430.36} & 26,186.48 & 26,170.18 & 26,170.18 & 26,170 \\
$\mathrm{3d^6(^3H)4s\ b\,^2H_{9/2}}$ & 26,352.77 & {21,581.61} & 26,331.50 & 26,352.77 & 26,352.77 & 26,353 \\
$\mathrm{3d^6(^3H)4p\ z\,^4H^o_{13/2}}$& 60,837.56 & 60,837.55 & 60,832.43 & & & \\
$\mathrm{3d^6(^3H)4p\ z\,^4H^o_{11/2}}$& 60,887.61 & & 60,880.58 & & & \\
$\mathrm{3d^6(^3H)4p\ z\,^4H^o_{9/2}}$ & 60,989.44 & & 61,012.21 & & & \\
$\mathrm{3d^6(^3H)4p\ z\,^4H^o_{7/2}}$ & 61,156.83 & 61,156.82 & 61,178.84 & & & \\
$\mathrm{3d^6(^3H)4p\ z\,^2H^o_{11/2}}$& 65,363.61 & 65,363.60 & 65,333.65 & & & \\
$\mathrm{3d^6(^3H)4p\ z\,^2H^o_{9/2}}$ & 65,556.27 & 65,556.26 & 65,551.82 & & & \\
$\mathrm{3d^6(^3G)4p\ y\,^4H^o_{13/2}}$& 66,411.71 & 66,411.70 & 66,306.76 & & & \\
$\mathrm{3d^6(^3G)4p\ y\,^4H^o_{11/2}}$& 66,463.54 & 66,463.54 & 66,602.84 & & & \\
$\mathrm{3d^6(^3G)4p\ y\,^4H^o_{9/2}}$ & 66,589.04 & 66,589.03 & 66,733.10 & & & \\
$\mathrm{3d^6(^3G)4p\ y\,^4H^o_{7/2}}$ & 66,672.34 & 66,672.32 & 66,705.11 & & & \\
$\mathrm{3d^6(^3G)4p\ y\,^2H^o_{11/2}}$& 67,516.33 & 67,516.32 & 67,942.69 & & & \\
$\mathrm{3d^6(^3G)4p\ y\,^2H^o_{11/2}}$& & {67,709.96} & & & & \\
$\mathrm{3d^6(^3G)4p\ y\,^2H^o_{9/2}}$ & 68,000.79 & 68,000.77 & 67,609.99 & & & \\
$\mathrm{3d^6(^3G)4p\ y\,^2H^o_{9/2}}$ & & {68,201.16} & & & & \\
$\mathrm{3d^6(^1G2)4p\ x\,^2H^o_{9/2}}$ & 72,130.38 & 72,130.36 & 72,407.94 & & & \\
$\mathrm{3d^6(^1G2)4p\ x\,^2H^o_{11/2}}$& 72,261.74 & 72,261.73 & 71,962.96 & & & \\
$\mathrm{3d^6(^1I)4p\ w\,^2H^o_{11/2}}$& 73,603.54 & 73,603.53 & 73,435.97 & & & \\
$\mathrm{3d^6(^1I)4p\ w\,^2H^o_{9/2}}$ & 73,751.28 & 73,751.27 & 73,868.28 & & & \\

			\bottomrule
		\end{tabularx}
	\end{adjustwidth}
\end{table}

\vspace{-6pt}

\begin{table}[H]
\caption{$A$-values for transitions arising from the $\mathrm{3d^64s\ a\,^4H_{J}}$ and $\mathrm{3d^64s\ b\,^2H_{J}}$ levels ($J=9/2$ and $11/2$) upper levels of [\feii] listed in different \texttt{atom} datasets, which show the large discrepancies in atom\_SRKFB19\_225.  \label{Htrans}}
\newcolumntype{C}{>{\centering\arraybackslash}X}
\begin{tabularx}{\textwidth}{CCCCCCC}
\toprule
\boldmath{$\lambda_\mathrm{air}$} \textbf{(\AA)} 
& \multicolumn{6}{c}{\boldmath{$A$}\textbf{-Value (s}\boldmath{$^{-1}$}\textbf{)}}\\
\midrule
  & \textbf{SRKFB19\_225} & \textbf{TZ18\_225} & \textbf{DH11\_52} & \textbf{B15\_52} & \textbf{QDZa96} & \textbf{QDZh96}\\
\midrule
4114.48 & $1.52\times 10^{-2}$ & $1.44\times 10^{-1}$& $1.61\times 10^{-1}$ & $1.05\times 10^{-1}$ & $7.80\times 10^{-2}$ & $1.03\times 10^{-1}$\\
4178.96	& $7.45\times 10^{-4}$ & $3.26\times 10^{-2}$& $2.02\times 10^{-2}$ & $1.89\times 10^{-2}$ & $9.95\times 10^{-3}$ & $1.56\times 10^{-2}$\\
4211.11	& $1.81\times 10^{-2}$ & $5.40\times 10^{-2}$& $5.93\times 10^{-2}$ & $3.15\times 10^{-2}$ & $3.37\times 10^{-2}$ & $4.44\times 10^{-2}$\\
4251.45	& $6.12\times 10^{-3}$ & $2.86\times 10^{-2}$& $1.96\times 10^{-2}$ & $1.33\times 10^{-2}$ & $1.28\times 10^{-2}$ & $1.87\times 10^{-2}$\\
5111.64	& $4.68\times 10^{-1}$ & $1.01\times 10^{-1}$& $1.07\times 10^{-1}$ & $1.19\times 10^{-1}$ & $1.20\times 10^{-1}$ & $1.31\times 10^{-1}$\\
5220.08	& $8.90\times 10^{-2}$ & $1.09\times 10^{-1}$& $1.11\times 10^{-1}$ & $1.32\times 10^{-1}$ & $1.34\times 10^{-1}$ & $1.44\times 10^{-1}$\\
5261.63	& $1.62\times 10^{+0}$ & $3.25\times 10^{-1}$& $3.39\times 10^{-1}$ & $3.95\times 10^{-1}$ & $4.01\times 10^{-1}$ & $4.29\times 10^{-1}$\\
5333.66	& $8.85\times 10^{-1}$ & $2.68\times 10^{-1}$& $2.78\times 10^{-1}$ & $3.23\times 10^{-1}$ & $3.32\times 10^{-1}$ & $3.51\times 10^{-1}$\\
\bottomrule
\end{tabularx}
\end{table}

\subsection{Observational and Spectroscopic~Benchmarks}

Due to the wide astrophysical interest in \feii, this ion presents interesting possibilities for implementing observational and spectroscopic benchmarks. There is interest not only in the forbidden-line spectrum but also in the allowed transition arrays involving levels from the odd-parity configurations $\mathrm{3d^64p}$ and $\mathrm{3d^54s4p}$, which may be populated by fluorescence pumping from the stellar continuum~\cite{rod99,ver00, har13}.  \feii\ ultraviolet emission in active galactic nuclei is not fully understood, impeding quasar classification schemes and Fe abundance estimates~\cite{sar21}.

\subsubsection{$A$-Value Ratio~Benchmark}\label{fe2_Aratio}


In Tables~\ref{Aratio_Fe2_HH202} and \ref{Aratio_Fe2_HH204}, we compare the observed line-intensity ratios in the spectra of HH\,202S and HH\,204 involving transitions from a common upper level with the corresponding $A$-value ratios (see Equation~(\ref{aratio})) from \texttt{atom} datasets B15\_52, DH11\_52, TZ18\_52, SRKFB19\_52, QDZa96, and~QDZh96. In~this comparison, we must point out the worrisome discrepancies displayed by some $A$-value ratios from atom\_SRKFB19\_52 with both the observed and other theoretical values; thus, the~integrity of this radiative dataset is~questionable.

\begin{table}[H]
\caption{Comparison of [\feii] line-intensity ratios for transitions arising from a common upper level in the HH\,202S spectrum (Obs) with those estimated from $A$-values in the following \texttt{atom} datasets: B15\_52 (T1); DH11\_52 (T2); TZ18\_52 (T3); SRKFB19\_52 (T4); QDZa96 (T5); QDZh96 (T6).  The~error of the least significant figure of the observed intensity ratio is indicated in~brackets. \label{Aratio_Fe2_HH202}}
	\begin{adjustwidth}{-\extralength}{0cm}
		\newcolumntype{C}{>{\centering\arraybackslash}X}
		\begin{tabularx}{\textwidth+\extralength}{CCCCCCCCCCCC}
		  \toprule
\multicolumn{3}{c}{\textbf{Line~1}} & \multicolumn{2}{c}{\textbf{Line~2}} & \textbf{Obs}& \textbf{T1}& \textbf{T2}&
          \textbf{T3}& \textbf{T4} & \textbf{T5} & \textbf{T6}\\
          \midrule
          \textbf{Upper}&\textbf{Lower}&\boldmath{$\lambda_\mathrm{air}$}\textbf{(\AA)}&\textbf{Lower}&\boldmath{$\lambda_\mathrm{air}$}\textbf{(\AA)}&\multicolumn{7}{c}{\textbf{Line Intensity Ratio}}\\
          \midrule
$\mathrm{b\ ^2H_{9/2}}$&$\mathrm{a\ ^4F_{7/2}}$&4178.96&$\mathrm{a\ ^4F_{5/2}}$&4251.45&1.3(9)&1.44&1.05&1.16&0.12&0.79&0.85 \\
$\mathrm{b\ ^2H_{11/2}}$&$\mathrm{a\ ^4F_{9/2}}$&4114.48&$\mathrm{a\ ^4F_{7/2}}$&4211.11&2.4(8)&3.42&2.78&2.73&0.86&2.37&2.37\\
$\mathrm{a\ ^4G_{7/2}}$&$\mathrm{a\ ^4F_{5/2}}$&4319.62&$\mathrm{a\ ^4F_{3/2}}$&4372.43&2.4(8)&2.00&1.95&1.97&1.88&1.98&1.96 \\
$\mathrm{a\ ^4G_{9/2}}$&$\mathrm{a\ ^4F_{9/2}}$&4177.20&$\mathrm{a\ ^4F_{7/2}}$&4276.84&0.28(8)&0.24&0.21&0.23&0.29&0.25&0.24\\
$\mathrm{a\ ^4G_{9/2}}$&$\mathrm{a\ ^4F_{7/2}}$&4276.84&$\mathrm{a\ ^4F_{5/2}}$&4352.78&2.1(6)&2.16&2.15&2.24&2.07&2.21&2.19 \\
$\mathrm{a\ ^4G_{11/2}}$&$\mathrm{a\ ^4F_{9/2}}$&4243.97&$\mathrm{a\ ^4F_{7/2}}$&4346.86&5(1)&4.40&4.54&4.67&4.24&4.60&4.59  \\
$\mathrm{a\ ^6S_{5/2}}$&$\mathrm{a\ ^6D_{9/2}}$&4287.39&$\mathrm{a\ ^6D_{7/2}}$&4359.33&1.3(3)&1.37&1.37&1.38&1.35&1.37&1.38 \\
$\mathrm{a\ ^6S_{5/2}}$&$\mathrm{a\ ^6D_{7/2}}$&4359.33&$\mathrm{a\ ^6D_{5/2}}$&4413.78&1.4(3)&1.44&1.43&1.44&1.42&1.42&1.44 \\
$\mathrm{a\ ^6S_{5/2}}$&$\mathrm{a\ ^6D_{5/2}}$&4413.78&$\mathrm{a\ ^6D_{3/2}}$&4452.10&1.6(4)&1.58&1.57&1.58&1.56&1.57&1.58 \\
$\mathrm{a\ ^6S_{5/2}}$&$\mathrm{a\ ^6D_{3/2}}$&4452.10&$\mathrm{a\ ^6D_{1/2}}$&4474.90&2.2(7)&2.07&2.06&2.06&2.05&2.06&2.06 \\
$\mathrm{b\ ^4F_{3/2}}$&$\mathrm{a\ ^6D_{3/2}}$&4509.61&$\mathrm{a\ ^4F_{5/2}}$&4950.76&0.4(2)&0.31&0.40&0.38&0.11&0.28&0.28 \\
$\mathrm{b\ ^4F_{3/2}}$&$\mathrm{a\ ^4F_{5/2}}$&4950.76&$\mathrm{a\ ^4F_{3/2}}$&5020.24&0.9(4)&0.96&0.97&0.98&0.95&0.97&0.97 \\
$\mathrm{b\ ^4F_{5/2}}$&$\mathrm{a\ ^6D_{7/2}}$&4432.45&$\mathrm{a\ ^6D_{5/2}}$&4488.75&0.6(3)&0.36&0.36&0.36&0.36&0.37&0.36 \\
$\mathrm{b\ ^4F_{5/2}}$&$\mathrm{a\ ^6D_{5/2}}$&4488.75&$\mathrm{a\ ^6D_{3/2}}$&4528.38&3(2)&3.44&3.44&3.43&3.38&3.42&3.45   \\
$\mathrm{b\ ^4F_{5/2}}$&$\mathrm{a\ ^6D_{3/2}}$&4528.38&$\mathrm{a\ ^4F_{7/2}}$&4874.50&0.3(2)&0.23&0.30&0.28&0.09&0.22&0.21 \\
$\mathrm{b\ ^4F_{5/2}}$&$\mathrm{a\ ^4F_{7/2}}$&4874.50&$\mathrm{a\ ^4F_{5/2}}$&4973.40&1.3(5)&1.25&1.24&1.27&1.26&1.26&1.24 \\
$\mathrm{b\ ^4F_{5/2}}$&$\mathrm{a\ ^4F_{5/2}}$&4973.40&$\mathrm{a\ ^4F_{3/2}}$&5043.53&1.5(7)&2.01&2.07&1.99&1.61&1.93&1.97 \\
$\mathrm{b\ ^4F_{7/2}}$&$\mathrm{a\ ^6D_{7/2}}$&4457.95&$\mathrm{a\ ^6D_{5/2}}$&4514.90&4(2)&4.30&4.35&4.32&4.22&4.30&4.34   \\
$\mathrm{b\ ^4F_{7/2}}$&$\mathrm{a\ ^6D_{5/2}}$&4514.90&$\mathrm{a\ ^4F_{9/2}}$&4774.73&0.6(3)&0.46&0.60&0.56&0.19&0.44&0.42 \\
$\mathrm{b\ ^4F_{7/2}}$&$\mathrm{a\ ^4F_{9/2}}$&4774.73&$\mathrm{a\ ^4F_{7/2}}$&4905.35&0.6(2)&0.58&0.58&0.60&0.72&0.59&0.59 \\
$\mathrm{b\ ^4F_{9/2}}$&$\mathrm{a\ ^6D_{9/2}}$&4416.27&$\mathrm{a\ ^6D_{7/2}}$&4492.64&7(2)&7.61&7.54&7.70&7.41&7.60&7.74   \\
$\mathrm{b\ ^4F_{9/2}}$&$\mathrm{a\ ^6D_{7/2}}$&4492.64&$\mathrm{a\ ^4F_{9/2}}$&4814.54&0.15(5)&0.14&0.19&0.17&0.07&0.13&0.12\\
$\mathrm{b\ ^4F_{9/2}}$&$\mathrm{a\ ^4F_{9/2}}$&4814.54&$\mathrm{a\ ^4F_{7/2}}$&4947.39&7(2)&7.57&6.35&7.00&0.88&7.09&7.17   \\
$\mathrm{b\ ^4F_{9/2}}$&$\mathrm{a\ ^4F_{7/2}}$&4947.39&$\mathrm{a\ ^4D_{7/2}}$&6809.24&4(2)&3.49&3.22&3.07&70.2&3.98&4.16  \\
$\mathrm{b\ ^4P_{3/2}}$&$\mathrm{a\ ^6D_{5/2}}$&4728.07&$\mathrm{a\ ^4F_{7/2}}$&5158.01&0.9(3)&1.23&1.83&2.00&0.57&1.29&1.19 \\
$\mathrm{b\ ^4P_{3/2}}$&$\mathrm{a\ ^4F_{7/2}}$&5158.01&$\mathrm{a\ ^4F_{5/2}}$&5268.89&3(1)&1.57&1.56&1.57&1.48&1.57&1.56   \\
$\mathrm{a\ ^4H_{7/2}}$&$\mathrm{a\ ^4F_{5/2}}$&5296.84&$\mathrm{a\ ^4F_{3/2}}$&5376.47&0.3(1)&0.32&0.34&0.34&0.13&0.34&0.34 \\
$\mathrm{a\ ^4H_{9/2}}$&$\mathrm{a\ ^4F_{7/2}}$&5220.08&$\mathrm{a\ ^4F_{5/2}}$&5333.66&0.4(1)&0.42&0.41&0.42&0.10&0.41&0.42 \\
$\mathrm{a\ ^4H_{11/2}}$&$\mathrm{a\ ^4F_{9/2}}$&5111.64&$\mathrm{a\ ^4F_{7/2}}$&5261.63&0.32(6)&0.31&0.32&0.32&0.30&0.31&0.31\\
$\mathrm{a\ ^2D2_{3/2}}$&$\mathrm{a\ ^6D_{3/2}}$&4889.71&$\mathrm{a\ ^4F_{5/2}}$&5412.68&3(1)&0.00&0.00&0.00&0.00&0.00&0.00\\
$\mathrm{a\ ^2D2_{3/2}}$&$\mathrm{a\ ^4F_{5/2}}$&5412.68&$\mathrm{a\ ^4F_{3/2}}$&5495.84&2(1)&1.89&1.91&1.90&1.89&1.89&1.90\\
$\mathrm{b\ ^4P_{5/2}}$&$\mathrm{a\ ^4F_{9/2}}$&5273.36&$\mathrm{a\ ^4F_{7/2}}$&5433.15&3(1)&3.32&2.88&3.19&3.27&3.28&3.26\\
$\mathrm{a\ ^2D2_{5/2}}$&$\mathrm{a\ ^4F_{7/2}}$&5527.35&$\mathrm{a\ ^4F_{5/2}}$&5654.87&10(4)&9.26&9.09&9.19&9.12&9.06&8.98\\
$\mathrm{a\ ^2G_{7/2}}$&$\mathrm{a\ ^4F_{7/2}}$&7172.00&$\mathrm{a\ ^4F_{5/2}}$&7388.17&1.4(3)&1.35&1.35&1.36&1.32&1.35&1.35\\
$\mathrm{a\ ^2G_{9/2}}$&$\mathrm{a\ ^4F_{9/2}}$&7155.17&$\mathrm{a\ ^4F_{7/2}}$&7452.56&3.1(5)&3.17&3.22&3.23&3.10&3.19&3.24\\
$\mathrm{a\ ^4P_{1/2}}$&$\mathrm{a\ ^4F_{5/2}}$&9033.49&$\mathrm{a\ ^4F_{3/2}}$&9267.55&0.8(2)&0.77&0.78&0.78&0.50&0.78&0.78\\
$\mathrm{a\ ^4P_{3/2}}$&$\mathrm{a\ ^6D_{5/2}}$&7686.93&$\mathrm{a\ ^4F_{7/2}}$&8891.93&0.27(7)&0.26&0.83&0.44&0.40&0.36&0.44\\
$\mathrm{a\ ^4P_{3/2}}$&$\mathrm{a\ ^4F_{7/2}}$&8891.93&$\mathrm{a\ ^4F_{5/2}}$&9226.63&1.7(3)&1.76&1.79&1.78&0.88&1.78&1.80\\
$\mathrm{a\ ^4P_{5/2}}$&$\mathrm{a\ ^6D_{7/2}}$&7637.52&$\mathrm{a\ ^4F_{7/2}}$&9051.95&0.5(1)&0.59&1.96&1.08&0.18&0.89&1.15\\
$\mathrm{a\ ^4P_{5/2}}$&$\mathrm{a\ ^4F_{7/2}}$&9051.95&$\mathrm{a\ ^4F_{5/2}}$&9399.04&7(2)&5.13&5.32&5.21&2.64&5.46&5.67\\
        \bottomrule
		\end{tabularx}
	\end{adjustwidth}
\end{table}

\begin{table}[H]
\caption{Comparison of [\feii] line-intensity ratios for transitions arising from a common upper level in the HH\,204 spectrum (Obs) with those estimated from $A$-values in the following \texttt{atom} datasets: B15\_52 (T1); DH11\_52 (T2); TZ18\_52 (T3); SRKFB19\_52 (T4); QDZa96 (T5); QDZh96 (T6).  The~error of the least significant figure of the observed intensity ratio is indicated in~brackets. \label{Aratio_Fe2_HH204}}
	\begin{adjustwidth}{-\extralength}{0cm}
		\newcolumntype{C}{>{\centering\arraybackslash}X}
		\begin{tabularx}{\textwidth+\extralength}{CCCCCCCCCCCC}
		  \toprule
\multicolumn{3}{c}{\textbf{Line~1}} & \multicolumn{2}{c}{\textbf{Line~2}} & \textbf{Obs}& \textbf{T1}& \textbf{T2}&
          \textbf{T3}& \textbf{T4} & \textbf{T5} & \textbf{T6}\\
          \midrule
          \textbf{Upper}&\textbf{Lower}&\boldmath{$\lambda_\mathrm{air}$}\textbf{(\AA)}&\textbf{Lower}&\boldmath{$\lambda_\mathrm{air}$}\textbf{(\AA)}&\multicolumn{7}{c}{\textbf{Line Intensity Ratio}}\\
          \midrule
$\mathrm{a\ ^2F_{7/2}}$&$\mathrm{a\ ^4D_{7/2}}$&5163.96&$\mathrm{a\ ^2G_{9/2}}$&8715.80&9(2)&7.50&10.5&11.1&2.81&6.73&7.48\\
$\mathrm{b\ ^2H_{9/2}}$&$\mathrm{a\ ^4F_{7/2}}$&4178.96&$\mathrm{a\ ^4F_{5/2}}$&4251.45&1.2(2)&1.44&1.05&1.16&0.12&0.79&0.85\\
$\mathrm{b\ ^2H_{11/2}}$&$\mathrm{a\ ^4F_{9/2}}$&4114.48&$\mathrm{a\ ^4F_{7/2}}$&4211.11&2.6(2)&3.42&2.78&2.73&0.86&2.37&2.37\\
$\mathrm{a\ ^4G_{5/2}}$&$\mathrm{a\ ^6D_{3/2}}$&3968.27&$\mathrm{a\ ^4F_{5/2}}$&4305.90&7(1)&0.00&0.00&0.00&0.00&0.00&0.00\\
$\mathrm{a\ ^4G_{5/2}}$&$\mathrm{a\ ^4F_{5/2}}$&4305.90&$\mathrm{a\ ^4F_{3/2}}$&4358.37&0.56(8)&0.44&0.42&0.44&0.48&0.46&0.45\\
$\mathrm{a\ ^4G_{7/2}}$&$\mathrm{a\ ^4F_{5/2}}$&4319.62&$\mathrm{a\ ^4F_{3/2}}$&4372.43&2.0(2)&2.00&1.95&1.97&1.88&1.98&1.96\\
$\mathrm{a\ ^4G_{9/2}}$&$\mathrm{a\ ^4F_{9/2}}$&4177.20&$\mathrm{a\ ^4F_{7/2}}$&4276.84&0.25(2)&0.24&0.21&0.23&0.29&0.25&0.24\\
$\mathrm{a\ ^4G_{9/2}}$&$\mathrm{a\ ^4F_{7/2}}$&4276.84&$\mathrm{a\ ^4F_{5/2}}$&4352.78&2.1(1)&2.16&2.15&2.24&2.07&2.21&2.19\\
$\mathrm{a\ ^4G_{11/2}}$&$\mathrm{a\ ^4F_{9/2}}$&4243.97&$\mathrm{a\ ^4F_{7/2}}$&4346.86&4.9(2)&4.40&4.54&4.67&4.24&4.60&4.59\\
$\mathrm{a\ ^6S_{5/2}}$&$\mathrm{a\ ^6D_{9/2}}$&4287.39&$\mathrm{a\ ^6D_{7/2}}$&4359.33&1.39(6)&1.37&1.37&1.38&1.35&1.37&1.38\\
$\mathrm{a\ ^6S_{5/2}}$&$\mathrm{a\ ^6D_{7/2}}$&4359.33&$\mathrm{a\ ^6D_{5/2}}$&4413.78&1.42(6)&1.44&1.43&1.44&1.42&1.42&1.44\\
$\mathrm{a\ ^6S_{5/2}}$&$\mathrm{a\ ^6D_{5/2}}$&4413.78&$\mathrm{a\ ^6D_{3/2}}$&4452.10&1.55(6)&1.58&1.57&1.58&1.57&1.57&1.58\\
$\mathrm{a\ ^6S_{5/2}}$&$\mathrm{a\ ^6D_{3/2}}$&4452.10&$\mathrm{a\ ^6D_{1/2}}$&4474.90&2.0(1)&2.07&2.06&2.06&2.05&2.06&2.06\\
$\mathrm{b\ ^4F_{3/2}}$&$\mathrm{a\ ^6D_{3/2}}$&4509.61&$\mathrm{a\ ^4F_{5/2}}$&4950.76&0.42(9)&0.31&0.40&0.38&0.11&0.28&0.28\\
$\mathrm{b\ ^4F_{3/2}}$&$\mathrm{a\ ^4F_{5/2}}$&4950.76&$\mathrm{a\ ^4F_{3/2}}$&5020.24&1.4(3)&0.96&0.97&0.98&0.95&0.97&0.97\\
$\mathrm{b\ ^4F_{5/2}}$&$\mathrm{a\ ^6D_{7/2}}$&4432.45&$\mathrm{a\ ^6D_{5/2}}$&4488.75&0.49(7)&0.36&0.36&0.36&0.36&0.37&0.36\\
$\mathrm{b\ ^4F_{5/2}}$&$\mathrm{a\ ^6D_{5/2}}$&4488.75&$\mathrm{a\ ^6D_{3/2}}$&4528.38&3.6(7)&3.44&3.44&3.43&3.38&3.42&3.45\\
$\mathrm{b\ ^4F_{5/2}}$&$\mathrm{a\ ^6D_{3/2}}$&4528.38&$\mathrm{a\ ^4F_{7/2}}$&4874.50&0.25(5)&0.23&0.30&0.28&0.09&0.22&0.21\\
$\mathrm{b\ ^4F_{5/2}}$&$\mathrm{a\ ^4F_{7/2}}$&4874.50&$\mathrm{a\ ^4F_{5/2}}$&4973.40&1.4(1)&1.25&1.24&1.27&1.26&1.26&1.24\\
$\mathrm{b\ ^4F_{5/2}}$&$\mathrm{a\ ^4F_{5/2}}$&4973.40&$\mathrm{a\ ^4F_{3/2}}$&5043.53&1.9(3)&2.01&2.07&1.99&1.61&1.93&1.97\\
$\mathrm{b\ ^4F_{7/2}}$&$\mathrm{a\ ^6D_{9/2}}$&4382.74&$\mathrm{a\ ^6D_{7/2}}$&4457.95&0.19(2)&0.20&0.21&0.20&0.20&0.21&0.20\\
$\mathrm{b\ ^4F_{7/2}}$&$\mathrm{a\ ^6D_{7/2}}$&4457.95&$\mathrm{a\ ^6D_{5/2}}$&4514.90&4.3(4)&4.30&4.35&4.32&4.22&4.30&4.34\\
$\mathrm{b\ ^4F_{7/2}}$&$\mathrm{a\ ^6D_{5/2}}$&4514.90&$\mathrm{a\ ^4F_{9/2}}$&4774.73&0.56(6)&0.46&0.60&0.56&0.19&0.44&0.42\\
$\mathrm{b\ ^4F_{7/2}}$&$\mathrm{a\ ^4F_{9/2}}$&4774.73&$\mathrm{a\ ^4F_{7/2}}$&4905.35&0.57(4)&0.58&0.58&0.60&0.72&0.59&0.59\\
$\mathrm{b\ ^4F_{9/2}}$&$\mathrm{a\ ^6D_{9/2}}$&4416.27&$\mathrm{a\ ^6D_{7/2}}$&4492.64&7.7(5)&7.61&7.54&7.70&7.41&7.60&7.74\\
$\mathrm{b\ ^4F_{9/2}}$&$\mathrm{a\ ^6D_{7/2}}$&4492.64&$\mathrm{a\ ^4F_{9/2}}$&4814.54&0.15(1)&0.14&0.19&0.17&0.07&0.13&0.12\\
$\mathrm{b\ ^4F_{9/2}}$&$\mathrm{a\ ^4F_{9/2}}$&4814.54&$\mathrm{a\ ^4F_{7/2}}$&4947.39&7.8(5)&7.57&6.35&7.00&0.88&7.09&7.17\\
$\mathrm{b\ ^4F_{9/2}}$&$\mathrm{a\ ^4F_{7/2}}$&4947.39&$\mathrm{a\ ^4D_{7/2}}$&6809.24&3.2(4)&3.49&3.22&3.07&70.21&3.98&4.16\\
$\mathrm{b\ ^4P_{3/2}}$&$\mathrm{a\ ^6D_{5/2}}$&4728.07&$\mathrm{a\ ^4F_{5/2}}$&5268.89&2.3(4)&1.93&2.85&3.14&0.84&2.02&1.85\\
$\mathrm{a\ ^4H_{7/2}}$&$\mathrm{a\ ^4F_{5/2}}$&5296.84&$\mathrm{a\ ^4F_{3/2}}$&5376.47&0.36(5)&0.32&0.34&0.34&0.13&0.34&0.34\\
$\mathrm{a\ ^4H_{9/2}}$&$\mathrm{a\ ^4F_{7/2}}$&5220.08&$\mathrm{a\ ^4F_{5/2}}$&5333.66&0.44(5)&0.42&0.41&0.42&0.10&0.41&0.42\\
$\mathrm{a\ ^4H_{11/2}}$&$\mathrm{a\ ^4F_{9/2}}$&5111.64&$\mathrm{a\ ^4F_{7/2}}$&5261.63&0.34(3)&0.31&0.32&0.32&0.30&0.31&0.31\\
$\mathrm{a\ ^2D2_{3/2}}$&$\mathrm{a\ ^6D_{3/2}}$&4889.71&$\mathrm{a\ ^4F_{3/2}}$&5495.84&12(2)&0.00&0.00&0.00&0.00&0.00&0.00\\
$\mathrm{b\ ^4P_{5/2}}$&$\mathrm{a\ ^4F_{9/2}}$&5273.36&$\mathrm{a\ ^4F_{7/2}}$&5433.15&2.7(2)&3.32&2.88&3.19&3.27&3.28&3.26\\
$\mathrm{b\ ^4P_{5/2}}$&$\mathrm{a\ ^4F_{7/2}}$&5433.15&$\mathrm{a\ ^4D_{7/2}}$&7764.71&7(1)&7.22&5.64&4.88&22.63&7.01&7.42\\
$\mathrm{a\ ^2G_{7/2}}$&$\mathrm{a\ ^4F_{9/2}}$&6896.17&$\mathrm{a\ ^4F_{7/2}}$&7172.00&0.10(1)&0.10&0.09&0.09&0.09&0.10&0.09\\
$\mathrm{a\ ^2G_{7/2}}$&$\mathrm{a\ ^4F_{7/2}}$&7172.00&$\mathrm{a\ ^4F_{5/2}}$&7388.17&1.34(9)&1.35&1.35&1.36&1.32&1.35&1.35\\
$\mathrm{a\ ^2G_{9/2}}$&$\mathrm{a\ ^4F_{9/2}}$&7155.17&$\mathrm{a\ ^4F_{7/2}}$&7452.56&3.2(2)&3.17&3.22&3.23&3.10&3.19&3.24\\
$\mathrm{a\ ^4P_{1/2}}$&$\mathrm{a\ ^6D_{3/2}}$&7665.28&$\mathrm{a\ ^6D_{1/2}}$&7733.13&2.8(7)&3.26&3.31&3.28&4.44&3.26&3.28\\
$\mathrm{a\ ^4P_{1/2}}$&$\mathrm{a\ ^6D_{1/2}}$&7733.13&$\mathrm{a\ ^4F_{5/2}}$&9033.49&0.11(2)&0.10&0.32&0.17&0.10&0.14&0.17\\
$\mathrm{a\ ^4P_{1/2}}$&$\mathrm{a\ ^4F_{5/2}}$&9033.49&$\mathrm{a\ ^4F_{3/2}}$&9267.55&0.75(7)&0.77&0.78&0.78&0.50&0.78&0.78\\
$\mathrm{a\ ^4P_{3/2}}$&$\mathrm{a\ ^6D_{5/2}}$&7686.93&$\mathrm{a\ ^6D_{1/2}}$&7874.23&7(1)&7.06&7.31&7.05&11.24&&\\
$\mathrm{a\ ^4P_{3/2}}$&$\mathrm{a\ ^6D_{1/2}}$&7874.23&$\mathrm{a\ ^4F_{7/2}}$&8891.93&0.03(1)&0.04&0.11&0.06&0.04&&\\
$\mathrm{a\ ^4P_{3/2}}$&$\mathrm{a\ ^4F_{7/2}}$&8891.93&$\mathrm{a\ ^4F_{5/2}}$&9226.63&1.7(2)&1.76&1.79&1.78&0.88&1.78&1.80\\
$\mathrm{a\ ^4P_{5/2}}$&$\mathrm{a\ ^6D_{3/2}}$&7926.88&$\mathrm{a\ ^4F_{7/2}}$&9051.95&0.05(1)&0.04&0.14&0.08&0.00&&\\
$\mathrm{a\ ^4P_{5/2}}$&$\mathrm{a\ ^4F_{7/2}}$&9051.95&$\mathrm{a\ ^4F_{5/2}}$&9399.04&5.4(6)&5.13&5.32&5.21&2.64&5.46&5.67\\          
        \bottomrule
		\end{tabularx}
	\end{adjustwidth}
\end{table}

For two cases, the~observed line-intensity ratios are very different from the corresponding theoretical values: $I(\lambda 3968.27)/I(\lambda 4305.90)$ and $I(\lambda 4889.71)/I(\lambda 5412.68)$, which are caused by the misidentification of the stronger [\feii] $\lambda\lambda 3968.66$, 4889.62 lines with the weaker [\feii] $\lambda\lambda 3968.27, 4889.71$, which are emitted from different atomic~transitions.

Excluding atom\_SRKFB19\_52, the~$A$-value ratios in Tables~\ref{Aratio_Fe2_HH202} and \ref{Aratio_Fe2_HH204} agree to $\sim 20{-}25\%$ except for two cases in HH\,202S, $I(\lambda 7637.52)/I(\lambda 9051.95)$ and $I(\lambda 7686.93)/I(\lambda 8891.93)$, and~one in HH\,204, $I(\lambda 7733.13)/I(\lambda 9033.49)$, which are caused by lines with $A$-values  of order $\lesssim10^{-3}$ subject to large numerical uncertainties. Most theoretical ratios are within the estimated observational error bars except for those involving $A$-values strongly affected by level mixing effects that are frequent in \feii\ 
\cite{deb10a}.

\subsubsection{[\feii] Spectrum~Fits \label{specfit_fe2}}

In a similar manner to [\feiii] (see Section~\ref{specfit}), we fit the reliable [\feii] lines of the high-resolution spectra of the bright objects HH\,202S and HH\,204 with {\tt PyNeb} theoretical emissivities optimized through a least-square procedure in terms of the electron temperature and density. The~theoretical emissivity and observed intensity of each line are, respectively, normalized to the sum of the emissivities and the sum of~intensities.

In the spectrum fit of HH\,202S and HH\,204, a~selection of lines were excluded mainly due to observational problems~\cite{mes09,men21b}:
$\lambda\lambda$4413.78, 4416.27, 4452.10, 4509.61, 4514.90, 4528.38, 4889.71, and~4905.35 in addition to $\lambda\lambda$4276.84, 4319.62, 4359.33, 4452.10, 7665.28, 9399.04, and~9997.49 in the latter source. The~line $\lambda$4382.74 is also questionable due to an incorrect~identification.

We show in Table~\ref{old_data_fe2} the results of the spectral fits with the atomic datasets available in {\tt PyNeb}~1.1.16. In~contrast to [\feiii] (see Section~\ref{specfit}), the~optimization procedure does not always converge to the desired accuracy (e.g., SRKFB19 for HH\,204), and~the density variation may not show a minimum or may depend on the initial input value. The~temperature scatter is worrisome, and the HH\,204 fits, in particular, show large values of $\chi^2$.

\begin{table}[H]
\caption{Temperature and density fits of the [\feii] observed spectra of HH\,202S and HH\,204 with the radiative ({\tt atom}) and collisional ({\tt coll}) datasets currently available in {\tt PyNeb} 1.1.16.\label{old_data_fe2}}
	\begin{adjustwidth}{-\extralength}{0cm}
		\newcolumntype{C}{>{\centering\arraybackslash}X}
		\begin{tabularx}{\textwidth+\extralength}{CCC>{\centering\arraybackslash}m{2cm}CC>{\centering\arraybackslash}m{2cm}C}
			\toprule
             \multicolumn{2}{c}{\textbf{Datasets}} & \multicolumn{3}{c}{\textbf{HH\,202S}} & \multicolumn{3}{c}{\textbf{HH\,204}}\\
            \midrule
			\textbf{Atom}&\textbf{Coll}& \boldmath{$T_e(10^3\,\textrm{K})$} & \boldmath{$n_e(10^4$} \textbf{cm}\boldmath{$^{-3})$} & \boldmath{$\chi^2$}& \boldmath{$T_e(10^3\,\textrm{K})$} & \boldmath{$n_e(10^4$} \textbf{cm}\boldmath{$^{-3}$})& \boldmath{$\chi^2$} \\		
			\midrule
			B15	     & B15      & 13.0 & 6.43 & 4.40  & 24.6 & 3.92 & 90.8 \\
			VVKFHF99 & VVKFHF99 & 9.87 & 1.52 & 7.80  & 12.8 & 0.98 & 105. \\
            SRKFB19  & SRKFB19  & 18.8 & 2.67 & 19.6  &      &      &  \\
			\bottomrule
		\end{tabularx}
	\end{adjustwidth}
\end{table}

The revised and new \feii\ atomic models comprise four radiative ({\tt atom}) and four collisional ({\tt coll}) files normalized to 52 levels. The~spectral fits for HH\,202S and HH\,204 are shown in Table~\ref{new_data_fe2}, where each collisional dataset appears to lead to a diverse temperature and density pair $(T_e,n_e)$. For~HH\,202S in units of $10^3$\,K and  $10^4$\,cm$^{-3}$, respectively,
\begin{description}
\item[coll\_B15\_52:] $(12.0\pm 0.1,6.3\pm 0.6)$
\item[coll\_TZ18\_52:] $(8.3\pm 0.4,3.9\pm 0.5)$
\item[coll\_SRKFB19\_52:] $(10.3\pm 0.6,2.4\pm 0.3)$
\item[coll\_VVKFHF99\_52:] $(10.2\pm 0.7,1.6\pm 0.2)$
\end{description}
where the scatter in both temperature and density due to the atomic data is surprisingly broad, and~only the temperature from coll\_TZ18\_52, $T_e=(8.3\pm 0.4)\times 10^3$\,K, agrees with that determined from the [\feiii] fit, $T_e=(7.9\pm 0.3)\times 10^3$\,K (see Section~\ref{specfit}). On~the other hand, the~density $n_e=(2.1\pm 0.3)\times 10^4$\,cm$^{-3}$ obtained in the [\feiii] fit matches the values by coll\_SRKFB19\_52, $(2.4\pm 0.3)\times 10^4$\,cm$^{-3}$, and~coll\_VVKFHF99\_52, $(1.6\pm 0.2)\times 10^4$\,cm$^{-3}$.

For HH\,204,
\begin{description}
\item[coll\_B15\_52:] $(18.4\pm 0.3,3.8\pm 0.2)$
\item[coll\_TZ18\_52:] $(11.9\pm 0.1,3.3\pm 0.6)$
\item[coll\_SRKFB19\_52:] $(15.7\pm 0.1,1.0\pm 0.2)$
\item[coll\_VVKFHF99\_52:] $(13.6\pm 0.2,0.85\pm 0.07)$
\end{description}
whereby the temperatures are significantly higher than the [\feiii] value $T_e=(8.9\pm 0.6)\times 10^3$\,K, while the [\feiii] density, $n_e=(1.5\pm 0.3)\times 10^4$\,cm$^{-3}$, favors coll\_SRKFB19\_52.

\begin{table}[t]
\caption{Temperature and density fits of the [\feii] observed spectra in HH\,202S and HH\,204 with the improved and new radiative ({\tt atom}) and collisional ({\tt coll}) datasets. \label{new_data_fe2}}
	\begin{adjustwidth}{-\extralength}{0cm}
		\newcolumntype{C}{>{\centering\arraybackslash}X}
		\begin{tabularx}{\textwidth+\extralength}{CCC>{\centering\arraybackslash}m{2cm}CC>{\centering\arraybackslash}m{2cm}C}
			\toprule
             \multicolumn{2}{c}{\textbf{Datasets}} & \multicolumn{3}{c}{\textbf{HH\,202S}} & \multicolumn{3}{c}{\textbf{HH\,204}}\\
            \midrule
			\textbf{Atom}&\textbf{Coll}& \boldmath{$T_e(10^3\,\textrm{K})$} & \boldmath{$n_e(10^4$} \textbf{cm}\boldmath{$^{-3})$} & \boldmath{$\chi^2$}& \boldmath{$T_e(10^3\,\textrm{K})$} & \boldmath{$n_e(10^4$} \textbf{cm}\boldmath{$^{-3}$})& \boldmath{$\chi^2$} \\		
			\midrule
            B15\_52     &B15\_52     &11.6&5.99&1.51&16.5&3.90&67.0\\
            DH11\_52    &B15\_52     &13.4&6.00&4.60&22.7&4.00&94.7\\
            TZ18\_52    &B15\_52     &11.4&5.98&1.94&17.1&3.69&70.1\\
            VVKFHF99\_52&B15\_52     &11.7&7.12&2.10&17.4&3.52&75.5\\
            TZ18\_52    &TZ18\_52    &7.94&3.29&7.90&11.2&2.63&154.\\
            B15\_52     &TZ18\_52    &8.14&4.10&7.37&11.5&3.62&145.\\
            DH11\_52    &TZ18\_52    &8.92&3.95&9.06&13.1&3.02&160.\\
            VVKFHF99\_52&TZ18\_52    &8.09&4.43&7.37&11.8&3.95&147.\\
            B15\_52     &SRKFB19\_52 &10.0&2.05&2.20&14.5&0.89&85.8\\
            DH11\_52    &SRKFB19\_52 &11.1&2.58&4.77&17.3&0.89&96.3\\
            TZ18\_52    &SRKFB19\_52 &9.91&2.21&2.75&15.2&1.15&84.7\\
            VVKFHF99\_52&SRKFB19\_52 &10.1&2.73&2.71&15.6&1.24&90.4\\
            VVKFHF99\_52&VVKFHF99\_52&9.92&1.66&7.79&12.7&0.89&106.\\
            B15\_52     &VVKFHF99\_52&9.85&1.43&7.19&12.1&0.81&88.3\\
            DH11\_52    &VVKFHF99\_52&11.2&1.75&12.5&16.7&0.93&164.\\
            TZ18\_52    &VVKFHF99\_52&9.94&1.47&7.70&13.0&0.79&101.\\
            TZ18\_173   &TZ18\_173   &7.81&3.29&7.91&10.4&1.51&40.2\\
            TZ18\_173   &SRKFB19\_173&9.75&2.32&2.84&15.5&1.19&64.2\\
			\bottomrule
		\end{tabularx}
	\end{adjustwidth}
\end{table}

\begin{figure}[H]
\includegraphics[width=13. cm]{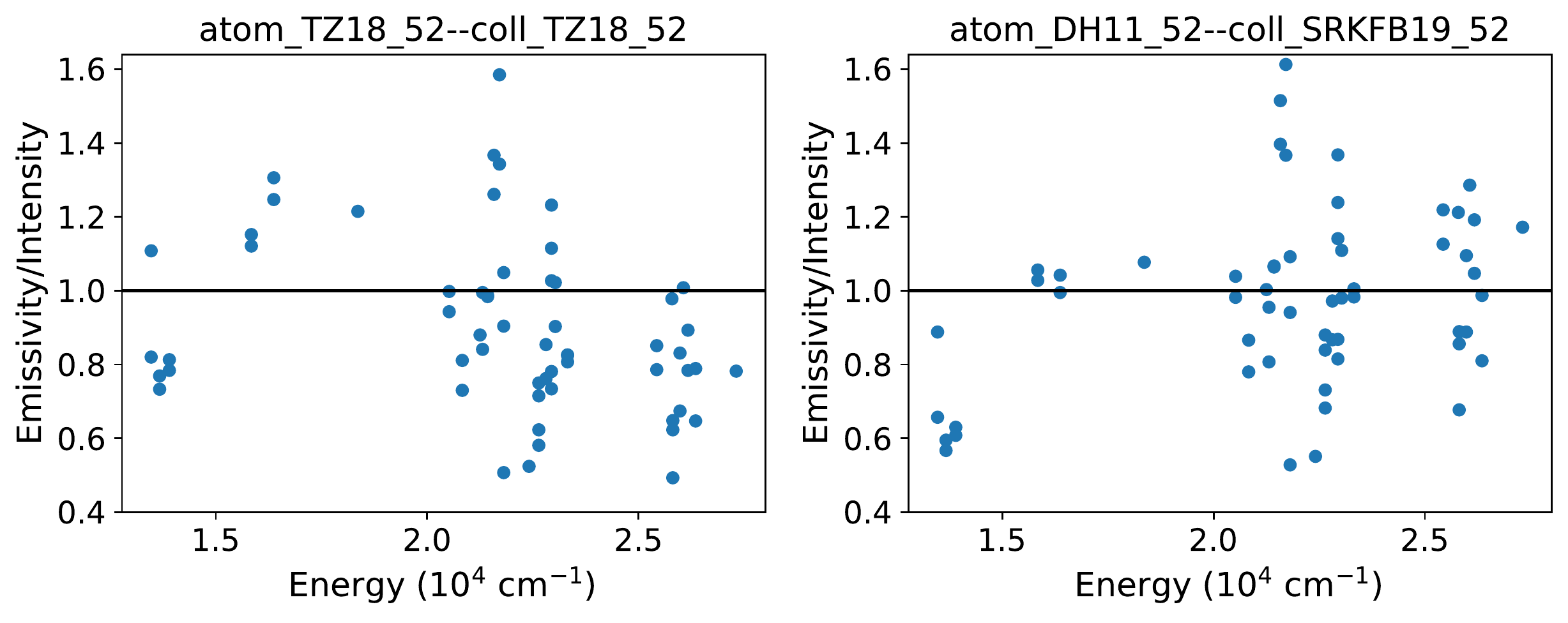}
\caption{Ratio of the normalized theoretical emissivity and observed intensity for lines in HH\,202S as a function of the upper-level energy of the transition. \textbf{Left panel}: spectral fit with datasets atom\_TZ18\_52 and coll\_TZ18\_52 at $T_e=7.94\times 10^3$\,K and $n_e=3.29\times10^4$\,cm$^{-3}$. \textbf{Right panel}: spectral fit with datasets atom\_DH11\_52 and coll\_SRKFB19\_52 at $T_e=11.1\times 10^3$\,K and $n_e=2.58\times10^4$\,cm$^{-3}$. \label{fe2_fluor}} 
\end{figure}

This confusing situation may lead to different analyses. For~instance, in~Figure~\ref{fe2_fluor}, we plot for lines in HH\,202S the ratio of the normalized theoretical emissivity to the normalized observed intensity  as a function of the upper-level energy of the transition. In~the left panel, the~spectrum was fitted with datasets on atom\_TZ18\_52 and coll\_TZ18\_52, resulting in the low-temperature  $T_e=7.94\times 10^3$\,K and density $n_e=3.29\times10^4$\,cm$^{-3}$. For~transitions with upper-level energies $E<2\times 10^4$\,cm$^{-1}$, the~number of ratios above and below 1.0 is roughly equal, while at the higher energies, the number of ratios above 1.0 progressively diminishes. This behavior is similar to that depicted in Figure~4 of~\cite{ver00}, interpreted as a signature of missing fluorescence pumping by the stellar continuum in the plasma model.

\noindent
In~the right panel, the~spectral fit was performed with datasets atom\_DH11\_52 and coll\_SRKFB19\_52, resulting in a higher temperature $T_e=11.1\times 10^3$\,K and comparable density $n_e=2.58\times10^4$\,cm$^{-3}$; however, the~missing continuum-pumping signature is not present. The~diverse, high temperatures of Table~\ref{new_data_fe2} might thus be interpreted as attempts to fit the spectra without accounting for all the level-populating~mechanisms. 

\begin{figure}[H]
\includegraphics[width=6.2 cm]{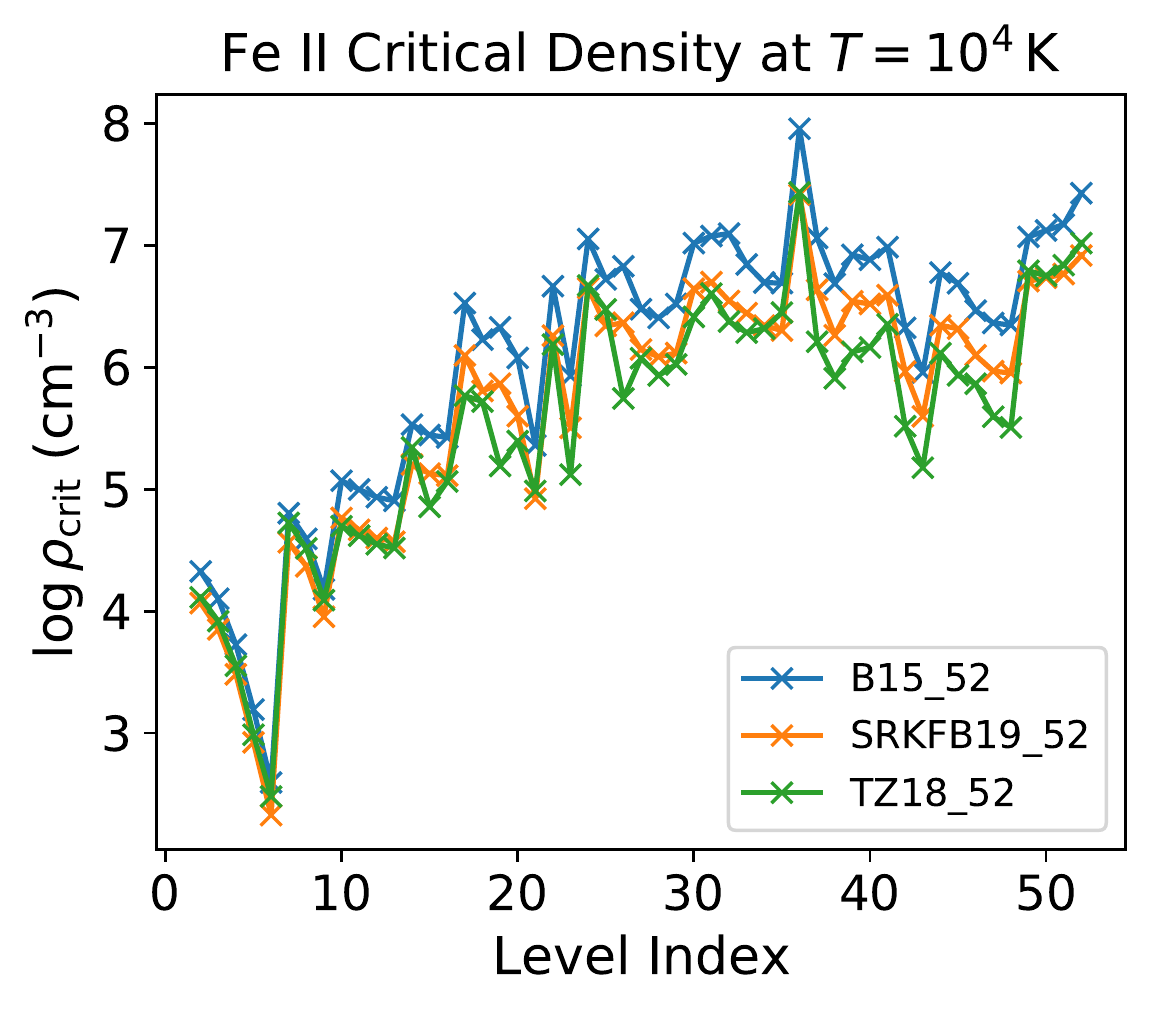} 
\caption{Critical densities for the lower 52 levels of [\feii] predicted at $T=10^4$\,K by using ECS from the \texttt{coll} datasets B15\_52, SRKFB19\_52, and~TZ18\_52  and $A$-values from atom\_TZ18\_52. \label{fe2_cd}}
\end{figure}

A second reading is to look at the critical densities of the levels. In~Figure~\ref{fe2_cd}, we plot the critical densities at $T_e=10^4$\,K for the lower 52 levels of [\feii] computed with dataset atom\_TZ18\_52 and three \texttt{coll} files: B15\_52, SRKFB19\_52, and~TZ18\_52. For~levels with index $i>10$, critical densities by coll\_B15\_52 are a factor of two higher, and~for $i>35$, they are all significantly different from each other. A~key feature is the low critical density ($\log\rho_\mathrm{crit}< 2.6$) of the $\mathrm{3d^7\ a\,^4F_{9/2}}$ metastable level (level $i=6$). For~an electron density $n_e=2\times 10^4$\,cm$^{-3}$, the~population of this level may be larger than that of the ground level for temperatures as low as $T_e\approx 7\times 10^3$\,K; therefore, this level plays a dominant role in the spectral formation of [\feii].

\begin{figure}[H]
\includegraphics[width=13.86cm]{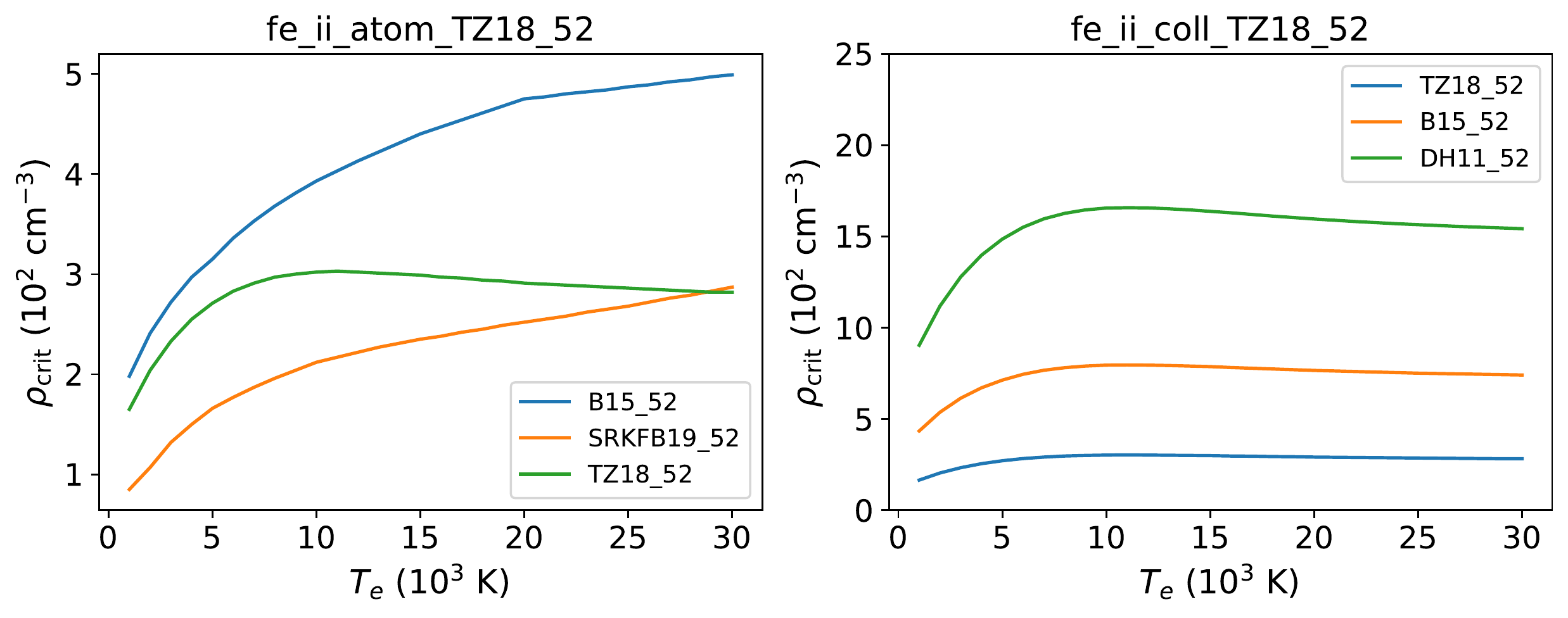}
\caption{Critical density of level $\mathrm{3d^7\ a\,^4F_{9/2}}$ of [\feii] as a function of temperature computed with different atomic datasets. \textbf{Left panel}: atom\_TZ18\_52 file and three \texttt{coll} files: B15\_52, SRKFB19\_52, and~TZ18\_52. \textbf{Right panel}: coll\_TZ18\_52 file and three \texttt{atom} files: TZ18\_52, B15\_52, and~DH11\_52. \label{fe2_cd2}}
\end{figure}

Moreover, in~Figure~\ref{fe2_cd2}, we plot the critical density of level $\mathrm{3d^7\ a\,^4F_{9/2}}$ as a function of temperature using different atomic datasets. In~the right panel, we show results computed with the dataset coll\_TZ18\_52 and three \texttt{atom} files (TZ18\_52, B15\_52, and~DH11\_52). Although~the sensitivity to the radiative data is extreme, for~$T_e> 5\times 10^3$\,K, it is approximately temperature independent. In~the left panel, the critical density is computed with the atom\_TZ18\_52  and three \texttt{coll} datasets (B15\_52, SRKFB19\_52, and~TZ18\_52). The~dependency on the collisional data is not as marked as on the radiative data, but it is temperature~dependent.

In Table~\ref{new_data_fe2}, we also list the fit results using the extended dataset pairs atom\_TZ18\_173--coll\_TZ18\_173 and atom\_TZ18\_173--coll\_SRKFB19\_173. For~the HH\,202S spectrum, the~resulting temperature (in $10^3$\,K units), density (in $10^4$\,cm$^{-3}$ units), and~$\chi^2$ values $(T_e,n_e,\chi^2)=(7.81,3.29,7.91)$ and  $(T_e,n_e,\chi^2)=(9.75,2.32,2.84)$, respectively, are, as~expected, close to those from the atom\_TZ18\_52--coll\_TZ18\_52 and atom\_TZ18\_52--coll\_SRKFB19\_52 pairs: $(T_e,n_e,\chi^2)=(7.94,3.29,7.90)$ and $(T_e,n_e,\chi^2)=(9.91,2.21,2.75)$. However, for~HH\,204, the smaller $\chi^2$ values in  atom\_TZ18\_173--coll\_TZ18\_173 and atom\_TZ18\_173--SRKFB19\_173 indicate improved fits:  $\chi^2= 40.2.$ and 64.2 as compared to $\chi^2= 154.$ and 84.7 in the pairs atom\_TZ18\_52--coll\_TZ18\_52 and atom\_TZ18\_52--coll\_SRKFB19\_52, respectively. Moreover, although~the temperatures and densities with coll\_SRKFB19\_52 and coll\_SRKFB19\_173 are close, the~densities with coll\_TZ18\_52 and coll\_TZ18\_173 differ substantially: 2.63 and 1.51, respectively.
 
\subsubsection{Line-Ratio~Diagnostics}\label{fe2_lrdiag}

An extensive discussion of [\feii] line-ratio diagnostics in the infrared, near-infrared, and~optical is given in~\cite{bau15}. Using lines from the HH\,202S and HH\,204 spectra, we have examined several line ratios to be used as density diagnostics involving ``safe'' low-energy levels to avoid predicted fluorescence effects~\cite{ver00}. Since temperature diagnostics inherently rely on levels at higher energies, we have refrained from their treatment in the present atomic data~comparisons. 

\begin{figure}[H]
\includegraphics[width=13.86cm]{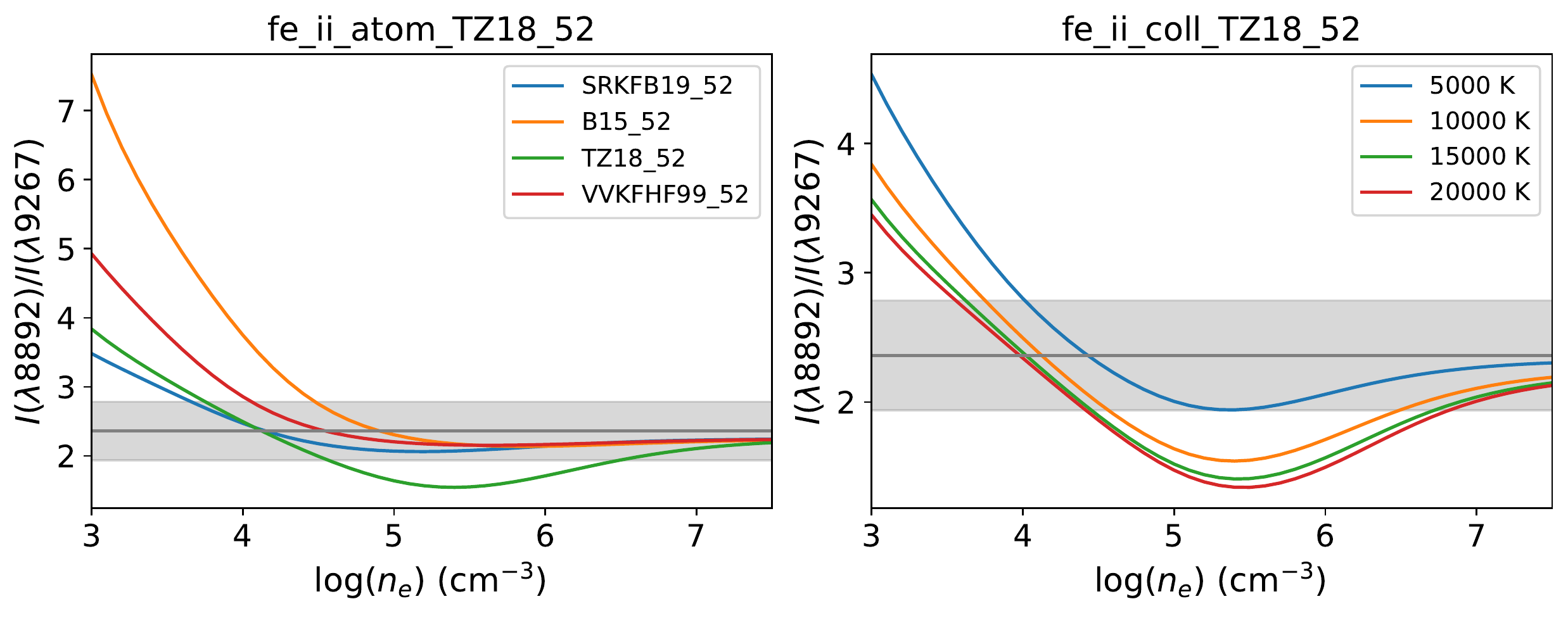}
\caption{Density behavior of the [\feii] $I(\lambda8892)/I(\lambda9267)$ emissivity ratio  using different atomic datasets. The~observed ratio in HH\,202S is $2.36\pm 0.42$, which is depicted as a gray band. \textbf{Left panel}: atom\_TZ18\_52 radiative dataset with four \texttt{coll} datasets (SRKFB19\_52, B15\_52, TZ18\_52, VVKFHF99\_52) at $T_e=10^4$~K. \textbf{Right panel}: atom\_TZ18\_52 and coll\_TZ18\_52 datasets at four electron~temperatures. \label{fe2_diag}} 
\end{figure}

We have selected as a showcase in Figure~\ref{fe2_diag} the emissivity ratio $I(\lambda 8892)/I(\lambda 9267)$ that is density sensitive in the range $3\leq \log n_e\leq 5$\,cm$^{-3}$. In~the left panel, we plot the emissivity ratio computed with the atom\_TZ18\_52 radiative dataset and several \texttt{coll} ECS files highlighting its dependency on the collisional datasets. We do not show its behavior below $\log n_e=3$ as it becomes more complicated, manifesting divergences also due to the radiative data. We also include in Figure~\ref{fe2_diag} the observed intensity ratio in HH\,202S, $I(\lambda 8892)/I(\lambda 9267)=2.36\pm 0.42$, which, except~for the curve computed with coll\_TZ18\_52 ECS, seems to degrade the diagnostic potential of this ratio for $\log n_e\gtrapprox 4.5$ since the density variations lie within its error band. On~the other hand, the~temperature dependency of the ratio above $T_e\approx 10^4$~K is generally weak, as shown in the right~panel.

If this observed line-intensity ratio in HH\,202S is used to determine the density, we obtain $7.94\times 10^4$, $1.32\times 10^4$, $1.41\times 10^4$, and~$3.55\times 10^4$\,cm$^{-3}$ for the \texttt{coll} datasets B15\_52, TZ18\_52, SRKFB19\_52, and~VVKFHF99\_52, respectively, which compare poorly with  estimates from the spectrum fits of Section~\ref{specfit_fe2}: respectively, $5.98\times 10^4$,  $3.29\times 10^4$,  $2.21\times 10^4$, and~ $1.47\times 10^4$\,cm$^{-3}$. Nonetheless, if~we take into consideration the gross uncertainties brought about by the collisional datasets, on~average, both methods coincide on a similar poor density diagnostic: $(4\pm 3)\times 10^4$\,cm$^{-3}$.

\subsection{Radiative Lifetimes of the $\mathrm{3d^64s}$, $\mathrm{3d^7}$, and~$\mathrm{3d^54s^2}$ Levels \label{rdl_fe2}}

\begin{figure}[H]
\includegraphics[width=7.5 cm]{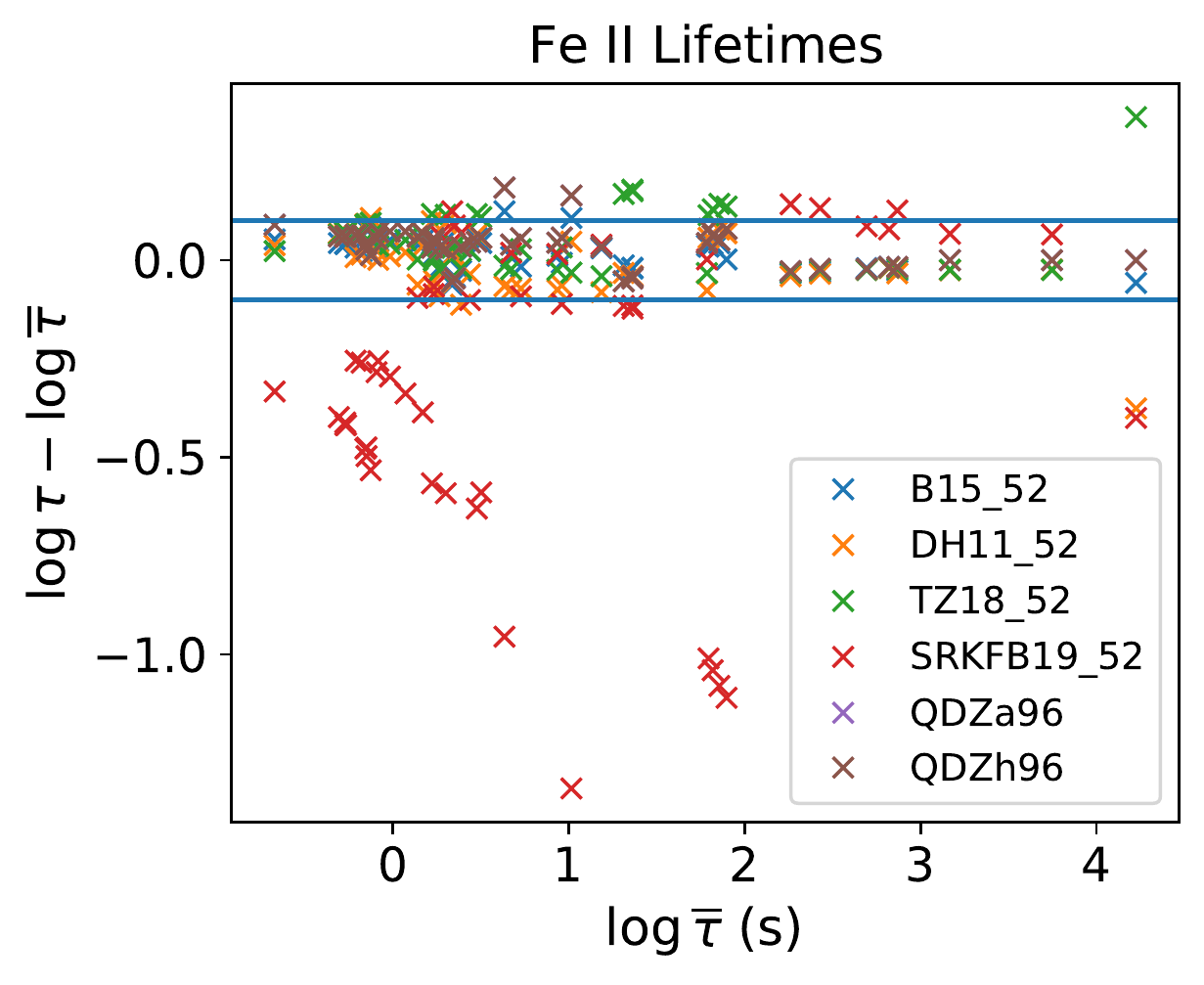}
\caption{[\feii] radiative lifetimes for levels within the $\mathrm{3d^64s}$, $\mathrm{3d^7}$, and~$\mathrm{3d^54s^2}$ configurations computed from the $A$-values in the \texttt{atom} datasets B15\_52, DH11\_52, TZ18\_52, SRKFB19\_52, QDZa96, and~QDZh96. For~each level, the~lifetime differences for these datasets with respect to the average value are plotted as a function of the average~value. \label{rdl_diff_fe2}}
\end{figure}

To compare the lifetimes of levels belonging to the $\mathrm{3d^64s}$, $\mathrm{3d^7}$, and~$\mathrm{3d^54s^2}$ configurations predicted by the \texttt{atom} datasets B15\_52, DH11\_52, TZ18\_52, SRKFB19\_52, QDZa96, and~QDZh96, we again determine an average lifetime for each level by comparing the respective differences for each dataset (see Figure~\ref{rdl_diff_fe2}). A~problematic and identifiable case in this plot is the lowest level of the first excited configuration, $\mathrm{3d^7\ a\,^4F_{9/2}}$, with~the longest lifetime ($\log \overline{\tau} = 4.22$~s) and large discrepant values ($\sim0.4$~dex). This is an important metastable level inasmuch as being dominant in the plasma radiative and collisional equilibrium and in opening the routes to continuum pumping. In~general, the scatter is $\sim 0.1$~dex, except for lifetimes derived from the $A$-values of the questionable atom\_SRKB19\_52~dataset. 

Lifetimes for a handful of metastable levels of [\feii] have been measured with a laser probing technique, and~$A$-values were derived using branching fractions from the spectra of the Eta Carinae ejecta taken with the \textit{Hubble Space Telescope}~\cite{ros01, har03, gur09}. A~comparison between the experimental lifetimes and those derived from the {\tt PyNeb} \texttt{atom} datasets is given in Table~\ref{lt_exp}. Lifetimes from atom\_SRKFB19\_52 are markedly discrepant from both the measured and other theoretical values and~are thus excluded from this table. The~theoretical lifetime of the $\mathrm{3d^64s\ b\,^2H_{11/2}}$ level shows a large scatter. Although~there is good agreement between theory and experiment for the $\mathrm{3d^54s^2\ a\,^6S_{5/2}}$ level, the~rest of the lifetimes from  B15\_52, DH11\_52, and~TZ18\_52 are generally longer (as long as 33\%) than experiment. The best overall agreement with the experiment ($\lesssim7\%$) is displayed by atom\_QDZh96.
In Table~\ref{lt_exp2}, we compare experimental and theoretical $A$-values. QDZh96 values are within the experimental error bars except for $A(\mathrm{a\,^4G_{9/2}, a\,^4F_{9/2}})$ and $A(\mathrm{b\,^4D_{7/2}, a\,^4F_{7/2}})$ for which the theoretical $A$-values agree to within 12\%. The~average agreement between the experiment and the rest of the theoretical data is around 20\%.

\subsection{Radiative Lifetimes of the $\mathrm{3d^6(^5D)4p}$ Odd-Parity Levels \label{rd2_fe2}}

Radiative lifetimes of the \feii\ odd-parity levels $\mathrm{3d^6(^5D)4p}$ have been measured with a time-resolved non-linear laser-induced fluorescence technique, whereby an improved signal-to-noise ratio reduces uncertainties to 2--3\%~\cite{sch04}. In~Table~\ref{lt_exp3}, we compare these measurements with lifetimes derived from $A$-values from atom\_TZ18\_173 and atom\_RKFB19\_173. Theoretical lifetimes agree to $\sim30\%$ but are,~on average, around 14\% shorter than measurements and well outside the aforementioned error~bars.

\begin{table}[t]
\caption{Comparison of [\feii] experimental lifetimes~(in s) with those derived from the revised {\tt PyNeb} \texttt{atom} datasets.  The~error of the least significant figure of the observed lifetime is indicated in~brackets.} \label{lt_exp}
	\begin{adjustwidth}{-\extralength}{0cm}
		\newcolumntype{C}{>{\centering\arraybackslash}X}
		\begin{tabularx}{\textwidth+\extralength}{CCCCCCCCC}
			\toprule
 &             \multicolumn{3}{c}{\textbf{Experiment}} & \multicolumn{5}{c}{\textbf{Theory}} \\
            \midrule
           \textbf{Level} & \textbf{Ref.~\cite{ros01}} & \textbf{Ref.~\cite{har03}} & \textbf{Ref.~\cite{gur09}}    &\textbf{B15\_52}&\textbf{DH11\_52}&\textbf{TZ18\_52}&
           \textbf{QDZa96}&\textbf{QDZh96}\\
            \midrule
$\mathrm{3d^54s^2\ a\,^6S_{5/2}}$      &0.23(3)&       &        &0.241&0.233&0.225&0.262&0.220\\
$\mathrm{3d^64s\ a\,^4G_{11/2}}$ &       &       &0.75(10)&0.909&0.960&0.934&0.774&0.706\\
$\mathrm{3d^64s\ a\,^4G_{9/2}}$  &       &0.65(2)&        &0.849&0.862&0.876&0.755&0.694\\
$\mathrm{3d^64s\ b\,^2H_{11/2}}$ &       &3.8(3) &        &5.31 &3.71 &4.18 &6.59 &5.20 \\
$\mathrm{3d^64s\ b\,^4D_{1/2}}$  &       &       &0.54(3) &0.592&0.627&0.632&0.616&0.550\\
$\mathrm{3d^64s\ b\,^4D_{7/2}}$  &0.53(3)&       &        &0.548&0.578&0.575&0.568&0.501\\
			\bottomrule
		\end{tabularx}
	\end{adjustwidth}
\end{table}


\begin{table}[t]
\caption{Comparison of experimental $A$-values (s$^{-1}$) for transitions in [\feii] with those listed in the revised {\tt PyNeb} \texttt{atom} datasets. The~error of the least significant figure of the observed $A$-value is indicated in~brackets.} \label{lt_exp2}
	\begin{adjustwidth}{-\extralength}{0cm}
		\newcolumntype{C}{>{\centering\arraybackslash}X}
		\begin{tabularx}{\textwidth+\extralength}{CCCCCCCCCC}
			\toprule
 &  &  &    \multicolumn{2}{c}{\textbf{Experiment}} & \multicolumn{5}{c}{\textbf{Theory}} \\
            \midrule
           \textbf{Upper} & \textbf{Lower} & \boldmath{$\lambda_\mathrm{air}$}\textbf{~(\AA)}& \textbf{Ref.~\cite{har03}} & \textbf{Ref.~\cite{gur09}}    &\textbf{B15\_52}&\textbf{DH11\_52}&\textbf{TZ18\_52}& \textbf{QDZa96}&\textbf{QDZh96}\\
            \midrule
$\mathrm{a\,^4G_{9/2}}$&$\mathrm{a\,^4F_{9/2}}$&4177.20&0.29(5)&        &0.16&0.14&0.14&0.18&0.19\\
                       &$\mathrm{a\,^4F_{7/2}}$&4276.84&0.83(7)&        &0.66&0.65&0.65&0.75&0.82\\
                       &$\mathrm{a\,^4F_{5/2}}$&4352.78&0.36(4)&        &0.31&0.31&0.29&0.35&0.38\\
$\mathrm{a\,^6S_{5/2}}$ &$\mathrm{a\,^6D_{9/2}}$&4287.39&1.53(22)&      &1.50&1.55&1.62&1.37&1.65\\
                       &$\mathrm{a\,^6D_{7/2}}$&4359.33&1.19(21)&       &1.11&1.15&1.19&1.02&1.22\\
                       &$\mathrm{a\,^6D_{5/2}}$&4413.78&0.84(13)&       &0.78&0.81&0.84&0.73&0.86\\
                       &$\mathrm{a\,^6D_{3/2}}$&4452.10&0.53(8)&        &0.50&0.52&0.54&0.47&0.55\\
                       &$\mathrm{a\,^6D_{1/2}}$&4474.90&0.26(4)&        &0.24&0.25&0.26&0.23&0.27\\
$\mathrm{b\,^4D_{7/2}}$ &$\mathrm{a\,^6D_{9/2}}$&3175.38&0.23(3)&       &0.21&0.24&0.21&0.19&0.21\\
                       &$\mathrm{a\,^4F_{9/2}}$&3376.20&0.96(10)&       &0.88&0.77&0.81&0.85&0.98\\
                       &$\mathrm{a\,^4F_{7/2}}$&3440.99&0.23(3)&        &0.30&0.27&0.28&0.29&0.33\\
                       &$\mathrm{a\,^4P_{5/2}}$&5551.31&0.18(4)&        &0.15&0.14&0.14&0.15&0.18\\
                       &$\mathrm{a\,^4P_{3/2}}$&5613.27&0.10(3)&        &0.078&0.069&0.075&0.081&0.093\\
$\mathrm{a\,^4G_{11/2}}$&$\mathrm{a\,^4F_{9/2}}$&4243.97&      &1.05(15)&0.85&0.81&0.84&1.02&1.12\\
                       &$\mathrm{a\,^4F_{7/2}}$&4346.85&       &0.25(5) &0.20&0.18&0.18&0.23&0.25\\
			\bottomrule
		\end{tabularx}
	\end{adjustwidth}
\end{table}

\begin{figure}[H]
\includegraphics[width=8.0 cm]{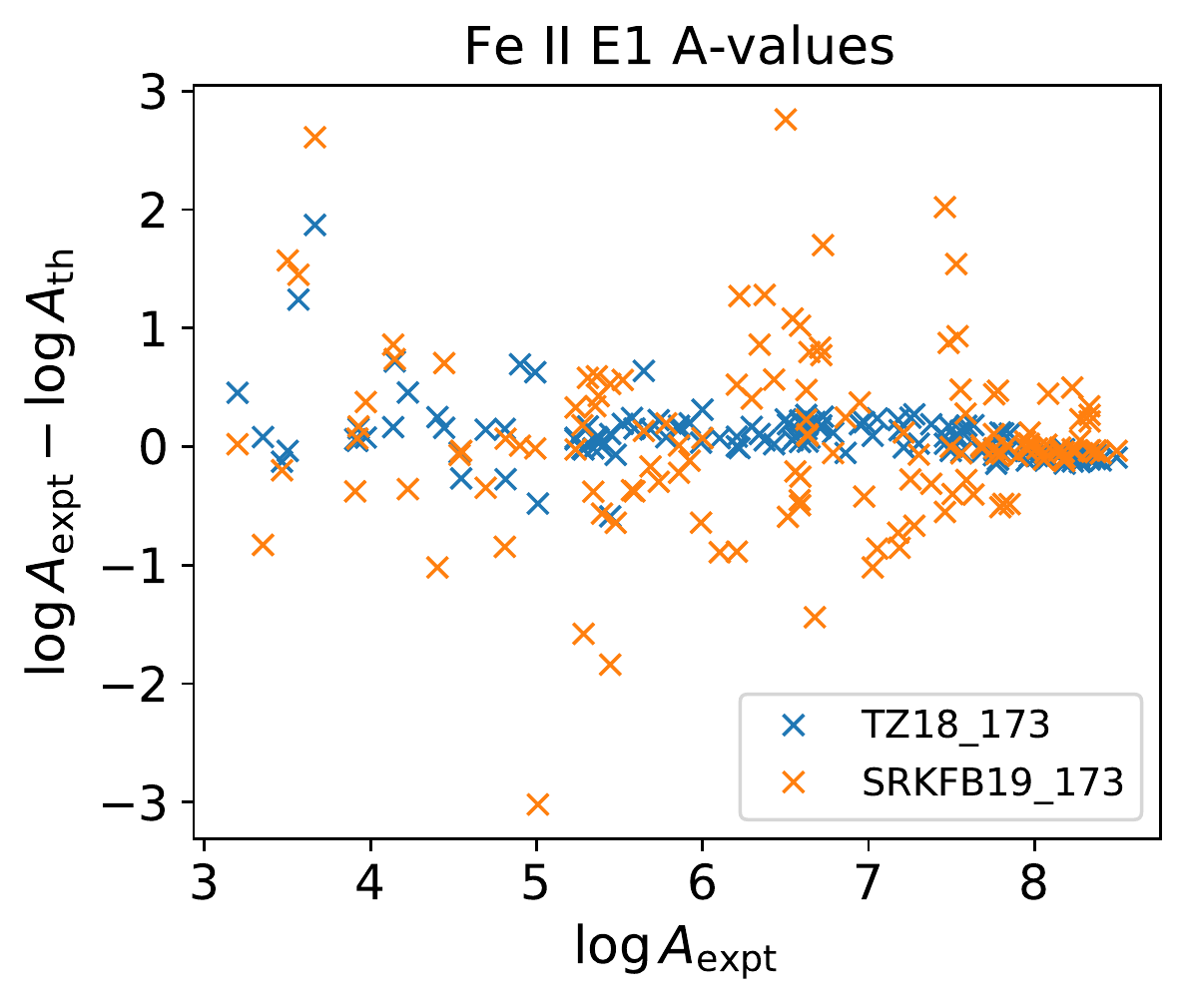}
\caption{Comparison of experimental $A$-values (s$^{-1}$) for E1 transitions arising from the $\mathrm{3d^6(^5D)4p}$ levels of \feii\ 
\cite{sch04} with those from the \texttt{atom} datasets TZ18\_173 and SRKFB19\_173. \label{A_expt_fe2}}
\end{figure}

Experimental E1 $A$-values were reported using the lifetimes in Table~\ref{lt_exp3} and branching fractions determined with a Fourier transform spectrometer and a high-resolution grating spectrometer to precisions of 6\% and 26\% for the strong and weak transitions, respectively~\cite{sch04}. We compare these measurements with $A$-values from atom\_TZ18\_173 and atom\_SRKFB19\_173 in Figure~\ref{A_expt_fe2}. $A$-values from the former dataset with $\log A>6$ agree with the experiment to within 0.3~dex, while large discrepancies (as large as 3~dex) are displayed by the~latter. 

\subsection{\feii\ $gf$-Values \label{gf_fe2}}

Oscillator strengths ($gf$-values) for 142 \feii\ lines ($\lambda\lambda 4000{-}8000$) have been derived from both laboratory and computed data and~benchmarked with accurate spectra from the Sun and metal-poor stars~\cite{mel09}. These $gf$-values have been compared with data computed with extensive CI using the \textsc{mcbp} code  {\sc civ3}, including fine-tuning~\cite{deb14}. This calculation included 262 fine-structure levels from the $\mathrm{3d^64s}$, $\mathrm{3d^7}$, $\mathrm{3d^54s^2}$, $\mathrm{3d^64p}$, and~$\mathrm{3d^54s4p}$ configurations. In~general, good agreement was found except for a few ill-behaved transitions susceptible to effects difficult to constrain computationally such as severe cancellation due to CI~mixing.

\begin{table}[H]
\caption{Comparison of  experimental lifetimes~(in ns) of the $\mathrm{3d^6(^5D)4p}$ levels of \feii\ with those derived from $A$-values listed in the revised {\tt PyNeb}  \texttt{atom} datasets. The~error of the least significant figure of the observed lifetime is indicated in~brackets. \label{lt_exp3}}
		\newcolumntype{C}{>{\centering\arraybackslash}X}
		\begin{tabularx}{\textwidth}{CCCCCCCCC}
			\toprule
          &\multicolumn{1}{c}{\textbf{Experiment}} & \multicolumn{2}{c}{\textbf{Theory}} \\
            \midrule
           \textbf{Level} & \textbf{Ref.~\cite{sch04}} &\textbf{TZ18\_173}& \textbf{SRKFB19\_173}\\
            \midrule
$\mathrm{z\,^6D^o_{9/2}}$ & 3.68(7)  &3.17&3.39\\
$\mathrm{z\,^6D^o_{7/2}}$ & 3.67(9)  &3.19&3.43\\
$\mathrm{z\,^6D^o_{5/2}}$ & 3.69(5)  &3.20&3.44\\
$\mathrm{z\,^6D^o_{3/2}}$ & 3.73(7)  &3.21&3.44\\
$\mathrm{z\,^6D^o_{1/2}}$ & 3.68(11) &3.21&3.45\\
$\mathrm{z\,^6F^o_{11/2}}$& 3.20(5)  &2.58&2.94\\
$\mathrm{z\,^6F^o_{9/2}}$ & 3.28(4)  &2.63&2.98\\
$\mathrm{z\,^6F^o_{7/2}}$ & 3.25(6)  &2.65&3.09\\
$\mathrm{z\,^6F^o_{5/2}}$ & 3.30(5)  &2.66&3.03\\
$\mathrm{z\,^6F^o_{3/2}}$ & 3.45(12) &2.67&3.04\\
$\mathrm{z\,^6P^o_{7/2}}$ & 3.71(4)  &2.93&3.53\\
$\mathrm{z\,^6P^o_{5/2}}$ & 3.75(10) &2.91&3.60\\
$\mathrm{z\,^6P^o_{3/2}}$ & 3.70(12) &2.89&3.57\\
$\mathrm{z\,^4D^o_{7/2}}$ & 2.97(4)  &2.74&2.10\\
$\mathrm{z\,^4D^o_{5/2}}$ & 2.90(6)  &2.76&2.15\\
$\mathrm{z\,^4D^o_{3/2}}$ & 2.91(9)  &2.74&2.13\\
$\mathrm{z\,^4F^o_{9/2}}$ & 3.72(10) &3.29&3.14\\
$\mathrm{z\,^4F^o_{7/2}}$ & 3.59(10) &3.14&3.00\\
$\mathrm{z\,^4F^o_{5/2}}$ & 3.55(8)  &3.16&2.99\\
$\mathrm{z\,^4P^o_{5/2}}$ & 3.27(6)  &3.20&2.74\\
$\mathrm{z\,^4P^o_{3/2}}$ & 3.23(9)  &3.21&2.55\\
			\bottomrule
		\end{tabularx}
\end{table}

\begin{figure}[H]
\includegraphics[width=8.0 cm]{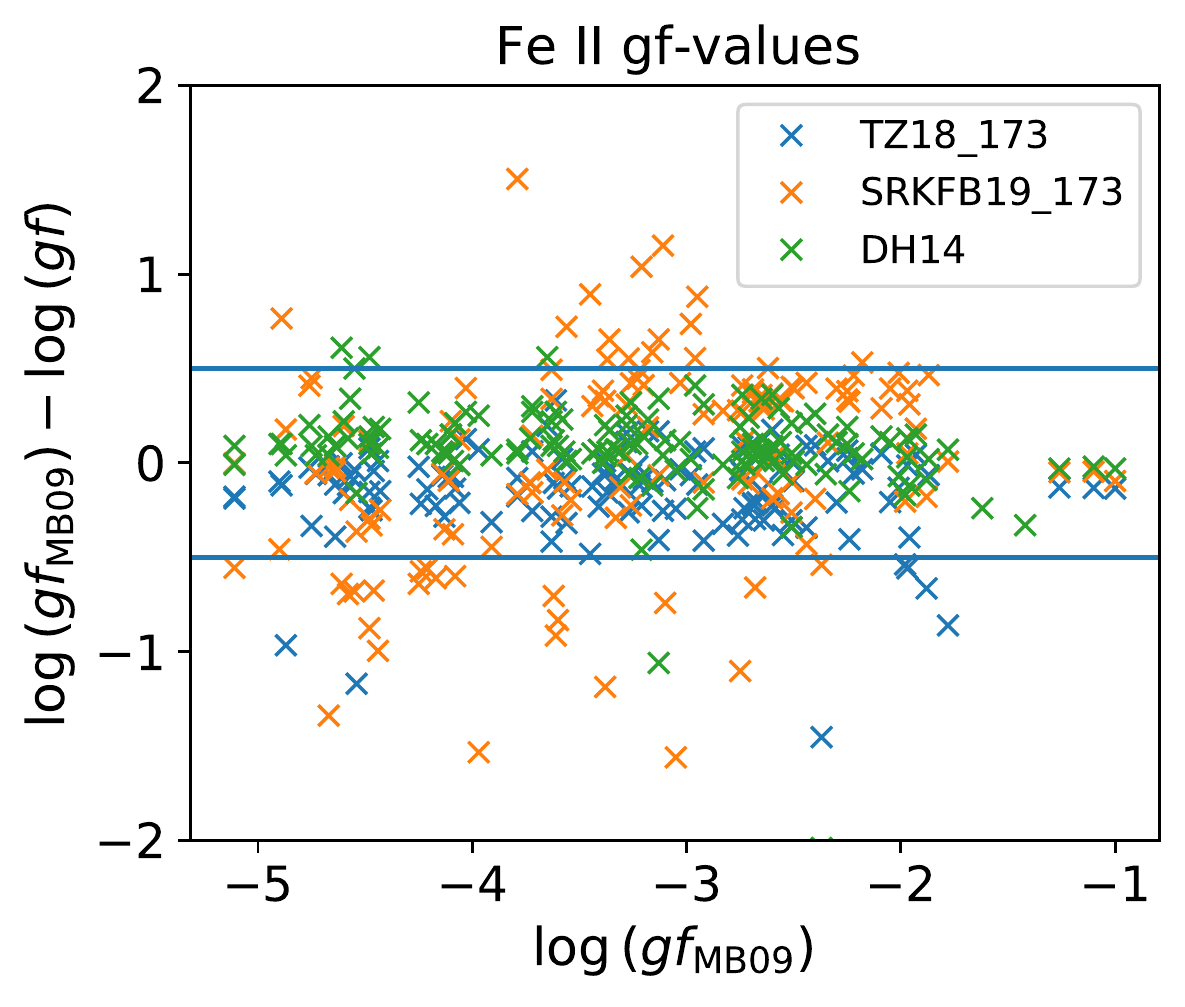}
\caption{Comparison of observed  $gf$-values for \feii\ (MB09)~\cite{mel09} with those calculated from atom\_TZ18\_173, atom\_SRKFB19\_173, and~Ref.~\cite{deb14} (DH14). \label{gf}}
\end{figure}

In Figure~\ref{gf}, we compare the observed $gf$-values (MB09)~\cite{mel09} with theoretical data from Ref.~\cite{deb14} (DH14), atom\_TZ18\_173, and~atom\_SRKFB19\_173. If~problematic transitions are excluded, the~average agreement of DH14 and atom\_TZ18\_173 with MB09 is within 0.2~dex. Dataset atom\_SRKFB19\_173, on~the other hand, contains several  transitions with discrepancies greater than 0.5~dex.

Table~\ref{gfproblem} shows three transitions for which DH14 shows sizable discrepancies with respect to MB09. Datasets atom\_TZ18\_173 and atom\_SRKFB19\_173 do not perform much better; thus, the~hindrance from cancellation effects in computational estimates put forward in Ref.~\cite{deb14} is reinforced. This table also lists three transitions we were not able to identify in atom\_TZ18\_173 or atom\_SRKFB19\_173 for which the agreement between DH14 and MB09 is~reasonable.

\begin{table}[H]
\caption{Problematic transitions in the $gf$-value comparison of MB09~\cite{mel09} with atom\_TZ18\_173, atom\_SRKFB19\_173, and~DH14~\cite{deb14} showing large discrepancies or unidentified~lines. \label{gfproblem}}
\newcolumntype{C}{>{\centering\arraybackslash}X}
\begin{tabularx}{\textwidth}{CCCCCCC}
\toprule
 & & &\multicolumn{1}{c}{\textbf{MB09}} & \textbf{TZ18\_173} & \textbf{SRKFB19\_173} & \textbf{DH14} \\
\midrule
Lower & Upper &$\lambda_\mathrm{air}$\,(\AA) & \multicolumn{4}{c}{$\log(gf)$}\\
\midrule
$\mathrm{3d^64p\ z\,^4F^o_{5/2}}$ & $\mathrm{3d^54s^2\  c\,^4D_{3/2}}$ &6508.13&$-$3.45&$-$3.93&$-$2.56&$-$7.90\\
$\mathrm{3d^64s\ c\,^4F_{9/2}}$   & $\mathrm{3d^64p\ x\,^4G^o_{9/2}}$  &6433.81&$-$2.37&$-$3.82&$-$2.91&$-$4.41\\
$\mathrm{3d^64p\ z\,^4F^o_{7/2}}$ & $\mathrm{3d^54s^2\ c\,^4D_{5/2}}$  &6371.13&$-$3.13&$-$3.54&$-$2.48&$-$4.19\\
  		                          &                                    &4635.32&$-$1.42&       &       &$-$1.75\\
 		                          &                                    &4625.89&$-$2.35&	   &       &$-$2.41\\
 		                          &                                    &4549.19&$-$1.62&       &	   &$-$1.86\\
\bottomrule
\end{tabularx}
\end{table}
\unskip

\subsection{\feii\ Branching~Fractions}\label{fe2_bf}

\begin{figure}[H]
\includegraphics[width=8.0 cm]{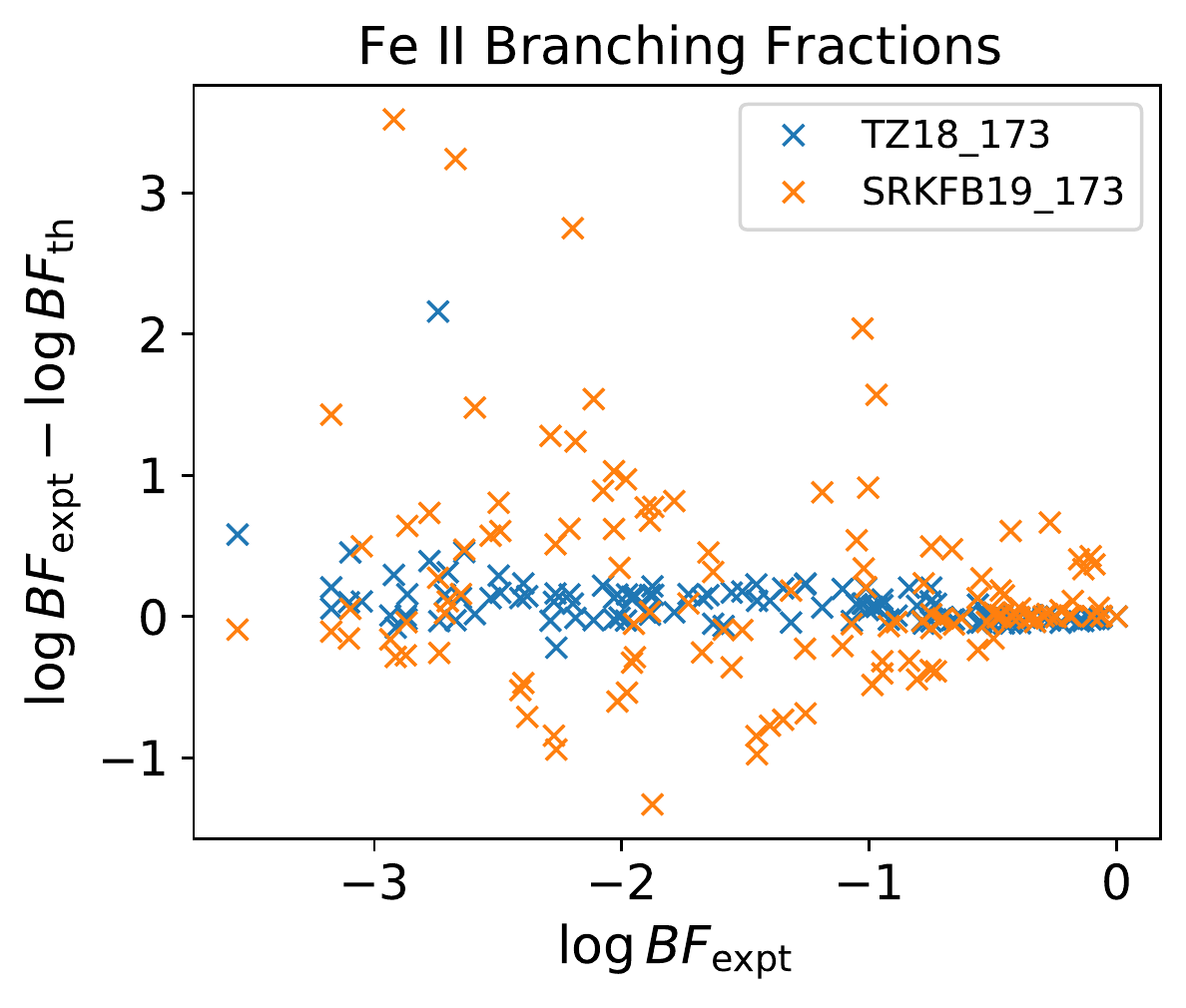}
\caption{Comparison of  \feii\ experimental branching fractions ($BF_\mathrm{expt}$)~\cite{den19} with theoretical values ($BF_\mathrm{th}$) derived from atom\_TZ18\_173 and atom\_SRKFB19\_173. \label{bfrac}}
\end{figure}

Branching fractions
\begin{equation}
BF_{ji} =\frac{I(j,i)}{\sum_i I(j,i)}=\frac{A(j,i)}{\sum_i A(j,i)}
\end{equation}
for 121  \feii\ UV lines originating from levels of the $\mathrm{3d^6(^5D)4p}$ configuration have been measured in low-current spectra from Fe--Ne and Fe--Ar hollow cathode discharge lamps. The~spectra were taken with a UW 3~m echelle spectrograph with a resolving power of 250,000~\cite{den19}. We compare these branching fractions in Figure~\ref{bfrac} with those derived from $A$-values of the atom\_TZ18\_173 and atom\_SRKFB19\_173 \texttt{atom} datasets. For~transitions with $\log(BF_\mathrm{expt})>-2$, the~branching fractions from atom\_TZ18\_173 agree with the experiment, on average, to better than 30\%, while significantly larger discrepancies are displayed by atom\_SRKFB19\_173. However, most theoretical branching fractions, in general, do not match the accuracy implicit in the experimental error~bars.

\unskip

\subsection{[\feii] Effective Collision~Strengths}\label{fe2_coll}

\begin{figure}[H]
\includegraphics[width=13.86 cm]{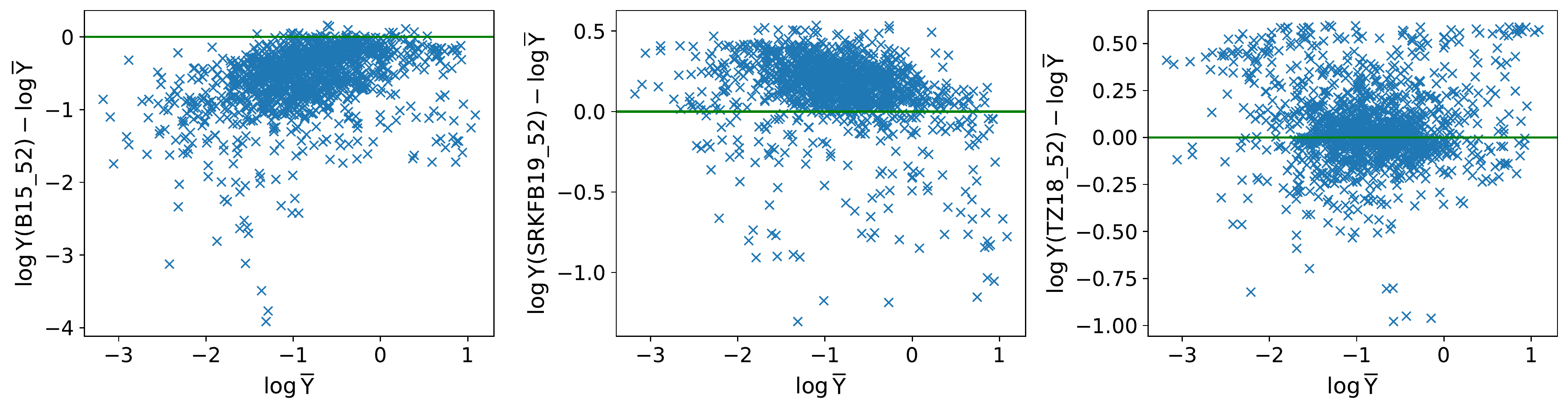}
\caption{[\feii] ECS logarithmic differences at $T=10^4$\,K for each transition in the \texttt{coll} datasets B15\_52, SRKFB19\_52, and~TZ18\_52 relative to the average~value. \label{fe2_ecs}}
\end{figure}

We have constructed three revised \texttt{coll} datasets containing ECS for the [\feii] 52-level atomic model of {\tt PyNeb}: B15\_52, TZ18\_52, and~SRKFB19\_52. We determine for each transition the average ECS from these three datasets at $T=10^4$~K, and~as depicted in Figure~\ref{fe2_ecs}, we compare the logarithmic differences with respect to this average.  ECS from coll\_B15\_52 appear to lie below average, displaying differences as large as $\Delta\log\Upsilon\approx -4$, while those from coll\_SRKFB19\_52 are above average with a scatter within 1.0~dex. The~most interesting result is coll\_TZ18\_52 with an ECS average well within 0.5~dex. This outcome supports the findings in the spectrum fits (see Table~\ref{new_data_fe2}), whereby the electron densities predicted by coll\_B15\_52 are higher, coll\_SRKFB19\_52 lower, and~coll\_TZ18\_52 in~between. 

\begin{figure}[H]
\includegraphics[width=13.86 cm]{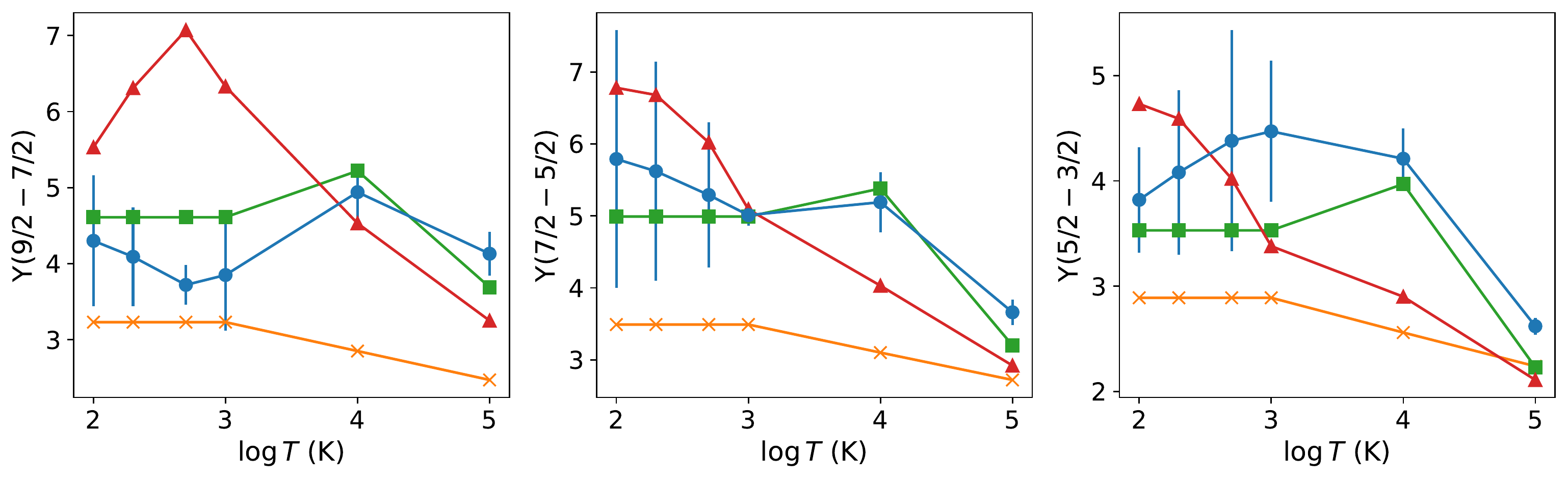}
\caption{ECS $\Upsilon(J-J')$ for the transitions $\mathrm{3d^6(^5D)4s\ a\,^6D}_J - \mathrm{a\,^6D}_{J'}$ within the ground term of [\feii]. Blue circles: Ref.~\cite{wan19}. Yellow crosses: coll\_B15\_52. Green squares: coll\_SRKFB19\_52. Red triangles: coll\_TZ18\_52. \label{fe2_ecs_ground}}
\end{figure}

Due to the diagnostic potential of [\feii] emission lines in interstellar-medium plasmas, accurate ECS at low temperatures were recently computed for transitions among levels of the $\mathrm{3d^6(^5D)4s\ a\,^6D}_J$ ground term~\cite{wan19}. To~obtain uncertainty estimates, different target models and relativistic $R$-matrix methods ({\sc bprm}, {\sc icft}, {\sc darc}) were used in these calculations. In~Figure~\ref{fe2_ecs_ground}, we compare selected ECS from~\cite{wan19} with those from \texttt{coll} datasets B15\_52, TZ18\_52, and~SRKFB19\_52. For~this purpose, we use the {\tt PyNeb} {\tt getOmega(T,j,i)} function that allows ECS linear interpolation at the input temperature \texttt{T}; if the latter lies outside the temperature range of the tabulated ECS, the~function lists the end value. This is the case of coll\_B15\_52 and coll\_SRKFB19\_52 for $\log T< 3$  (see Figure~\ref{fe2_ecs_ground}). As~expected, the~best agreement is with coll\_SRKFB19\_52, although there are points outside the error bars. Large discrepancies are also found with coll\_B15\_52 and coll\_TZ18\_52 at  higher temperatures ($\log T> 4$).

\unskip

\subsection{\feii\ Discussion}

In the implementation of new datasets for [\feii], we found term misassignments for levels with total orbital angular momentum quantum number $L=5$ in the atomic model of atom\_SRKFB19\_225 (see Table~\ref{srkfb19}). A~further comparison in Table~\ref{Htrans} of $A$-values for transitions arising from the $\mathrm{3d^64s\ a\,^4H_{J}}$ and $\mathrm{3d^64s\ b\,^2H_{J}}$ ($J=9/2$ and $11/2$) levels indicate large discrepancies with respect to other \texttt{atom} datasets (TZ18\_225, DH11\_52, B15\_52, QDZa96, and~QDZh96). This is an indication of faulty data that can be corroborated with the discrepancies displayed by dataset atom\_SRKFB19\_52 in the comparison of observed and computed $A$-value ratios carried out in Section~\ref{fe2_Aratio}. 

This comparison also brings out observed lines with questionable identifications (e.g., $\lambda\lambda 3968.27,4889.71$) and unreliable line-intensity ratios---e.g., $I(\lambda 7637.52)/I(\lambda 9051.95)$, $I(\lambda 7686.93)/\allowbreak I(\lambda 8891.93)$, and~$I(\lambda 7733.13)/I(\lambda 9033.49)$---due to $A$-values with magnitude $\log A\lesssim -3$. Theoretical $A$-value ratios agree to within 20--25\% and, with~respect to the observed line-intensity ratios, to~within the error bars except for some outliers subject to strong level mixing, as pointed out in~\cite{deb10a}.

We have found atomic data files, namely atom\_B15 and coll\_B15, with~faulty data caused again by the incorrect mapping to atomic models with different level numbering. This problem is exacerbated by data producers who use several multi-configuration methods implementing a variety of relativistic Hamiltonians (Dirac--Fock, Breit--Pauili, Pauli). This multi-code approach may enable accuracy estimates of the radiative and collisional data, but~it often leads to level structures with conflicting numbering and identifications that are difficult to remap. The~use of fine-tuning in atomic calculations to match the experimental level order does mitigate this~problem.

In contrast to [\feiii], for~which the spectrum fits using different atomic datasets resulted in acceptable temperature (6\%) and density (20\%) uncertainties thus sustaining their reliability, for~[\feii], the collisional datasets lead to temperature variations of around 15--20\% and density discrepancies as large as a factor of four. However, the~reasonable temperature and density accord between coll\_SRKFB19\_52 and coll\_VVKFHF99\_52 must be noted. These results appear to indicate that, for~[\feii], the~electron density is not a good fitting parameter or, as~previously proposed~\cite{rod99,ver00, har13}, fluorescence continuum pumping might be a level-populating mechanism we have not taken into account. Furthermore, the~extremely low critical density of level $\mathrm{3d^7\ a\,^4F_{9/2}}$ (see Figure~\ref{fe2_cd}) and its sensitivity to both the radiative and collisional datasets (Figure~\ref{fe2_cd2}) give this level a protagonistic role in the [\feii] spectrum formation, notwithstanding the poor accuracy of its small $A$-value.

The potential of line-ratio density diagnostics in [\feii] is compromised by the mismatch with the spectrum fits, and~the temperature diagnostics are discouraged by the looming role of fluorescence pumping. In~Section~\ref{fe2_lrdiag}, we use the $I(\lambda 8892)/I(\lambda 9267)$ ratio to illustrate a density diagnostic problem: the density variations in the range of interest ($3\leq\log n_e\leq 5$\,cm$^{-3}$) lie within the observed line-intensity error band. As~performed in Section~\ref{diagnostics_fe3} for the analysis of the [\feiii] density diagnostics, an~observational sample of low-density ionized nebulae with [\feii] detections would be helpful to constrain the predictions of both the radiative and collisional parameters. The~lack of such a sample in the literature prevents us from drawing further~conclusions.

We have performed extensive comparisons of computed and measured radiative data for both the $\mathrm{3d^64s}$, $\mathrm{3d^7}$, and~$\mathrm{3d^54s^2}$ even-parity levels and the $\mathrm{3d^64p}$ and $\mathrm{3d^54s4p}$ odd-parity levels. Regarding even-parity levels (see Section~\ref{rdl_fe2}), the~computed radiative lifetimes agree to within 0.1~dex except for the singular $\mathrm{3d^7\ a\,^4F_{9/2}}$ metastable level with $\log\tau\sim 4$\,s and an uncertainty of 0.4~dex. It must be noted that the $A$-values in atom\_SRKB19\_52 display questionable lifetimes for some of these levels (see Figure~\ref{rdl_diff_fe2}). The~best overall agreement between theory and experiment ($\sim7\%$)  is with the atom\_QDZh96 dataset, while other \texttt{atom} datasets (B15\_52, DH11\_52, and~TZ18\_52) in general yield longer (as long as 33\%) lifetimes than experiment. As~discussed in~\cite{deb10a}, the~large uncertainty of the $\mathrm{3d^64s\ b\,^2H_{11/2}}$ lifetime is caused by imprecise level mixing. $A$-values derived from the experimental lifetimes, on average, agree to around 20\% with those in the theoretical~datasets. 

Accurate (2--3\%) radiative lifetimes have been measured for the $\mathrm{3d^6(^5D)4p}$ odd-parity levels~\cite{sch04}, which have been compared in Section~\ref{rd2_fe2} with those derived from the atom\_TZ18\_173 and atom\_SRKFB19\_173 datasets. Theoretical lifetimes agree to within 30\%, but~on average they are around 14\% shorter than measurements. E1 $A$-values have also been derived from these lifetime measurements, and~those with $\log A > 6$ agree with atom\_TZ18\_173 to within 0.3~dex while larger discrepancies (as large as 3~dex) are found in atom\_SRKFB19\_173. Moreover, the~comparison of the observed $gf$-values~\cite{mel09} in Section~\ref{gf_fe2} with those derived from the theoretical datasets shows good agreement (0.2~dex) with atom\_TZ18\_173 and discrepancies larger than 0.5~dex with atom\_SRKFB19\_173. A~similar outcome is obtained in the comparison of measured branching ratios~\cite{den19} in Section~\ref{fe2_bf}, whereby those derived from $A$-values in atom\_TZ18\_173 with $\log(BF_\mathrm{expt})>-2$ agree to better than 30\%, while some from atom\_SRKFB19\_173 show larger discrepancies. In~general, the~theoretical radiative datasets do not reach anywhere near the accuracy quoted for the measured lifetimes, $A$-values, $gf$-values, and~branching~fractions.

The comparison of collisional datasets in Figure~\ref{fe2_ecs} of Section~\ref{fe2_coll} gives a macro-measure of the overall ECS magnitudes that has a direct impact on the critical densities. In~this comparison, we determined for each transition an average ECS using three \texttt{coll} datasets (B15\_52, SRKFB19\_52, and~TZ18\_52) and~then plotted the dispersion for each dataset. Figure~\ref{fe2_ecs} shows that the smallest dispersion (within 0.5~dex) is by coll\_TZ18\_52 while differences in coll\_B15\_52 and coll\_SRKFB19\_52 are larger and, respectively, mostly negative and positive. The~fairly large ECS discrepancies found in Figure~\ref{fe2_ecs_ground} for transitions within the $\mathrm{3d^6(^5D)4s\ a\,^6D}$ ground term at low temperatures gives a further indication of the patchy quality of the collisional data for \feii.

From the extensive data tests hereby carried out on the \feii\ datasets, we find the radiative data in atom\_SRKFB19\_52 to be faulty, and~this could indeed have an impact on the coll\_SRKFB19\_52 data. We would therefore select as the \texttt{PyNeb} 1.1.17 defaults the atom\_TZ18\_173 and coll\_ TZ18\_173~datasets.  




\section{Conclusions}\label{conc}

We have performed extensive data evaluations of the \texttt{levels}, \texttt{atom}, and~\texttt{coll} files for [\feiii] and [\feii] in \texttt{PyNeb} 1.1.16 to examine their worthiness in nebular spectral modeling and to select the package defaults. In~most cases, we have reconstructed the datasets from the sources and introduced newly published data, thus leading to a more reliable atomic database for these species in the new release \texttt{PyNeb} 1.1.17. 

The present data assessment comprises: fits of high-resolution spectra from the HH\,202S and HH\,204 Herbig--Haro objects taken with the Ultraviolet Visual Echelle Spectrograph~\cite{mes09,men21b}; comparisons of theoretical and observed line-intensity ratios; and comparisons with measurements of radiative lifetimes, $A$-values, $gf$-values, and~branching ratios. Observed lines may suffer from misidentifications, line blending, and~telluric or instrumental contamination. On~the other hand, the~computation of accurate radiative data for lowly ionized ions of the iron group with open $\mathrm{3d}^n$ shells is hindered by strong configuration mixing, double-excitation representations, and~subtle core--valence correlation involving closed and open subshells; thus, broadly speaking, we would not expect an $A$-value accuracy of better than 20\%. Similarly, the~ECS at low temperatures are affected by complex resonance structures sensitive to target models and channel convergence that seriously compromise accuracy~ratings.

For \feiii, we introduced a new \texttt{levels} file from the NIST database with more complete level assignments and accurate energies. The~previous version of the \texttt{levels} file is in the \texttt{deprecated} directory, renamed as \texttt{fe\_iii\_levels\_2023.dat}.
As described in Section~\ref{newdata_fe3}, we implemented the new \texttt{atom} files  Q96\_34, DH09\_34, BBQ10\_34, BB14\_34, and~BB14\_144 and the \texttt{coll} files Z96\_34, BBQ10\_34, BB14\_34, Z96\_144, and~BB14\_144. In~\texttt{PyNeb} 1.1.17 atom\_BBQ10\_34, atom\_BB14\_34, coll\_Z96\_34 are not included as they are, respectively, made redundant by atom\_BBQ10, atom\_BB14\_144, and~coll\_Z96\_144 while coll\_BB14 and coll\_BBQ10 are~deprecated.

A comparison of the observed line-intensity ratios  for transitions arising from the same upper level in [\feiii] with the corresponding $A$-value ratios constrains the overall accuracy of the latter to around 20--30\%. Spectrum fits with the new datasets led to temperatures and densities within 6\% and 20\%, respectively, validating their usefulness. Radiative lifetimes computed with these \texttt{atom} datasets agree to within 0.1~dex. A~revision of \texttt{coll} files brings out the incompleteness of the Z96\_34 dataset and its poor accuracy for ECS with $\log\Upsilon < -1$. For~[\feiii], we have selected the atom\_BB14\_144 and coll\_BB14\_144 datasets as the defaults in \texttt{PyNeb} 1.1.17.

For \feii, we have implemented the \texttt{atom} files VVKFHF99\_52, DH11\_52, B15\_52, TZ18\_52, SRKFB19\_52, TZ18\_173,  SRKFB19\_173 and \texttt{coll} files VVKFHF99\_52, B15\_52, TZ18\_52, SRKFB19\_52, TZ18\_173, and~SRKFB19\_173. After~extensive tests and much pondering, the~atom\_SRKFB19\_52 and atom\_SRKFB19\_173 datasets were found to contain faulty $A$-values, while atom\_VVKFHF99\_52 and coll\_VVKFHF99\_52 have missing data; therefore, these files are not included in \texttt{PyNeb} 1.1.17. Apart from this drawback, the~theoretical $A$-value ratios for transitions arising from a common upper level derived from the rest of the \texttt{atom} files agree to within 20--25\%, and most lie within the experimental error bars. The~spectrum fits yielded temperature and density variations of around 20\% and a factor of 4, respectively, which appear to indicate that the density is not a good fitting parameter and fluorescence continuum pumping might be a level-populating contributor not taken into account. However, the~poor theoretical characterization of the key level $\mathrm{3d^7\ a\,^4F_{9/2}}$, inasmuch as its inaccurate $A$-value and extremely low critical density sensitive to both the radiative and collisional datasets, deters us from more conclusive remarks. Extensive comparisons of the radiative data for both forbidden and allowed transitions with experiment bring out the computational difficulties to match the measured accuracy (better than 10\%); thus, a~global theoretical $A$-value accuracy of 20--30\% would be statistically sound. Finally, the~examined \texttt{coll}  datasets lead to different critical densities around $T_e=10^4$\,K, a~situation that is difficult~ to analyze. 

In  \texttt{PyNeb} 1.1.17, the~\texttt{atom} files VVKFHF99\_52, TZ18\_52 and \texttt{coll} files VVKFHF99\_52, SRKFB19\_52, SRKFB19\_173 are made redundant by atom\_VVKFHF99, atom\_TZ18\_173, \\   coll\_VVKFHF99, and~coll\_SRKFB19. Furthermore, atom\_B15, coll\_B15, atom\_VVKFHF99, coll\_VVKFHF99 are deprecated. We select atom\_TZ18\_173 and coll\_TZ18\_173 as~defaults. 

The status of all the data files for \feiii\ and \feii\ in \texttt{PyNeb} 1.1.16 and 1.1.17 are summarized in Table~\ref{tab:pyneb} of Appendix~\ref{A}. 

A final recommendation emerging from the present work must be put forward regarding the curation of the atomic databases associated with spectral modeling codes. The~potential of the latter in astrophysical analyses relies in great part on the integrity (accuracy and completeness) of their tabulations of radiative and collisional rates, which,  as shown here, demand detailed evaluation and upgrading. For~complex ionic spectra such as [\feiii] and [\feii], this may imply a lengthy~endeavor.


\vspace{6pt}




\authorcontributions{C.M. (Claudio Mendoza) conceptualized and coordinated the full paper. J.E.M.-D. and J.G.-R. provided the observational data and extensive discussions about their validity and provenance. M.A.B. provided data, performed calculations, and~led discussions about fluorescence continuum pumping in [\feii]. C.M. (Christophe Morisset) researched data, performed ancillary calculations, and~software support regarding the \texttt{PyNeb} package. Investigation, methodology, analysis, and~validation: C.M. (Claudio Mendoza), J.E.M.-D., J.G.-R., M.A.B., C.M. (Christophe Morisset). Writing and original draft preparation: C.M. (Claudio Mendoza), J.E.M.-D. Review and editing: C.M. (Claudio Mendoza), J.E.M.-D., C.M. (Christophe Morisset), J.G.-R., M.A.B. Visualization: C.M. (Claudio Mendoza), J.E.M.-D. All authors have read and agreed to the published version of the manuscript.}

\funding{J.E.M.-D. acknowledges funding from the Deutsche Forschungsgemeinschaft (DFG, German Research Foundation) in the form of an Emmy Noether Research Group (grant number KR4598/2-1, PI Kathryn Kreckel). Ch.M. acknowledges the support from UNAM/DGAPA/PAPIIT grant IN101220. J.G.-R. acknowledges support from an Advanced Fellowship under the Severo Ochoa excellence program CEX2019-000920-S; financial support from the Canarian Agency for Research, Innovation, and Information Society (ACIISI) of~the Canary Islands Government and~the European Regional Development Fund (ERDF) under~grant with reference ProID2021010074; and support under grant P/308614 financed by funds transferred from the Spanish Ministry of Science, Innovation, and Universities (MCIU) charged to the General State Budgets and with funds transferred from the General Budgets of the Autonomous Community of the Canary Islands by the MCIU.}

\dataavailability{The data files revised and compiled for this revision are publicly available in \texttt{PyNeb} 1.1.17.}

\acknowledgments{We acknowledge extensive private communications and data exchange with Catherine Ramsbottom (Queen's University Belfast, UK) regarding the radiative and collisional data for \feii. We are indebted to  Gra\.{z}yna Stasi\'nska (Observatoire de Paris, France) for data exchange and useful discussions.}

\conflictsofinterest{The authors declare no conflict of~interest.}


\abbreviations{Abbreviations}{
The following abbreviations are used in this manuscript:\\

\noindent
\begin{tabular}{@{}ll}
CDS & Centre de Donn\'ees astronomiques de Strasbourg \\
NIST & National Institute os Standards and Technology \\
\end{tabular}
}

\appendixtitles{no} 
\appendixstart
\appendix
\section[\appendixname~\thesection]{\label{A}}

\begin{table}[H]
\caption{Status of the \feiii\ and \feii\ atomic data files in the \texttt{PyNeb} library for versions 1.1.16 and 1.1.17. `X' means the file is present; `*' means the file is present and is the default; `D' means the file is deprecated (but may still be accessible with the command
\texttt{pn.atomicData.includeDeprecatedPath()}).}\label{tab:pyneb}
\newcolumntype{C}{>{\centering\arraybackslash}X}
\begin{tabularx}{\textwidth}{lcclcc}
\toprule
\textbf{\feiii\ File Name} & \textbf{v1.1.16} & \textbf{v1.1.17} & \textbf{\feii\ File Name} & \textbf{v1.1.16} & \textbf{v1.1.17}\\
\midrule
fe\_iii\_atom\_BB14\_144.dat &   & * & fe\_ii\_atom\_TZ18\_173.dat  &   & * \\
fe\_iii\_atom\_BBQ10.dat     & X & X & fe\_ii\_atom\_B15.dat        & * & D \\   
fe\_iii\_atom\_DH09\_34.dat  &   & X & fe\_ii\_atom\_B15\_52.dat    &   & X \\   
fe\_iii\_atom\_NP96.dat      & X & X & fe\_ii\_atom\_DH11\_52.dat   &   & X \\   
fe\_iii\_atom\_Q96\_34.dat   &   & X & fe\_ii\_atom\_SRKFB19.dat    & X & D \\   
fe\_iii\_atom\_Q96\_J00.dat  & * & X & fe\_ii\_atom\_VVKFHF99.dat   & X & D \\   
\midrule                                                                
fe\_iii\_coll\_BB14\_144.dat &   & * & fe\_ii\_coll\_TZ18\_173.dat  &   & * \\
fe\_iii\_coll\_BB14.dat      & X & D & fe\_ii\_coll\_B15.dat        & * & D \\   
fe\_iii\_coll\_BBQ10.dat     & X & D & fe\_ii\_coll\_B15\_52.dat    &   & X \\   
fe\_iii\_coll\_BBQ10\_34.dat &   & X & fe\_ii\_coll\_B15\_old.dat   & X & D \\   
fe\_iii\_coll\_Q96.dat       & X & X & fe\_ii\_coll\_SRKFB19.dat    & X & X \\   
fe\_iii\_coll\_Z96.dat       & * & X & fe\_ii\_coll\_VVKFHF99.dat   & X & D \\   
fe\_iii\_coll\_Z96\_144.dat  &   & X &                              &   &   \\
\bottomrule
\end{tabularx}
\end{table}

\begin{adjustwidth}{-\extralength}{0cm}
\printendnotes[custom]

\reftitle{References}

\PublishersNote{}
\end{adjustwidth}
\end{document}